\documentclass{ws-ijmpb}

\makeatletter
\def\leaderfill{\leaders\hbox to 1em{\hss.\hss}\hfill}
\newcommand\@pnumwidth{2.55em}
\newcommand\@tocrmarg{2.55em}
\newcommand\@dotsep{4.5}
\newcommand\tableofcontents{%
{\global
\@topnum\z@
\@afterindentfalse
\if@twocolumn
\@restonecoltrue
\onecolumn
\else
\@restonecolfalse
\fi
\vspace*{10pt}
\noindent
{\bf Contents}\par
\vskip1em
\nobreak}
{\small
\@starttoc{toc}%
}\if@restonecol
\twocolumn
\fi}
\newcommand*\l@section[2]{%
\ifnum \c@tocdepth >\z@
\addpenalty\@secpenalty
\setlength\@tempdima{1.4em}
\begingroup
\parindent \z@
\rightskip
\@pnumwidth \parfillskip -\@pnumwidth
\leavevmode
\advance\leftskip\@tempdima
\hskip -\leftskip
#1\nobreak\leaderfill\nobreak
\hb@xt@\@pnumwidth{\hss #2}\par
\endgroup
\fi}
\newcommand*\l@subsection{\@dottedtocline{2}{1.4em}{2.2em}}
\newcommand*\l@subsubsection{\@dottedtocline{2}{3.6em}{3em}}

\makeatother

\begin{document}

\markboth{K. Kasamatsu, M Tsubota and M. Ueda}
{Vortices in multicomponent BECs}

\catchline{}{}{}{}{}

\title{VORTICES IN MULTICOMPONENT BOSE-EINSTEIN CONDENSATES}

\author{KENICHI KASAMATSU}
\address{Department of General Education, Ishikawa National College of Technology, Tsubata, Ishikawa 929-0392, Japan\\
kenichi@ishikawa-nct.ac.jp}

\author{MAKOTO TSUBOTA}
\address{Department of Physics, Osaka City University, Sumiyoshi-Ku, Osaka 558-8585, Japan\\
tsubota@sci.osaka-cu.ac.jp}

\author{MASAHITO UEDA}
\address{Department of Physics, Tokyo Institute of Technology, Meguro-ku, Tokyo 152-8551, Japan\\
ueda@ap.titech.ac.jp}

\maketitle

\begin{history}
\received{Day Month Year}
\revised{Day Month Year}
\end{history}

\begin{abstract}
We review the topic of quantized vortices in multicomponent Bose-Einstein condensates of dilute atomic gases, with an emphasis on that in two-component condensates. First, we review the fundamental structure, stability and dynamics of a single vortex state in a slowly rotating two-component condensates. To understand recent experimental results, we use the coupled Gross-Pitaevskii equations and the generalized nonlinear sigma model. An axisymmetric vortex state, which was observed by the JILA group, can be regarded as a topologically trivial skyrmion in the pseudospin representation. The internal, coherent coupling between the two components breaks the axisymmetry of the vortex state, resulting in a stable vortex molecule (a meron pair). We also mention unconventional vortex states and monopole excitations in a spin-1 Bose-Einstein condensate. Next, we discuss a rich variety of vortex states realized in rapidly rotating two-component Bose-Einstein condensates. We introduce a phase diagram with axes of rotation frequency and the intercomponent coupling strength. This phase diagram reveals unconventional vortex states such as a square lattice, a double-core lattice, vortex stripes and vortex sheets, all of which are in an experimentally accessible parameter regime. The coherent coupling leads to an effective attractive interaction between two components, providing not only a promising candidate to tune the intercomponent interaction to study the rich vortex phases but also a new regime to explore vortex states consisting of vortex molecules characterized by anisotropic vorticity. A recent experiment by the JILA group vindicated the formation of a square vortex lattice in this system. 
\end{abstract}

\keywords{Bose-Einstein condensation; superfluidity; quantized vortex; vortex lattice; binary system; rotation; spinor condensate; spin texture; relative phase}

\tableofcontents


\section{Introduction}
Due to their being a common feature of fluid motion, vortices appear in diverse fields of science. In a superfluid, that is, a fluid without viscosity, the flow has quantized circulation which originates in quantized vortices\cite{Doneley}. A quantized vortex is commonly viewed as a topological defect of the order parameter in Bose-Einstein condensation. Since the gradient of the phase of the order parameter is proportional to the superfluid velocity, quantized vortices play a crucial role in understanding superfluidity. Moreover, the structure of the order parameters in multicomponent Bose-Einstein condensates, hereafter BECs, are analogous to other studies including superfluid $^{3}$He\cite{Salomaa}, unconventional superconductors\cite{Joynt}, quantum Hall systems\cite{Girvin}, nonlinear optics\cite{Chen}, nuclear physics\cite{Skyrmi}, and cosmology\cite{cosmo}. Therefore, the investigation of unconventional topological defects in such multicomponent BECs is of great importance and of broad interest.

While quantized vortices have been extensively studied in the field of superfluid helium\cite{Doneley}, they have seen a remarkable resurgence of interest since the realization of Bose-Einstein condensation in dilute alkali-atomic gases\cite{Pethickbook,Pitaevskiibook}. This Bose-condensed system has some distinct advantages for the study of quantized vortices. First, the BECs of alkali-atomic gases are so dilute that the interatomic interaction can be accurately parametrized in terms of the $s$-wave scattering length. Experimentally, the diluteness leads to a relatively large healing length that characterizes the vortex core size, thus enabling us to directly visualize the quantized vortices by optical techniques. Also, because the gas is dilute, the Gross-Pitaevskii equation gives an accurate description of static and dynamic properties of the atomic condensate within a mean-field approximation. This situation differs from superfluid $^4$He in which the high density and the strong interaction complicate the theoretical treatments. Second, rapidly developing techniques of atomic, molecular and optical physics allow us to carry out quantum-state engineering of the condensates in a well-controlled manner, thereby opening a new direction for the study of coherent quantum phenomena. For example, since alkali atoms have internal degrees of freedom attributed to the hyperfine spin, an external field can be used to induce conversions between spin states and thus it can be used to control the spatial variation of the wave functions. Moreover, a field-induced Feshbach resonance allows one to tune of the atom-atom interaction and create of a quasi-bound molecular state\cite{Duine}. 

The first experimental detection of a vortex in a dilute alkali-atomic gas BEC was made by Matthews {\it et al.} in 1999\cite{Matthews}. Their study involves condensates of $^{87}$Rb atoms residing in two hyperfine states. The generation of a vortex was achieved by a phase imprinting technique proposed by Williams and Holland\cite{Williams}, in which the interconversion between two components was controlled spatially and temporally by an external coupling field. After turning off the coupling, the vortex state is formed by one circulating component surrounding a nonrotating core of the other component. Leanhardt {\it et al.}\cite{Leanhardt} used ``topological phase imprinting"\cite{Nakahara} to create a vortex in a spin-1 BEC. These two methods use internal degrees of freedom of the condensate to create vortices, but a simple way to create vortices is to stir a condensate mechanically with a rotating ``bucket". Following the latter method, the ENS group observed the formation of a single vortex and multiple vortices in a single-component elongated condensate\cite{Madison}. The condensate was trapped in a static axisymmetric magnetic trap which was deformed by a nonaxisymmetric attractive dipole potential created with stirring laser beams. This combined potential produces a cigar-shaped harmonic trap with a slightly anisotropic transverse profile. By rotating the orientation of the transverse anisotropy at frequency $\Omega$, they observed the dynamic formation of a vortex above a certain critical value of $\Omega$. A lattice involving more vortices appeared when $\Omega$ was further increased. In contrast to indirect visualization methods used in superfluid helium systems, the quantized vortices were directly visualized as hollows in the transverse density profile of the time-of-flight image. Later, groups at MIT\cite{Abo}, JILA\cite{Haljan}, and Oxford\cite{Hodby} observed vortex lattices with the first two groups succeeding in creating 100 or more vortices. 

Theoretical studies mainly used the mean-field Gross-Pitaevskii theory to describe the main features of the vortex states (see Ref. \refcite{Fetterrev} for a review), and several predictions have been shown to agree with experiments. Some of the important problems are concerned with the equilibrium properties of a single vortex, including its structures and dynamics\cite{Feder}$^{-}$\cite{Rosenbusch}, the critical frequency and the surface-mode instability for vortex nucleation\cite{Ishosima}$^{-}$\cite{Kawaja1}, and the nonlinear dynamics of vortex lattice formation\cite{Madison2}$^{-}$\cite{Lobo}. Finally, experiments on a rapidly rotating BEC\cite{Engels}$^{-}$\cite{Bretin2} have motived theoretical studies of collective oscillations of a vortex lattice\cite{Baym}$^{-}$\cite{TPSimula}, new vortex phases in an anharmonic potential\cite{Fetteranh}$^{-}$\cite{Saito}, and a strongly-correlated fractal quantum Hall phase via quantum melting of a vortex lattice\cite{Cooper}$^{-}$\cite{UWEFI}. 

Another important issue in vortex physics is to elucidate the vortex phase in multicomponent BECs. Multicomponent order parameters allow the formation of various unconventional topological defects with complex properties that arise from interactions between different order-parameter components\cite{Pismen}. Since it is possible to load and cool atoms in more than one hyperfine spin state or more than one atomic element in the same trap, multicomponent condensates can be realized experimentally\cite{Myatt}$^{-}$\cite{Modugno2}. Such systems offer an ideal testing ground for the study of such topological defects; moreover the microscopic characterization of these systems should in principle be possible because this system is free from impurities and well controlled by optical techniques. 

In this paper we review recent studies on quantized vortices in multicomponent BECs, especially focusing on those in {\it two-component BECs}. To the best of our knowledge, there are no reviews of two-component BECs despite the large number of studies of this system\cite{Ho2}$^{-}$\cite{Savage}. Also, while a number of interesting phenomena have been reported on vortices of a spin-1 BEC, we mention this system only briefly and would like to refer the reader to references \refcite{Ohmi}-\refcite{Savage2} for more details. 

This paper is organized as follows. After a brief introduction of quantized vortices in a single-component BEC in Sec. \ref{singlevor}, we formulate in Sec. \ref{basic} a mean-field theory that describes two-component atomic-gas BECs and give a brief review on the ground state structure without vortices. In Sec. \ref{slow}, after describing experiments on the creation and observation of vortices in two-component BECs\cite{Matthews}, we discuss the static structure and dynamic properties of the single-vortex state in two-component BECs in a slowly rotating regime. By introducing the ``pseudospin" that describes the two-component BECs, we discuss the properties of topologically ``trivial" and ``nontrivial" spin textures. In particular, we predict a vortex molecule in a two-component BEC with an internal coherent coupling. We also mention briefly unconventional vortex states in a spin-1 BEC. In Sec. \ref{fast}, we consider the vortex states of two-component BECs under a rapid rotation. Combination of an analytical approach and numerical simulations reveals a rich variety of vortex states, which is not found in a single-component BEC. We also discuss the effect of the internal coherent coupling on the structure of vortex states. Finally, we conclude this review article in Sec. \ref{conclusion}.

\section{What is a quantized vortex?}\label{singlevor}
Below a critical temperature in an ideal Bose gas, a finite fraction of the particles occupies the same single-particle ground state and forms a BEC. When the particles interact, the single-particle density matrix $n^{(1)}({\bf r},{\bf r}') = \langle \hat{\Psi}^{\dagger}({\bf r}) \hat{\Psi}({\bf r}) \rangle$ is still defined, where $\hat{\Psi}^{\dagger}$ ($\hat{\Psi}$) is the boson creation (anihilation) operator. When a single-particle state is occupied by a macroscopic number of particles, $n^{(1)}({\bf r},{\bf r}') $ has a macroscopic eigenvalue. We may define a condensate wave function $\Psi({\bf r},t)$ as the corresponding eigenstates\cite{Onsager}. In the absence of interactions this wave function reduces to that of the macroscopically occupied single particle state. When a BEC is held in an external potential $V({\bf r})$, the dynamics of $\Psi({\bf r},t)$ is described by the Gross-Pitaevskii (GP) equation\cite{Pethickbook,Pitaevskiibook}
\begin{equation}
i \hbar \frac{\partial \Psi({\bf r},t)}{\partial t} = \biggl( - \frac{\hbar ^2}{2m}\nabla^2 + V({\bf r}) +g |\Psi({\bf r},t)|^{2} \biggr) \Psi({\bf r},t),
\label{gpequa}
\end{equation}
where $g=4\pi \hbar^2 a/m$ represents the strength of interaction characterized by the $s$-wave scattering length $a$ and particle mass $m$. The condensate wave function is normalized by the total particle number $N$ as $\int d^{3}r |\Psi|^{2} = N$. Writing $\Psi({\bf r},t) = \sqrt{\rho ({\bf r},t)} \exp (i \theta ({\bf r},t))$, the absolute squared amplitude $\rho({\bf r},t)=|\Psi({\bf r},t)|^{2}$ gives the condensate density and the gradient of the phase $\theta({\bf r},t)$ gives the superfluid velocity ${\bf v}_s({\bf r},t) = (\hbar/m) \nabla \theta({\bf r},t)$. As a result, the circulation of ${\bf v}_s$ around a closed path ${\cal C}$ in the fluid is quantized as
\begin{equation}
\oint_{\cal C} d{\bf s} \cdot {\bf v}_s = \frac{\hbar}{m} \oint_{\cal C} d{\bf s} \cdot \nabla \theta
= n \kappa  \quad (n=0, \pm 1, \pm 2, \ldots) 
\label{circuletion}
\end{equation}
with the quantum of circulation $\kappa =h/m$. Such a vortex with a quantized circulation is called a ``quantized vortex".

Let us review the fundamental properties of quantized vortices. Consider a single straight vortex line along the symmetry axis of a cylindrical vessel that can rotate at frequency ${\bf \Omega} = \Omega \hat{\bf z}$. In this case, the velocity field of the vortex line is 
\begin{equation}
{\bf v}_{\rm s} = \frac{n \kappa}{2 \pi r} \hat{\theta}, 
\end{equation}
where $r$ is the radial distance from the line and $\theta$ is the azimuthal angle. The corresponding ``vorticity" is 
\begin{equation}
\nabla \times {\bf v}_{\rm s} = \kappa \delta^{(2)}({\bf r}_{\perp}) \hat{\bf z}. \label{vorticirtyt}
\end{equation}
There is a singularity on the vortex line ($r=0$), where $v_{\rm s}$ diverges. Near $r=0$ the superfluid density is significantly supressed, forming the vortex core. In the case of superfluid helum, the size of the vortex core $r_{c}$ is of the order of the interatomic distance, while in a dilute gas this size equals the healing length $r_{c} \simeq \xi$, where 
\begin{equation}
\xi \equiv \frac{\hbar}{\sqrt{2 m g \rho}}. 
\end{equation}
The angular momentum $L_{z}$ of the fluid due to the vortex line is  
\begin{equation}
L_{z} = \int \rho v_{\rm s} r d {\bf r} = n \bar{\rho} \frac{\kappa L R^{2}}{2}, 
\label{angleu}
\end{equation}
where $L$ is the height of the vessel, $R$ is its radius, and $\bar{\rho}$ is the value of the superfluid density far from the core. In deriving the last equality we assumed $r_{c} \ll R$. 

The energy associated with the vortex line is dominated by the kinetic energy term. A straight calculation yields
\begin{equation}
E_{\rm v} = \int \frac{1}{2} \rho v_{\rm s}^{2} d {\bf r} \simeq \frac{\bar{\rho}}{2} L \int^{R}_{r_{c}} v_{\rm s}^{2} 2 \pi r d r = n^{2} L \frac{\bar{\rho} \kappa^{2}}{4 \pi} \ln \left( \frac{R}{r_{c}} \right). \label{onevorenrgy}
\end{equation} 
Since $E_{\rm v} \propto n^{2}$, vortices with $n>1$ are energetically unfavorable. That is, the energy cost to create one $n=2$ vortex is higher than that to create two $n=1$ vortices, even though they give the same angular momentum for $R \gg 1$ ($L_{z} \propto n$). Therefore, a stable quantized vortex always has $n=1$; the disintegration of a multiply quantized vortex has been intensely studied\cite{Puspil}$^{-}$\cite{Kawaspil} and it was observed experimentally by Shin {\it et al.} in an atomic-gas BEC\cite{Shinspil}.  

Using Eqs. (\ref{angleu}) and (\ref{onevorenrgy}) one can easily find the critical frequency of rotation $\Omega_{c}$ for the existence of an energetically-stable vortex line. In a frame rotating with ${\bf \Omega}$, the free energy is given by $F=E-\Omega L_{z}$ and the presence of a vortex increases $E$ by $E_{\rm v}$. For a vortex with $n=1$, the critical frequency is obtained by imposing that the change in the energy $\Omega L_{z}$ due to the creation of the vortex be equal to $E_{\rm v}$ as
\begin{equation}
\Omega_{c} = \frac{E_{\rm v}}{L_{z}} = \frac{\kappa}{2 \pi R^{2}} \ln \left( \frac{R}{r_{c}} \right).
\end{equation}
For $\Omega > \Omega_{c}$ the free energy of the system with one vortex is lower than that without a vortex. When the rotation frequency is significantly higher than $\Omega_{c}$, more vortices will appear in the form of a triangular lattice of vortices. For very large $\Omega$ the rotation of the superfluid will then look similar to the one of a rigid body characterized by $\nabla \times {\bf v}_{\rm s} = 2 {\bf \Omega}$. Using Eq. (\ref{vorticirtyt}), one finds that the average vorticity per unit area is given by $\nabla \times {\bf v}_{\rm s} = \kappa \bar{n}_{v} \hat{\bf z}$, where $\bar{n}_{v}$ is the number of vortices per unit area, so that the density of vortices is related to the rotation frequency $\Omega$ as 
\begin{equation}
\bar{n}_{v} = \frac{2 \Omega}{\kappa}.
\end{equation}
This relation, often referred to as Feynman's rule\cite{Feynman}, can be used to determine the maximum possible number of vortices in a given area as a function of $\Omega$.  

The above formalism is extended to quantized vortices in a trapped BEC by including the explicit form of the trapping potential $V({\bf r})$. This potential leads to an inhomogeneous condensate density, which plays an important role in the stability and dynamical behavior of the vortices. As an example, consider the structure of a single vortex in a condensate trapped by the axisymmetric harmonic potential $V(r,z)=m \omega_{\perp}^{2} (r^{2}+\lambda^{2}z^{2})/2$ with the aspect ratio $\lambda=\omega_{z}/\omega_{\perp}$. The equilibrium state is determined by the time-independent GP equation
\begin{equation}
\left(-\frac{\hbar^{2}}{2m}\nabla^{2} + V({\bf r}) + g |\Psi_{0} ({\bf r})| ^{2}\right) \Psi_{0}({\bf r}) = \mu \Psi_{0}({\bf r}), \label{singletimeindGP}
\end{equation}
where the time dependence of $\Psi_{0}$ is fixed by the chemical potential $\mu$ as $\Psi({\bf r},t) = \Psi_{0}({\bf r}) e^{-i \mu t/\hbar}$. The value of $\mu$ is determined by the normalization condition $\int |\Psi_{0}({\bf r})|^{2} = N $. The condensate wave function with a singly-quantized vortex line located along the $z$-axis takes the form $\Psi_{0}({\bf r}) = f(r,z) e^{i\theta}$ with the cylindrical coordinate $(r,\theta,z)$, where $f$ is a real function that is related to the condensate density as $\rho_{0} = f^{2}$. Then, Eq. (\ref{singletimeindGP}) becomes 
\begin{equation}
\left[ -\frac{\hbar^{2}}{2m} \left( \frac{\partial^{2}}{\partial r^{2}} + \frac{1}{r} \frac{\partial}{\partial r} + \frac{\partial^{2}}{\partial z^{2}} \right) + \frac{\hbar^{2}}{2mr^{2}} + V(r,z) +g \rho_{0} \right] f
= \mu f. \label{singletimeindGPrz}
\end{equation}
Here, the centrifugal term $\hbar^{2}/2mr^{2}$ arises from the azimuthal motion of the condensate which gives rise to a kinetic energy density $\rho_{0} m v_{\rm s}^{2} /2 = \rho_{0} \hbar^{2} /2mr^{2}$. 

The trapping potential defines the characteristic length scale $b_{\rm ho}=\sqrt{\hbar/m\omega_{\perp}}$. Without an interparticle interaction ($g=0$), the scale $b_{\rm ho}$ determines the spatial extent of the wave function. An increase in the repulsive atomic interaction expands the condensate, so that the kinetic energy associated with the density variation can be made negligible compared with the trap energy and the interaction energy. The ratio of the interaction energy to the harmonic-oscillator energy $g \rho_{0}/\hbar \omega_{\perp} \sim N a/b_{\rm ho}$ serves to quantify the effect of the interaction. In the limit $N a /b_{\rm ho} \gg 1$, which is applicable to current experiments on trapped BECs, one can omit the terms involving derivatives with respect to $r$ and $z$ in Eq. (\ref{singletimeindGPrz}). This approximation is called the Thomas-Fermi approximation\cite{Baym3}, and the density profile for the vortex state is then given by
\begin{eqnarray}
\rho_{0}(r,z) &=& \frac{1}{g} \left( \mu - \frac{ m \omega_{\perp}^{2}}{2} (r^{2} + \lambda^{2} z^{2}) - \frac{\hbar^{2}}{2mr^{2}} \right) \Theta \left( \mu - \frac{m \omega_{\perp}^{2}}{2} (r^{2} + \lambda^{2} z^{2}) - \frac{\hbar^{2}}{2mr^{2}} \right) \nonumber \\
 &=& \rho_{0}(0) \left( 1 - \frac{r^{2}}{R_{\perp}^{2}} - \frac{z^{2}}{R_{z}^{2}} - \frac{\xi(0)^{2}}{r^{2}} \right) \Theta \left( 1 - \frac{r^{2}}{R_{\perp}^{2}} - \frac{z^{2}}{R_{z}^{2}} - \frac{\xi(0)^{2}}{r^{2}}  \right), \label{Thoamvor}
\end{eqnarray}
where $\Theta(x)$ is a step function and $\rho_{0}(0)$ is the density at the center of the {\it vortex-free} Thomas-Fermi profile. We define the Thomas-Fermi radius $R_{\perp} = \sqrt{2 \mu / m \omega_{\perp}^{2}}$ and $R_{z} = \sqrt{2 \mu / m \omega_{z}^{2}}$ and the healing length $\xi(0) = \sqrt{\hbar^{2}/2mg\rho_{0}(0)}$. Due to the centrifugal term $\xi(0)^{2}/r^{2}$, the condensate density vanishes at the center out to a distance of order $\xi$, whereas the density in the outer region has the form of an upward-oriented parabola. The Thomas-Fermi approximation is valid in the region $\xi \ll r< R_{\perp}$. The asymptotic form of $\rho_{0}(r)$ for $r \ll \xi$ is\cite{JacksonAdam} $\rho_{0}(r) = \rho_{0}(0) (r/\xi(0))^{2}$ whose interpolation to Eq. (\ref{Thoamvor}) gives 
\begin{equation}
\rho_{0}(r,z) \simeq \rho_{0}(0) \left( \frac{r^{2}}{r^{2} + \xi^{2}} \right) \left( 1 - \frac{r^{2}}{R_{\perp}^{2}} - \frac{z^{2}}{R_{z}^{2}} \right) \Theta \left( 1 - \frac{r^{2}}{R_{\perp}^{2}} - \frac{z^{2}}{R_{z}^{2}} \right). \label{Thoamvorimprove}
\end{equation}
This form agrees fairly well with the profile of the single vortex state obtained by the numerical calculation of the GP equation\cite{JacksonAdam}.

For a detailed account of the single vortex state, we refer to a review article by Fetter and Svidzinsky\cite{Fetterrev} (see also Chap.9 in Ref.\refcite{Pethickbook} and Chap.14 in Ref.\refcite{Pitaevskiibook}). 

\section{Formulation of two-component Bose-Einstein condensates}\label{basic}
\subsection{Coupled Gross-Pitaevskii equations}
We now consider two-component BECs as described by the condensate wave functions for the two hyperfine spin states $\Psi_{1}$ and $\Psi_{2}$\cite{Matthews,Myatt,Hall1}. These wave functions can also describe a system of two atomic species\cite{Modugno2,Riboli,Mudrich} or two isotopes of an atomic element. The theory for the latter case is described in Refs. \refcite{Burke} and \refcite{Shchesnovich}, and experiments are described in Ref. \refcite{IBloch}. In addition, the condensates are confined by trapping potentials $V_{i}({\bf r})$ ($i=1,2$), which is assumed to rotate at a rotation frequency $\Omega$ about the $z$ axis as ${\bf \Omega}=\Omega \bf{\hat{z}}$. Viewed from the rotating frame, the GP energy functional of the two-component BEC is\cite{Pethickbook,Pitaevskiibook}
\begin{eqnarray}
E[\Psi_{1},\Psi_{2}] = \int d^{3} r \biggl[ \sum_{i=1,2} \Psi_{i}^{\ast} \biggl( - \frac{\hbar^{2} \nabla^{2}}{2m_{i}} + V_{i}  -\Omega L_{z} + \frac{g_{i}}{2} |\Psi_{i}|^{2} \biggr) \Psi_{i} + g_{12} |\Psi_{1}|^{2} |\Psi_{2}|^{2} \biggr]. \nonumber \\ 
\label{energyfunctio2}
\end{eqnarray}
Here, $m_{i}$ is the mass of the $i$-th atom and $\Omega L_{z} = i \hbar \Omega (y \partial_{x} -x \partial_{y})$ is a centrifugal term. The coefficients $g_{i}$ ($i=1,2$) and $g_{12}$ represent the atom-atom interaction between atoms iin the same hyperfine state, and atoms in different hyperfine states, respectively. They are expressed in terms of the corresponding $s$-wave scattering lengths $a_{1}$, $a_{2}$, and $a_{12}$ as 
\begin{eqnarray}
g_{i} = \frac{4 \pi \hbar^{2} a_{i}}{m_{i}},  \\
g_{12} = \frac{2 \pi \hbar^{2} a_{12}}{m_{12}},
\end{eqnarray}
where $m_{12}^{-1}=m_{1}^{-1}+m_{2}^{-1}$ is the reduced mass. 

When the trapping potential $V_{i}({\bf r})$  is produced by a magnetic field, each component has its own potential energy because the potential depends on a magnetic moment of the atom. For example, the trapping potential for atoms with the hyperfine states $| F_{i},M_{Fi} \rangle$ is expressed as a function of the magnetic field $|B({\bf r})|$ as\cite{Riboli}
\begin{eqnarray}
V_{i}({\bf r}) &=&  \mu_{B} G_{i} M_{F i} |B({\bf r})| 
\simeq \mu_{B} G_{i} M_{F_{i}} \biggl[ B_{0} + \frac{1}{2} (K_{x}x^{2}+K_{y}y^{2}+K_{z}z^{2}) \biggr] \nonumber \\
 & \equiv & V_{i}^{0} + \frac{1}{2} m_{i} ( \omega_{ix}^{2}  x^{2} + \omega_{iy}^{2}  y^{2} + \omega_{iz}^{2}  z^{2} ), 
\label{trap2compda}
\end{eqnarray}
where $\mu_{B}$ is the Bohr magneton, $G_{i}$ is the $g$-factor of the $i$-th atom, $V_{i}^{0}=\mu_{B} G_{i} M_{F i} B_{0}$, and $\omega_{ik}$ ($k=x,y,z$) is the trapping frequency satisfying the relation
\begin{equation}
m_{i}\omega_{ik}^{2} = \mu_{B} G_{i} M_{F i} K_{k}.
\label{massfre}
\end{equation} 
The potential minima $V_{i}^{0}$ for the BECs with different values of $G_{i} M_{F i}$ do not coincide. This positional shift is also influenced by a gravitational effect, by the difference of nuclear magnetic moments, and by nonlinearity in the Zeeman shift\cite{Myatt,Hall1}. If the potential is created by a pure optical laser\cite{Stenger,Barrett,Schmaljohann,Chang}, all atoms share the same harmonic potential. 

The time-dependent coupled GP equations for two-component BECs can be obtained by using a variational procedure $i \hbar \partial \Psi_{i} = \delta E / \delta \Psi_{i}^{\ast}$ as 
\begin{eqnarray}
i \hbar \frac{\partial \Psi_{1}}{\partial t} = -\frac{\hbar^{2}}{2m_{1}} \nabla^{2} \Psi_{1} + V_{1}({\bf r}) \Psi_{1} + g_{1}|\Psi_{1}|^{2}\Psi_{1}+g_{12}|\Psi_{2}|^{2}\Psi_{1} -\Omega L_{z} \Psi_{1}, \label{bingp1td}\\
i \hbar \frac{\partial \Psi_{2}}{\partial t} = -\frac{\hbar^{2}}{2m_{2}} \nabla^{2} \Psi_{2} + V_{2}({\bf r}) \Psi_{2} + g_{2}|\Psi_{2}|^{2}\Psi_{2}+g_{12}|\Psi_{1}|^{2}\Psi_{2} -\Omega L_{z} \Psi_{2}. \label{bingp2td}
\end{eqnarray}
Substituting $\Psi_{i} ({\bf r},t) = \Psi_{i}({\bf r}) e^{-i \mu_{i} t / \hbar}$ yields the time-independent coupled GP equations 
\begin{eqnarray}
-\frac{\hbar^{2}}{2m_{1}} \nabla^{2} \Psi_{1} + V_{1}({\bf r}) \Psi_{1} + g_{1}|\Psi_{1}|^{2}\Psi_{1}+g_{12}|\Psi_{2}|^{2}\Psi_{1} -\Omega L_{z} \Psi_{1} =\mu_{1} \Psi_{1}, \label{bingp1}\\
-\frac{\hbar^{2}}{2m_{2}} \nabla^{2} \Psi_{2} + V_{2}({\bf r}) \Psi_{2} + g_{2}|\Psi_{2}|^{2}\Psi_{2}+g_{12}|\Psi_{1}|^{2}\Psi_{2} -\Omega L_{z} \Psi_{2} =\mu_{2} \Psi_{2}, \label{bingp2}
\end{eqnarray}
where we have introduced the Lagrange multiplier $\mu_{i}$ which represents the chemical potential and is determined so as to satisfy the conservation of particle number $N_{i}= \int d^{3} r |\Psi_{i}|^{2}$. In the two-component BECs, each component interacts with the other through the intercomponent mean-field coupling $\propto g_{12}$, which yields structures and dynamics not found in a single-component BEC.

\subsection{Ground state structure}
It is instructive to start with the ground state of two-component BECs without rotation ($\Omega=0$). The equilibrium solutions of the coupled GP equations (\ref{bingp1}) and (\ref{bingp2}) show a rich variety of stable structures\cite{Ho2,Shchesnovich,Riboli,Esry2}$^{-}$\cite{Svidzinsky}, depending on the various parameters of the system. In particular, the intercomponent interaction $g_{12}$ plays an important role in determining the structure of the ground state; when the intercomponent interaction is strongly repulsive, the two components phase separate\cite{Tim,Ao}. The qualitative features of the ground state for a trapped two-component BEC can be understood by using the Thomas-Fermi approximation\cite{Ho2,Riboli,Trippenbach} in which one neglects the quantum pressure terms $(\hbar^{2}/ 2m_{i}\sqrt{\rho_{i}}) \nabla^{2} \sqrt{\rho_{i}} $ with the condensate density $\rho_{i} = |\Psi_{i}|^{2}$ in Eqs. (\ref{bingp1}) and (\ref{bingp2}). First consider the simple case $G_{1}M_{F_{1}}=G_{2} M_{F_{2}}$ in Eq. (\ref{massfre}) (i.e., $m_{1} \omega_{1k}^{2} = m_{2} \omega_{2k}^{2}$). The equilibrium density distributions $\rho_{i}^{(0)} $ in the spatial region where the two components coexist ($\rho_{1}^{(0)} \neq 0$ and $\rho_{2}^{(0)} \neq 0$) are 
\begin{eqnarray}
\rho_{1}^{(0)} = \frac{g_{2} \mu_{1} -g_{12} \mu_{2}-g_{2} V_{1} + g_{12} V_{2}}{g_{1}g_{2} - g_{12}^{2}} 
= \frac{1}{g_{1} \Gamma_{12}} \biggl[ \mu_{1} - \frac{g_{12}}{g_{2}} \mu_{2} - \Gamma_{2} V_{1}({\bf r}) \biggr],  \label{2compTFden1} \\
\rho_{2}^{(0)} = \frac{g_{1} \mu_{2} -g_{12} \mu_{1}-g_{1} V_{2} + g_{12} V_{1}}{g_{1}g_{2} - g_{12}^{2}} 
= \frac{1}{g_{2} \Gamma_{12}} \biggl[ \mu_{2} - \frac{g_{12}}{g_{1}} \mu_{1} - \Gamma_{1} V_{2}({\bf r}) \biggr],
\label{2compTFden2}
\end{eqnarray}
where, we defined the following dimensionless parameters 
\begin{eqnarray}
\Gamma_{12} &\equiv& 1-\frac{m_{1}+m_{2}}{4m_{12}} \frac{a_{12}^{2}}{a_{1}a_{2}}, \label{condito12}\\
\Gamma_{1} &\equiv& 1-\frac{m_{1}}{2m_{12}} \frac{a_{12}}{a_{1}}, \\
\Gamma_{2} &\equiv& 1-\frac{m_{2}}{2m_{12}} \frac{a_{12}}{a_{2}}. 
\label{condito}
\end{eqnarray}
For the two components to coexist, the right-hand sides of Eqs. (\ref{2compTFden1}) and (\ref{2compTFden2}) must be positive. Then, the parameter $\Gamma_{12}$ must be positive, that is, 
\begin{equation}
g_{1}g_{2}-g_{12}^{2}>0. 
\end{equation}
The coexisting region is given by the region delimited by the surface defined as the solution to the equation $\mu_{i} - g_{12} \mu_{j} / g_{j}  - \Gamma_{j} V_{i}({\bf r})=0$\cite{Riboli}. Otherwise, that is when
\begin{equation}
g_{1}g_{2}-g_{12}^{2}<0, \label{sepcond}
\end{equation}
there is no coexisting region, which implies that the two components phase-separate. 

For $\Gamma_{12}>0$, there are two possible density profiles that depend on the signs of the parameters $\Gamma_{1}$ and $\Gamma_{2}$. These signs determine the curvature of the density profile at the center of the trapping potential because $\Gamma_{1}$ and $\Gamma_{2}$ are the coefficients of the $x^{2}$, $y^{2}$ and $z^{2}$ terms in Eqs. (\ref{2compTFden1}) and (\ref{2compTFden2}). For $\Gamma_{1}>0$ and $\Gamma_{2}>0$, both density profiles are written by upward-convex parabolas in the coexisting region. For $\Gamma_{1}>0$ and $\Gamma_{2}<0$ ($\Gamma_{1}<0$ and $\Gamma_{2}>0$), one density profile is an upward-oriented parabola, but the other is a usual parabola. In the peripheral region in which the density of one component vanishes, the other component follows the single-component Thomas-Fermi profile $\rho_{i}^{(0)}=(\mu_{i}-V_{i}({\bf r}))/g_{i}$. This profile is connected to the density profile of Eqs. (\ref{2compTFden1}) and (\ref{2compTFden2}) by adjusting the chemical potential $\mu_{i}$ ($i=1$ or $2$) subject to the normalization condition $\int d^{3} r \rho_{i}^{(0)} = N_{i}$\cite{Ho2}. 

For $\Gamma_{12}<0$ the two components phase-separate as the coexisting regions do not exist. In the region in which only one density is nonvanishing ($\rho_{i}^{(0)} \neq 0$ and $\rho_{k}^{(0)} = 0$, for $i \neq k$), the density profile is given by the single-component Thomas-Fermi profile $\rho_{i}^{(0)}=(\mu_{i}-V_{i}({\bf r}))/g_{i}$. Since the Thomas-Fermi approximation fails to describe the domain boundary region, where the quantum pressure term cannot be neglected, numerical analysis of Eqs. (\ref{bingp1}) and (\ref{bingp2}) is necessary to determine the explicit density profile in this regime\cite{Pu}. The numerical calculation shows that there are two characteristic configurations of the phase-separated two-component BEC in the trapping potential. One configuration consists of a core of one component at the center of the trap and a shell of the other around the one component, a structure that preserves the spatial symmetry of the trapping potential\cite{Ho2,Pu}. The other configuration breaks the spatial symmetry by displacing the center of each condensate from the center of the trapping potential\cite{Ohberg1,Esry}. Whether the ground state takes the former or the latter configuration depends not only on the intercomponent interaction but also on the particle number, intracomponent interaction, and the shape of the trapping potential\cite{Ohberg3,KasamatsuMQT,Svidzinsky}. 

Compared to single-component BECs, two-component BECs exhibit a rich variety of ground state structures. These structures and their stability depend upon various parameters, particularly on the condition of the phase separation given by Eq. (\ref{sepcond}). Two-component mixtures of BECs with different spin states have been created in the laboratory for the systems of $^{87}$Rb (Refs. \refcite{Myatt,Hall1}) and $^{23}$Na (Ref. \refcite{Miesner}). For the case of $^{87}$Rb, the nearly-equal scattering lengths ($a_{1} = 5.67$ nm, $a_{2} = 5.34$ nm, and $a_{12} = 5.50$ nm) and the small relative displacement of the trapping minima lead to a weak separation between the two components. In contrast, for the $^{23}$Na-mixture, the value $a_{1}=2.75$ nm, $a_{2}=2.65$ nm, and $a_{12}=2.75$ nm causes a spin domain to form\cite{Stenger,Miesner}. Current experiments are aiming to create two-component BECs with two atomic species\cite{Mudrich,Modugnopre}; a mixture of the $^{41}$K and $^{87}$Rb BEC has been observed\cite{Modugno2}. In this case, the intercomponent scattering length $a_{12}$ may take values different from the intracomponent ones $a_{1}$ or $a_{2}$. Furthermore, we should note that an additional possibility to change the value of $a_{12}$ by the use of the Feshbach resonance\cite{Burke,Simoni}$^{-}$\cite{Inouye}. 

\section{Vortex states in slowly rotating two-component BECs}\label{slow}
\subsection{Creation of a quantized vortex}\label{experim}
Although most theoretical studies on quantized vortices have focused on single-condensate systems, the first observation of a quantized vortex in a gaseous BEC was achieved in a two-component system\cite{Matthews,Williams}. In the JILA experiment\cite{Matthews}, the vortex was created in a two-component BEC consisting of $^{87}$Rb atoms with the hyperfine spin states $|F=1,M_{F}=-1 \rangle \equiv |1 \rangle$ and $|F=2,M_{F}=1 \rangle \equiv |2 \rangle$. Since the scattering lengths between atoms of $ |1 \rangle$ and $|1 \rangle$, $ |2 \rangle$ and $|2 \rangle$, and $|1 \rangle$ and $|2 \rangle$ are all different, the two states are not equivalent, and the two-component condensate is characterized by the two-component order parameter. 

The advantage of using the $|1\rangle$ and $|2\rangle$ states of $^{87}$Rb atoms to perform quantum state engineering is that, first, their magnetic moments are nearly equal, so that they can be confined simultaneously in almost identical magnetic potentials. This is advantageous to investigate the intrinsic interaction phenomena between two condensates. Second, nearly equal s-wave scattering lengths for two $^{87}$Rb atoms in $| 1 \rangle$ and $| 2 \rangle$ result in the fortuitously small inelastic spin-exchange rate\cite{Julienne}, which provides a stable binary system with a long lifetime. In contrast, the inelastic rate is very large in $^{23}$Na atoms. Third, we can use a microwave rf-pulse to change the populations of atoms from the $| 1 \rangle$ state to the $| 2 \rangle$ state\cite{Matthews,Hall2,Matthews3}. If a microwave is applied continuously, this system is referred to as an ``internal" Josephson effect because the coexisting components are coupled through their internal degrees of freedom\cite{Williams2}$^{-}$\cite{Leggett}. Thus, this system offers new possibilities not only for quantum-mechanical state engineering, but also for exploring new vortex structures in coherently coupled multicomponent BECs as described in Sec.\ref{meronsect}.  

In experiments at JILA experiments\cite{Hall1}, atoms are initially trapped in one state, say, the $|1\rangle$ state, and then cooled below the Bose-Einstein transition temperature. When the atoms in the $|1\rangle$ state forms a condensate, a two-photon microwave field is applied, inducing transitions between the $|1\rangle $ state and the $|2\rangle$ state. As a result, the atoms undergo coherent oscillations between the two hyperfine levels with an effective Rabi frequency $\omega_{\rm R}^{\rm eff}=\sqrt{\omega_{\rm R}^{2}+\Delta^{2}}$, where the Rabi frequency $\omega_{\rm R}$ gives the rate at which population would oscillate between the two states when $\Delta=0$ and depends on the microwave power and the atom-field coupling constant. The detuning parameter $\Delta$ denotes the difference between the frequency of the coupling rf field and the frequency difference between the two internal atomic states. 

The effect of internal coupling can be formulated by introducing the energy density  
\begin{equation}
E_{\rm C}=-\hbar \omega_{\rm R} (\Psi_{2}^{\ast} \Psi_{1} e^{-i \Delta t} + \Psi_{1}^{\ast} \Psi_{2} e^{i \Delta t}), \label{intjosenerg}
\end{equation}
which allows atoms to change their internal state coherently. Since the microwave coupling fields are time dependent, it is useful to use the frame in which the field is constant (a frame rotating with the laser field). Then, the coupled time-dependent GP equations read 
\begin{eqnarray}
 i\hbar \frac{\partial \Psi_{1}}{\partial t} = \left( -\frac{\hbar^{2} \nabla^{2}}{2m} + V_{1} + g_{1} |\Psi_{1}|^2 + g_{12} |\Psi_{2}|^2 + \frac{\hbar \Delta}{2} \right) \Psi_{1} - \hbar \omega_{\rm R} \Psi_{2}, \label{2TDCGP1} \\
 i\hbar \frac{\partial \Psi_{2}}{\partial t} = \left( -\frac{\hbar^{2} \nabla^{2}}{2m} + V_{2} + g_{2} |\Psi_{2}|^2 + g_{12} |\Psi_{1}|^2 - \frac{\hbar \Delta}{2} \right) \Psi_{2} - \hbar \omega_{\rm R} \Psi_{1} \label{2TDCGP2}.
\end{eqnarray}
The last terms in Eqs. (\ref{2TDCGP1}) and (\ref{2TDCGP2}) cause interconversions between the components. For a homogenous system without $V_{i}$ but having uniform phases of the condensate wave functions for both components, the interconversion takes place at the same rate everywhere. However, the time variation of the spatially inhomogeous potential $V_{i}$ changes the way the interconversion is made, which is a key ingredient of the phase imprinting method for vortex creation\cite{Williams}.  

To proceed, let us assume that the external potential consists of a radially isotropic harmonic potential and a time-dependent perturbation $V_{1}({\bf r}, t)=m \omega_{\perp}^{2} r^{2}/2 + \hat{H}_{1}$ and $V_{2}({\bf r}, t)=m \omega_{\perp}^{2} r^{2}/2 - \hat{H}_{1}$ with  
\begin{equation}
\hat H_{1}=\kappa [f({\bf r})\cos (\Omega' t)+g({\bf r})\sin (\Omega' t)],
\label{m5}
\end{equation}
where $\kappa$ characterizes the strength of the perturbation and $f({\bf r})$ and $g({\bf r})$ describe the spatial dependence of  the perturbation. The form of $\hat H_{1}$ can be determined from the symmetry of the quantum state to be prepared. To create a vortex state with one unit of angular momentum, one can take $\kappa = m \omega_{\perp}^2 r_{0}$, $f({\bf r})=x$, and $g({\bf r})=y$ in Cartesian coordinates\cite{Williams}. This form of perturbation effectively confines the two hyperfine states in separate axisymmetric harmonic potentials with the same trap frequency $\omega_{\perp}$. The trap centers are offset spatially in the $x$-$y$ plane by a distance $r_{0}$ from the center and rotate about the symmetry axis at angular frequency $\Omega'$. 

The underlying physics for creating a vortex can be understood by considering the co-rotating frame with the trap centers at the angular frequency $\Omega'$ of the perturbation $\hat{H}_{1}$. In this frame, the energy of a vortex with one unit of angular momentum is shifted by $\hbar \Omega'$ relative to its value in the laboratory frame. When this energy shift is compensated for by the sum of the detuning energy $\hbar \Delta$ and the small chemical potential difference between vortex and non-vortex states, a resonant transfer of population can occur. From an initial nonrotating component $|1\rangle$, the $|2\rangle$ component is transferred to the state with unit angular momentum by controlling precisely the time of turning off of the coupling drive; this scheme was confirmed to work by numerical simulation of Eqs. (\ref{2TDCGP1}) and (\ref{2TDCGP2})\cite{Williams}. A typical structure shows that the $|2\rangle$ component has a vortex with a single quantized circulation at the center, whereas the nonrotating $|1\rangle$ component is located at the center and provides a ``pinning" potential that stabilizes the vortex core [see Figs. \ref{JILAvortexdensity} and \ref{axisymvor}]. 

\begin{figure}[btp]
\begin{center}
\includegraphics[height=0.28\textheight]{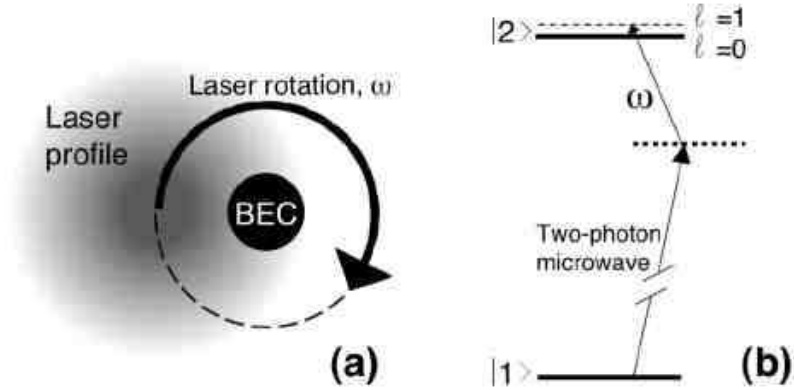}
\end{center}
\caption{(a) The experimental technique used to create a vortex at JILA. An off-resonant laser provides a rotating gradient in the AC Stark shift across the condensate as a microwave that causes detuning is applied. (b) The level diagram shows the microwave transition to very near the $| 2 \rangle$ state and the modulation due to the laser rotation frequency that couples only to the angular momentum $\ell=1$ state when $\omega=\Delta$ plus a small chemical potential difference between the vortex $(\ell=1)$ and nonvortex $(\ell=0)$ state, where $\omega$ corresponds to $\Omega'$ in the text. In the figure, the energy splitting ($< 1$ Hz) between the $l=1$ and $l=0$ states is exaggerated. [Matthews {\it et al}., Phys. Rev. Lett. {\bf 83} 2498 (1999), reproduced with permission. Copyright (1999) by the American Physical Society.]}
\label{JILAschem}
\end{figure}
Following the above method, Matthews {\it et al.} created a vortex for the first time in trapped BECs\cite{Matthews}. To realize the above configuration, they used a microwave and a movable laser beam with a spatially inhomogeneous profile on the condensate\cite{Matthews}. The laser beam was rotated around the initial condensate as in Fig. \ref{JILAschem}(a), which reproduces the perturbation given by Eq. (\ref{m5}). The resonant transfer of population from the nonrotating condensate into the vortex state was accomplished by detuning the microwave frequency and rotating the laser beam at an appropriate frequency for the resonant coupling. As shown in Fig. \ref{JILAschem}(b) for large detunings, the energy resonance condition means that atoms can change only the internal state through the coupling of the time-varying perturbation, and therefore must follow any selection rule that the spatial symmetry of that perturbation might impose. It should be emphasized that it is not simply a mechanical force of the optical field that excites the vortex. If one changes the sign of detuning $\Delta$ at a fixed trap rotation frequency, a vortex with opposite circulation will be created. Vortices with opposite circulations experience opposite energy shifts in transforming to the rotating frame, and therefore they must be detuned in an opposite sense to obtain resonant coupling.

\begin{figure}[btp]
\begin{center}
\includegraphics[height=0.45\textheight]{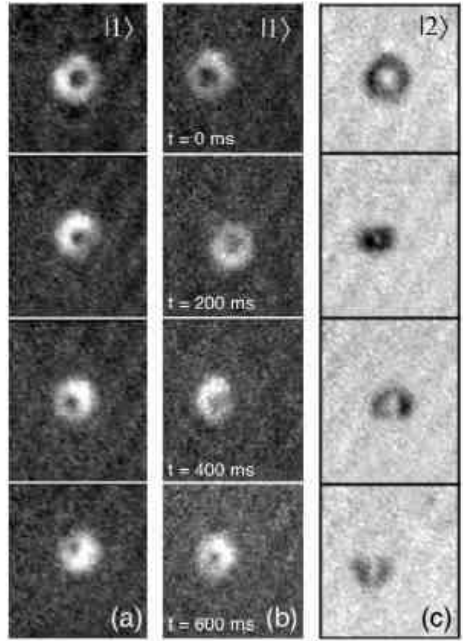}
\end{center}
\caption{(a,b) Two examples of the free evolution of a magnetically trapped vortex in the $| 1 \rangle$ state. The lifetime of this configuration was much longer than the trap oscillation period (128 ms).  (c) The free evolution of a vortex in the $ | 2 \rangle$ state. This vortex exhibits unstable dynamics. The vortex first contracts and then splits into fragments. Each column is obtained from a single run of the experiment, where the origin of time $t$ was chosen to be the time when the vortex was created ($t$ is common to every row).  The $| 1 \rangle$ and $| 2 \rangle$ state images appear different due to different signs of probe detuning. [Matthews {\it et al}., Phys. Rev. Lett. {\bf 83} 2451 (1999), reproduced with permission. Copyright (1999) by the American Physical Society.]}
\label{JILAwatch}
\end{figure}

Experimentally, it is possible to put the initial condensate into either the $| 1 \rangle $ or $| 2 \rangle $ state, and then make a vortex in the $| 2 \rangle $ or $| 1 \rangle $ state. In the case of $^{87}$Rb atoms, the intracomponent and intercomponent scattering lengths are in the proportion $a_{1}:a_{12}:a_{2}=1.03:1:0.97$ with the average of the three being 55 \AA. For the case without vorticity, the atoms with the larger scattering length $a_{1}$ in the state $| 1 \rangle$ formed a lower-density shell around the atoms with the smaller scattering length $a_{2}$\cite{Hall1} [see Fig. \ref{JILAvortexdensity}(a)]. The stability properties of the vortex states are strongly dependent on the relation of these scattering lengths. The time evolution of the vortex can be observed over time scales from milliseconds to seconds in the experiment. In Ref. \refcite{Matthews}, the vortex was stable in only the state with the vortex in the $|1\rangle $ state with the $|2\rangle $ state in the core [Fig. \ref{JILAwatch}(a,b)]. The other state [Fig. \ref{JILAwatch}(c)], in which the vortex was in the $|2\rangle $ state, showed an instability with the $|2 \rangle$ vortex sinking in toward the trap center and breaking up. Further discussion of dynamical stability is in Sec. \ref{stabdyn}.

\subsection{Structure of axisymmetric vortex states} \label{axisym}
\subsubsection{Thomas-Fermi approximation}
First, we consider the equilibrium structures of single vortex states in two-component BECs. As observed in Ref. \refcite{Matthews}, the simplest vortex structure consists of one circulating component that surrounds the other nonrotating component. Here, both wave functions have axisymmetric profiles. The authors in Ref. \refcite{Ho2,Chui,Jezek} used the Thomas-Fermi approximation to extensively analyze this system. Here we summarize their findings. 

The starting point for the calculation is the coupled GP equations (\ref{bingp1}) and (\ref{bingp2}). Let us assume that the condensate wave functions have one quantized vortex at the center as $\Psi_{i}({\bf r})= f_{i}(r,z) e^{i \gamma_{i} \theta}$; then Eqs. (\ref{bingp1}) and (\ref{bingp2}) become 
\begin{eqnarray}
\biggl[ -\frac{\hbar^{2}}{2m_{1}} \biggl( \frac{\partial^{2}}{\partial r^{2}} + \frac{1}{r} \frac{\partial}{\partial r} + \frac{\partial^{2}}{\partial z^{2}} - \frac{\gamma_{1}}{r^{2}} \biggr) + V_{1} + g_{1} \rho_{1} + g_{12} \rho_{2} - \gamma_{1} \Omega \biggr] f_{1} = \mu_{1} f_{1},  \label{axibingp1} \\
\biggl[ -\frac{\hbar^{2}}{2m_{2}} \biggl( \frac{\partial^{2}}{\partial r^{2}} + \frac{1}{r} \frac{\partial}{\partial r} + \frac{\partial^{2}}{\partial z^{2}} - \frac{\gamma_{2}}{r^{2}} \biggr) + V_{2} + g_{2} \rho_{2} + g_{12} \rho_{1} - \gamma_{2} \Omega \biggr] f_{2} = \mu_{2} f_{2}, \label{axibingp2} 
\end{eqnarray} 
where $\rho_{i}=f_{i}^{2}$ is the condensate density and the trapping potential is given by $V_{i} = m_{i} (\omega_{i \perp}^{2} r^{2} + \omega_{i z}^{2} z^{2}) / 2$. The axisymmetric vortex state is obtained by solving Eqs. (\ref{axibingp1}) and (\ref{axibingp2}) with $(\gamma_{1}, \gamma_{2}) = (1,0)$ or $(0,1)$. 

\begin{figure}[btp]
\begin{center}
\includegraphics[height=0.60\textheight]{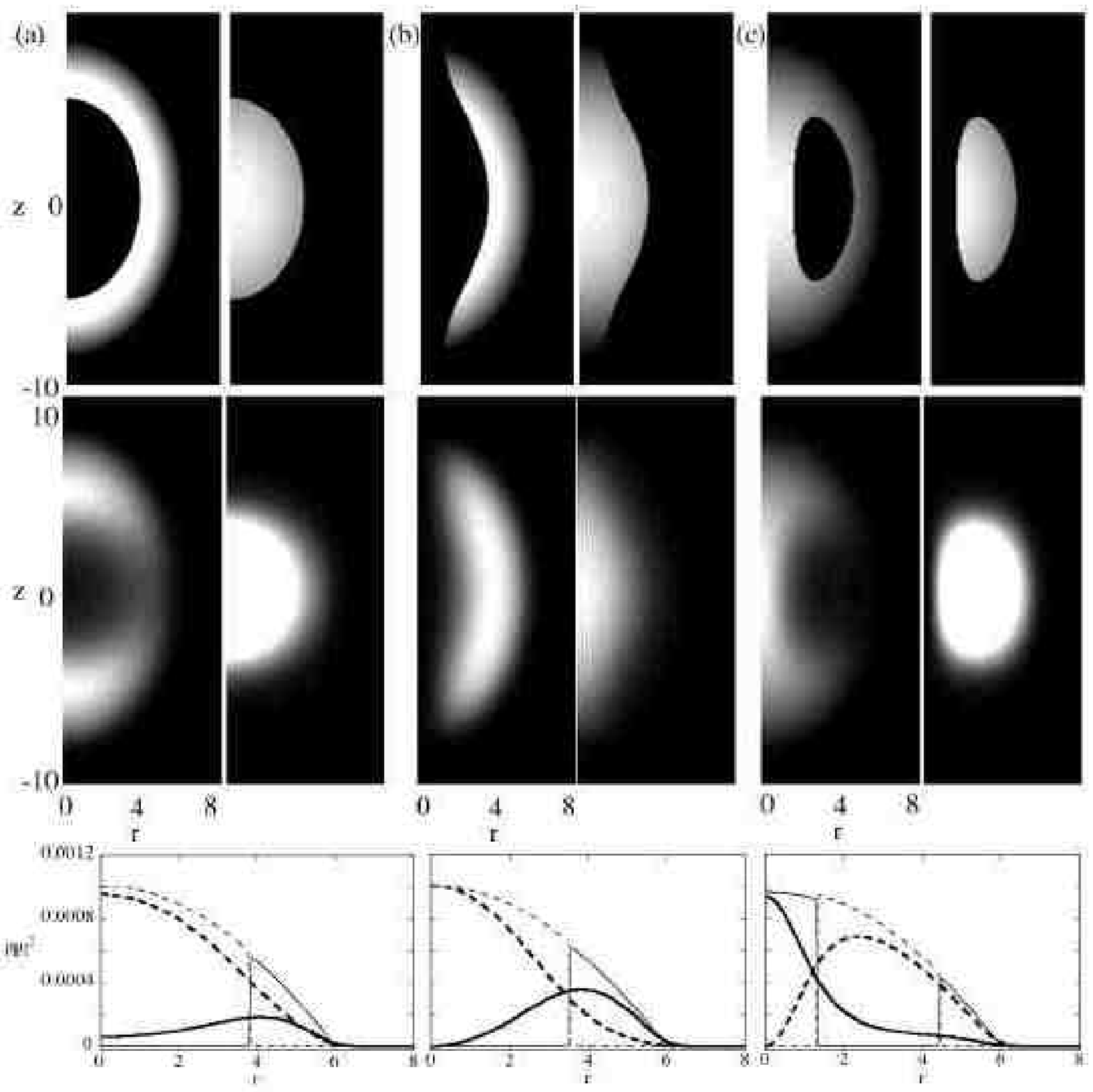}
\end{center}
\caption{Particle density of a vortex state in a trapping potential with parameters $a_{1}=5.67$ nm, $a_{2}=5.34$ nm, $a_{12}=5.50$ nm, $\omega_{\perp} = 7.8$Hz, and $\omega_{z}/\omega_{\perp} =1/\sqrt{8}$, and $N_{1}=N_{2}=5 \times 10^{5}$, taken from the JILA experiments. The wave function is normalized as $\int d^{3} r |\Psi_{i}|^{2}=1$. The contour plots in (a)-(c) show the density profile $\rho_{i}=|\Psi_{i}|^{2}$ ($i=1,2$) in the ($r$, $z$) plane obtained using the Thomas-Fermi approximation (top) and numerical simulations (middle) for (a) ($\gamma_{1},\gamma_{2}$)=(0,0), (b) (1,0) and (c) (0,1). The white (black) region represents the high (low) density region. The bottom plots show the corresponding radial density profiles $\rho_{1}$ (solid curve) and $\rho_{2}$ (dashed curve) for the equilibrium state at $z=0$. The thin and bold curve represent the results obtained with the Thomas-Fermi approximation and numerical simulations, respectively.}
\label{JILAvortexdensity}
\end{figure}
Under the Thomas-Fermi approximation, the quantum pressure terms (derivatives with respective to the coordinates $r$ and $z$) in Eqs. (\ref{axibingp1}) and (\ref{axibingp2}) are neglected. Then, the density profiles of the condensates can be calculated in a manner similar to that for the non-vortex case shown in Sec. \ref{basic}. In the region where only one wave function is nonvanishing ($\rho_{j} \neq 0$ and $\rho_{k} = 0$ for $j \neq k$), Eqs. (\ref{axibingp1}) and (\ref{axibingp2}) are decoupled and the solution is 
\begin{equation}
\rho_{i} = \frac{\mu_{i} - V_{i} - \hbar^{2} \gamma_{i}/(2 m_{i} r^{2})}{g_{i}}  \hspace{3mm} (i=1,2).
\end{equation}
 On the other hand, in the region where the wave functions overlap ($f_{i}>0$), one obtains 
\begin{eqnarray}
\rho_{1} = \frac{1}{g_{1} \Gamma_{12}} \biggl[ \mu_{1} - V_{1} - \frac{\hbar^{2}}{2 m_{1}} \frac{\gamma_{1}}{r^{2}} - \frac{g_{12}}{g_{2}} \biggl( \mu_{2} - V_{2} - \frac{\hbar^{2}}{2 m_{2}} \frac{\gamma_{2}}{r^{2}} \biggr) \biggr] ,  \label{axivorTFden1} \\
\rho_{2} = \frac{1}{g_{2} \Gamma_{12}} \biggl[ \mu_{2} - V_{2} - \frac{\hbar^{2}}{2 m_{2}} \frac{\gamma_{2}}{r^{2}} - \frac{g_{12}}{g_{1}} \biggl( \mu_{1} - V_{1} - \frac{\hbar^{2}}{2 m_{1}} \frac{\gamma_{1}}{r^{2}} \biggr) \biggr] .
\label{axivorTFden2}
\end{eqnarray}
Here, the chemical potential has been rewritten as $\mu_{i} - \gamma_{i} \Omega \rightarrow \mu_{i}$. The term $\hbar^{2}\gamma_{i}/(2m_{i}r^{2})$ represents the density singularity of the vortex core. To obtain the density profile, we have to calculate the chemical potentials $\mu_{i}$ from the normalization condition $\int d^{3} r \rho_{i} = N_{i}$. This leads to a problem involving two coupled integral equations that can be solved numerically. Choosing the parameters corresponding to those of the JILA experiment\cite{Matthews}, namely, $a_{1}=5.67$ nm, $a_{2}=5.34$ nm, $a_{12}=5.50$ nm, $\omega_{\perp} = 7.8$Hz, $\omega_{z}/\omega_{\perp} =1/\sqrt{8}$, and $N_{1}=N_{2}=5 \times 10^{5}$, we calculate the density profile for $(\gamma_{1},\gamma_{2})=(0,0)$ as well as $(1,0)$ and $(0,1)$ by both using the Thomas-Fermi approximation and numerical simulation, and the results are shown in Fig. \ref{JILAvortexdensity}. When neither component has nonzero vorticity [Fig. \ref{JILAvortexdensity}(a)], the components phase separate due to the intercomponent repulsion. Specifically, to decrease the intracomponent mean field energy, the $\Psi_{1}$ component with the larger intracomponent scattering length forms a shell outside the $\Psi_{2}$ component in the central region. The density contours in the ($r$, $z$) plane show that $|\Psi_{1}|^{2}$ extends more along the $z$ axis than the $r$ axis. When the $\Psi_{1}$ component has a vortex, a density hole associated with the vortex core appears in the density of $\Psi_{1}$ and penetrates through the $z$-axis as shown in Fig. \ref{JILAvortexdensity}(b). Then, the $\Psi_{2}$ component fills the vortex core because of the intercomponent repulsion, forming a ``cigar-shaped" profile. As a result, the characteristic scale of the core size becomes larger than the healing length $\xi=\hbar/(2 m g_{1} \rho_{1})$, which characterizes the core size of a single-component BEC. In this case, the centrifugal force associated with the vortex makes the $\Psi_{1}$ component expand radially, which allows a decrease in the intracomponent mean-field energy of the $\Psi_{1}$ component rather than that of the nonvortex state. On the other hand, when the $\Psi_{2}$ component with the smaller intracomponent scattering length has a vortex, the structure is different from the $(\gamma_{1},\gamma_{2})=(1,0)$ solution as shown in Fig. \ref{JILAvortexdensity}(c). First, the core size is smaller than that of the $(1,0)$ solution. Second, while a fraction of the $\Psi_{1}$ component fills the vortex core, the rotating $\Psi_{2}$ component is surrounded by a shell of the excessive $\Psi_{1}$ component, despite the centrifugal force. As discussed below, the latter configuration is dynamically unstable\cite{Ripoll3,Ripoll31,Skrybin}. The above consideration shows that the Thomas-Fermi approximation captures some qualitative features of the solutions\cite{Ho2,Chui,Jezek}. However, one cannot neglect the quantum pressure term for the description of the vortex states in this system; in particular, the quantum pressure is necessary to determine the boundary profile between segregated components\cite{Tim,Ao}.

\subsubsection{Pseudospin representation: Nonlinear sigma model}
To analyze the vortex structure, we present a 2-D (two-dimensional) model\cite{Kasamatsuskyrm} that is based on the concept of ``pseudospin"\cite{VLeonhardt,Kasamatsupre,Mueller,Matthews3}. The spinor order parameter of two-component BECs allows us to analyze this system as a spin-1/2 BEC\cite{VLeonhardt,Kawaja2,Ruostekoski,Mueller,Matthews3}. To simplify the problem, we assume that all atoms have the same mass ($m_{1}=m_{2}=m$) and the trapping potentials are equal ($V_{1}=V_{2}=V$). We map the Hilbert space of the system into pseudospin space by introducing the normalized complex-valued spinor $\chi({\bf r})=[\chi_{1}({\bf r}), \chi_{2}({\bf r})]^{T}=[|\chi_{1}|e^{i\theta_{1}}, |\chi_{2}|e^{i\theta_{2}}]^{T}$. Also, we decompose the wave function as $\Psi_{i}=\sqrt{\rho_{\rm T}({\bf r})} \chi_{i}({\bf r})$. Here, $\rho_{\rm T}=\rho_{1}+\rho_{2}$ is the total density, and the spinor thus satisfies
\begin{equation}
|\chi_{1}|^{2}+|\chi_{2}|^{2}=1.
\label{normalspinor}
\end{equation}
In terms of the spinor, the spin density is defined as ${\bf S}=\bar{\chi}({\bf r}) \bm{\sigma} \chi({\bf r})$, where $\bm{\sigma}$ is the Pauli matrix. Explicit expressions of ${\bf S}=(S_{x},S_{y},S_{z})$ read 
\begin{eqnarray}
S_{x} &=& (\chi_{1}^{\ast} \chi_{2} + \chi_{2}^{\ast} \chi_{1}) = 2 |\chi_{1}||\chi_{2}| \cos (\theta_{1} - \theta_{2} ), \label{spincomponx} \\
S_{y} &=& - i (\chi_{1}^{\ast} \chi_{2} - \chi_{2}^{\ast} \chi_{1}) = -2 |\chi_{1}||\chi_{2}| \sin (\theta_{1} - \theta_{2} ), \label{spincompony} \\
S_{z} &=& (|\chi_{1}|^{2} - |\chi_{2}|^{2} ), \label{spincomponz}
\end{eqnarray}
where the modulus of the total spin is $|{\bf S}|=1$ everywhere. We assume that $\Psi_{1}$ and $\Psi_{2}$ represent the up and down components of the spin-1/2 spinor, respectively. The nonzero spin projection on the $x$-$y$ plane implies a relative phase coherence between the up and down components. 

The transformation of Eq. (\ref{energyfunctio2}) to the pseudospin representation is straightforward\cite{Kasamatsuskyrm}. The result is  
\begin{eqnarray}
E = \int  d^{3} r  \biggl[ \frac{\hbar^{2}}{2m}(\nabla \sqrt{\rho_{\rm T}})^{2}+ \frac{\hbar^{2} \rho_{\rm T}}{8 m} (\nabla {\bf S})^{2} + \frac{m \rho_{\rm T}}{2} ( {\bf v}_{\rm eff} - {\bf \Omega} \times {\bf r} )^{2} \nonumber \\
+ V \rho_{\rm T} + \frac{\rho_{\rm T}^{2}}{2} ( c_{0} + c_{1} S_{z} + c_{2} S_{z}^{2}) \biggr],  \label{nonsigmamod}
\end{eqnarray}
where we define new coupling constants as
\begin{eqnarray}
c_{0} &\equiv& \frac{g_{1}+g_{2}+2g_{12}}{4},  \nonumber \\
c_{1} &\equiv& \frac{g_{1}-g_{2}}{2},  \\
c_{2} &\equiv& \frac{g_{1}+g_{2}-2g_{12}}{4}, \nonumber
\end{eqnarray}
and an effective velocity field as
\begin{eqnarray}
{\bf v}_{\rm eff} &=& \frac{\hbar}{2im} (\chi_{1}^{\ast} \nabla \chi_{1} - \chi_{1} \nabla \chi_{1}^{\ast} + \chi_{2}^{\ast} \nabla \chi_{2} - \chi_{2} \nabla \chi_{2}^{\ast} )  \nonumber \\
&=& \frac{\hbar}{2m} \biggl[ \nabla \Theta + \frac{S_{z} (S_{y} \nabla S_{x} - S_{x} \nabla S_{y})}{ S_{x}^{2}+S_{y}^{2} } \biggr] \label{effectivevelo}
\end{eqnarray}
which depends on the gradient of the total phase $\Theta=\theta_{1}+\theta_{2}$ and that of the pseudospin. 

The form of Eq. (\ref{nonsigmamod}) is analogous to the classical {\it nonlinear sigma model} (NL$\sigma$M) for Heisenberg ferromagnets in which only the $(\nabla {\bf S})^{2}$ term appears. A NL$\sigma$M allows one to study a various type of ``topological excitations" such as {\it monopoles}, {\it domain walls}, and {\it skyrmions}\cite{Rejan}. Because we have changed merely the representation of the Gross-Pitaevskii energy functional, Eq. (\ref{nonsigmamod}) is still an exact description of the two-component BEC. The four degrees of freedom of the original condensate wave functions $\Psi_{1}$ and $\Psi_{2}$ (their amplitudes and phases) have now been expressed in terms of the total density $\rho_{\rm T}$, the total phase $\Theta$, and two of the spin density components ($S_{x},S_{y},S_{z}$) (one of which is determined besides its sign by the other two because of $|{\bf S}|=1$). 

Unlike the classical NL$\sigma$M case, the system here has several unique features that are revealed in Eq. (\ref{nonsigmamod}): (i) the total density $\rho_{\rm T}$, which is a prefactor of the $(\nabla {\bf S})^{2}$ term and effectively `stiffens' the pseudospin, is spatially inhomogeneous because of the trapping potential; (ii) there is a hydrodynamic kinetic-energy term $m \rho_{\rm T} ({\bf v}_{\rm eff} - {\bf \Omega} \times {\bf r})^{2}/2$, which is associated with the topological excitations; (iii) if $g_{1} \neq g_{2} \neq g_{12}$, the anisotropic terms involving the coefficients $c_{1}$ and $c_{2}$ effectively break the ``SU(2)-invariance"\cite{VLeonhardt,Kasamatsuskyrm,Kawaja2,Kuklov} because then the rotation of ${\bf S}$ removes the energy degeneracy. The coefficient $c_{1}$ can be interpreted as a longitudinal (pseudo)magnetic field that tends to align the spin along the $z$-axis. The term with the coefficient $c_{2}$ may determine the spin-spin interaction associated with $S_{z}$, which is antiferromagnetic for $c_{2}>0$ and ferromagnetic for $c_{2}<0$ \cite{Kasamatsu3}. If $m_{1} \neq m_{2}$ or $V_{1} \neq V_{2}$, other anisotropic terms appear in Eq. (\ref{nonsigmamod}).

In the following, we confine ourselves to the 2-D problem for simplicity. Two-dimensional calculations of the GP equations (\ref{bingp1}) and (\ref{bingp2}) are carried out by separating the degrees of freedom of the original wave function as $\Psi_{i}({\bf r})=\psi_{i}(x,y)\phi(z)$. This separation yields the dimensionless 2-D coupling constants $u_{i}=4 \pi a_{i} \eta N$ and $u_{12}=4 \pi a_{12} \eta N$ with $\eta=\int dz |\phi(z)|^{4} / \int dz |\phi(z)|^{2}$ [see Ref.\refcite{Kasamatsu} for more details], where $N=N_{1}+N_{2}$ is the total particle number. The dimensionless GP equations become
\begin{eqnarray}
\biggl( -\frac{1}{2} \nabla^{2} + \frac{r^{2}}{2} + u_{1} |\psi_{1}|^{2} + u_{12} |\psi_{2}|^{2} - \Omega L_{z} \biggr) \psi_{1} = \mu_{1} \psi_{1} , \label{nondimgpeq1} \\
\biggl( -\frac{1}{2} \nabla^{2} + \frac{r^{2}}{2} + u_{2} |\psi_{2}|^{2} + u_{12} |\psi_{1}|^{2} - \Omega L_{z} \biggr)  \psi_{2} = \mu_{2} \psi_{2}, \label{nondimgpeq2}
\end{eqnarray}
where the length and time are measured in units of $b_{\rm ho}=\sqrt{\hbar/m\omega_{\perp}}$ and $\omega_{\perp}^{-1}$, which are the characteristic scales of the radial trapping potential $V=m\omega_{\perp}^{2} r^{2}/2$. Also, the NL$\sigma$M (\ref{nonsigmamod}) becomes 
\begin{eqnarray}
E = \int  d^{2} r  \biggl[ \frac{1}{2}(\nabla \sqrt{\rho_{\rm T}})^{2}+ \frac{\rho_{\rm T}}{8} (\nabla {\bf S})^{2} + \frac{\rho_{\rm T}}{2} ( {\bf v}_{\rm eff} - {\bf \Omega} \times {\bf r} )^{2} \nonumber \\
+ \frac{r^{2}}{2} \rho_{\rm T} + \frac{\rho_{\rm T}^{2}}{2} ( c_{0} + c_{1} S_{z} + c_{2} S_{z}^{2}) \biggr]  \label{sigmamoddimless}
\end{eqnarray}
with $c_{0} = (u_{1}+u_{2}+2u_{12})/4$, $c_{1} = (u_{1}-u_{2})/2$, and $c_{2} = (u_{1}+u_{2}-2u_{12})/4$.

\subsubsection{A skyrmion: Numerical analysis}
\begin{figure}[btp]
\begin{center}
\includegraphics[height=0.45\textheight]{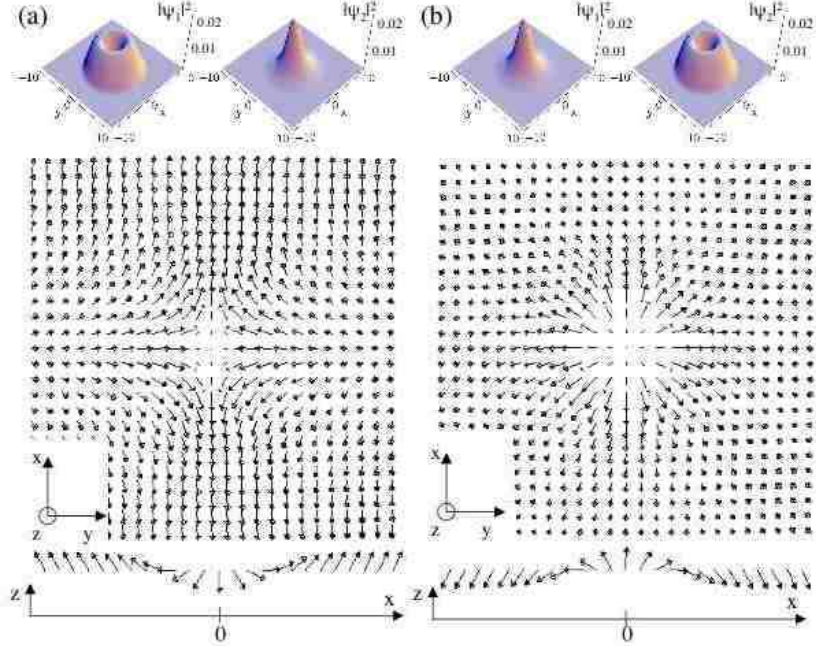}
\end{center}
\caption{Results of simulations of 2-D coreless vortex states. Top panels: the density profiles consisting of (a) the rotating $\psi_{1}$ component and the nonrotating $\psi_{2}$ component and (b) the nonrotating $\psi_{1}$ component and the rotating $\psi_{2}$ component. The parameters are $u_{1}=u_{2}=u_{12}$ ($c_{0}=1000$, $c_{1}=c_{2}=0$) and $\Omega=0.15$. For the calculation, we fixed only the total particle number $N=N_{1} + N_{2}$. Then, the solution converges to $N_{1}/N_{2}=2.465$ in (a) and 0.406 in (b). Middle panels: vector plots of the spin texture projected onto the $x$-$y$ plane corresponding to (a) (left) and (b) (right). The region is [$-5 \leq x, y \leq 5$]. The bottom panels: the cross-sections of the spin texture along the $x$-axis at $y=0$ corresponding to the vector plots of the middle panels.} 
\label{axisymvor}
\end{figure}
First, we consider the axisymmetric vortex state for the case $g_{1}=g_{2}=g_{12}$ ($c_{1}=c_{2}=0$), in which case the system possesses SU(2) symmetry. In the pseudospin picture, the axisymmetric vortex observed in Ref. \refcite{Matthews} can be interpreted as a type of spin texture called a ``{\it skyrmion}"\cite{VLeonhardt,Mueller}. Typical examples are shown in Fig. \ref{axisymvor}. In Fig. \ref{axisymvor}(a), we assume that the $\psi_{1}$ component has one singly-quantized vortex at the center of the trap. At the center of the atomic cloud, the $\psi_{1}$ component vanishes, so that the pseudospin points down according to the definition of the spin $S_{z}$ in Eq. (\ref{spincomponz}). The spin aligns with a hyperbolic distribution as $(S_{x},S_{y}) \propto (-x,y) $ around the singularity at the center. At the edge of the atomic cloud, the $\psi_{2}$ component vanishes, and the pseudospin points up. In between, the pseudospin continuously rolls from down to up as shown in the bottom of Fig. \ref{axisymvor}. This ``cross-disgyration" texture is often referred to as a skyrmion in analogy to the work of Skyrme\cite{Skyrmi}. In the field of superfluid $^{3}$He, this skyrmion texture is referred to as an Anderson-Toulouse vortex\cite{VLeonhardt}. If $\psi_{2}$ component has a vortex, the texture exhibits a ``radial-disgyration" as illustrated in the middle panel of Fig. \ref{axisymvor}(b), in which the spin in the $x$-$y$ plane aligns as $(S_{x},S_{y}) \propto (x,y) $ around the singularity. These two skyrmions are degenerate in the SU(2) symmetric case because they are interconverted by the overall spin; in fact, all configurations obtained by any global spin rotation are degenerate. 

The size of the skyrmion is determined by the interaction coefficients and the rotation frequency\cite{Kasamatsuskyrm}, and is related to the ratio $N_{1}/N_{2}$. When we minimize the energy to determine the ground state, we fix the total particle number, not each particle number. This procedure gives us the {\it true} thermodynamically stable configuration of the skyrmion. Although each particle number is conserved in experiments performed to date (this case will be discussed below), the population imbalance determined as above can be realized by controlling the strength and time of the coupling drive in the phase imprinting method. 

\begin{figure}[btp]
\begin{center}
\includegraphics[height=0.30\textheight]{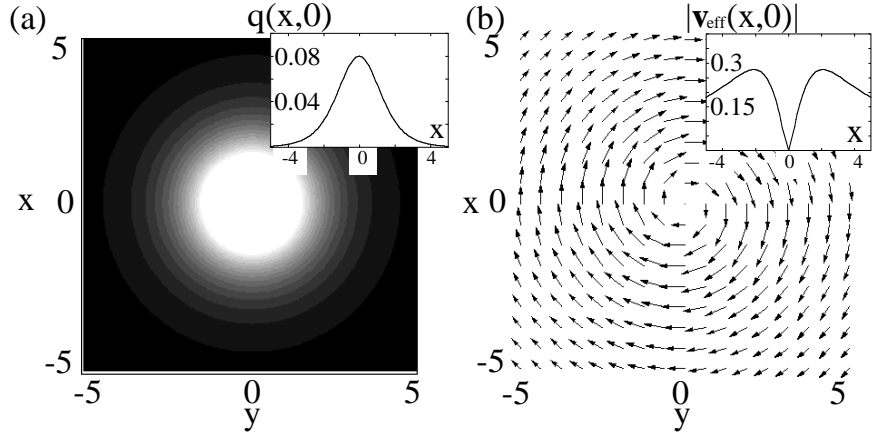}
\end{center}
\caption{(a) Topological charge density $q({\bf r})$ defined by Eq. (\ref{topologicaldensity}). (b) The effective velocity field ${\bf v}_{\rm eff}$ defined by Eq. (\ref{effectivevelo}), corresponding to the solution shown in Fig. \ref{axisymvor}. The insets are cross-sections of $q({\bf r})$ and $|{\bf v}_{\rm eff}|$ along the $y=0$ line within the range $-5 \leq x \leq 5$.} 
\label{skyrmtopological}
\end{figure}

It is known that the skyrmion in a 2-D system has a topological invariant defined as
\begin{equation}
Q \equiv \frac{1}{8 \pi} \int d^{2} r \epsilon^{ij} {\bf S} \cdot \partial_{i} {\bf S} \times \partial_{j} {\bf S},
\label{topologicalnumber}
\end{equation}
which is known as the topological charge or the Pontryagian index \cite{Rejan}. The skyrmion with any spin profile is shown to have $Q= \pm 1$, whose sign depends on the direction of the circulation. The integrand of Eq. (\ref{topologicalnumber}) is the topological charge density $q({\bf r})$ associated with the vorticity. The topological charge density can be derived from the effective velocity ${\bf v}_{\rm eff}$ as\cite{Kasamatsuskyrm,Mueller,Mizushima4}: 
\begin{equation}
q({\bf r}) \equiv \frac{1}{8 \pi} \epsilon^{ij} {\bf S} \cdot \partial_{i} {\bf S} \times \partial_{j} {\bf S} = \frac{1}{2\pi} (\nabla \times {\bf v}_{\rm eff} )_{z},
\label{topologicaldensity}
\end{equation}
where we used the relation $\sum_{i} \chi_{i} \nabla \chi_{i}^{\ast} = - \sum_{i} \chi_{i}^{\ast} \nabla \chi_{i}$ $(i=1,2)$. The topological charge density $q({\bf r})$ characterizes the spatial distribution of the skyrmion. Figure \ref{skyrmtopological} shows the spatial distribution of $q({\bf r})$ and the corresponding ${\bf v}_{\rm eff}$-field for the solution of Fig. \ref{axisymvor}(a). The topological charge is distributed around the center and, contrary to the case of a conventional vortex in a single-component condensate, $|{\bf v}_{\rm eff}|$ vanishes at the center. This makes a coreless vortex (skyrmion) without a density dip in the total density. 

\subsubsection{A skyrmion: Variational analysis}
To study the physical properties of the skyrmion in more detail, we used a variational analysis on the NL$\sigma$M described by Eq. (\ref{sigmamoddimless}). In the following, we confine ourselves to the case in which the $\psi_{1}$ component has a vortex. The original NL$\sigma$M, where only the ($\nabla {\bf S})^{2}$ term is present in Eq. (\ref{sigmamoddimless}), admits a skyrmion solution whose explicit analytic expression is known\cite{Rejan}. Here we seek for a more general form of the skyrmion solution of Eq. (\ref{sigmamoddimless}) as a variational function, which can reproduce well the numerical solution. The skyrmion solution of Fig. \ref{axisymvor}(a) may be parametrized as\cite{Sinova2} 
\begin{equation}
S_{x} = \frac{4 \lambda x e^{-\alpha r^{2}/2}}{r^{2}+4 \lambda^{2} e^{-\alpha r^{2}}}, \hspace{3mm}
S_{y} = \frac{-4 \lambda y e^{-\alpha r^{2}/2}}{r^{2}+4 \lambda^{2} e^{-\alpha r^{2}}},  \hspace{3mm}
S_{z} = \frac{r^{2}-4\lambda^{2} e^{-\alpha r^{2}}}{r^{2}+4\lambda^{2} e^{-\alpha r^{2}}} \label{skyrmansatz}
\end{equation}
with $|{\bf S}|=1$. Equations (\ref{skyrmansatz}) with $\alpha=0$ correspond to the explicit skyrmion solution of the classical NL$\sigma$M\cite{Rejan}. The variational parameters $\lambda$ and $\alpha$ determine the size and the shape of the skyrmion. Typically, $\lambda$ represents a characteristic size of the domain over which the direction of the spin is maintained and $\alpha$ is the degree of spatial variation of the spin inversion. If we take both $\lambda$ and $\alpha$ as variable parameters, the number of particles in each pseudospin component is not conserved, but the total particle number is conserved. In the limit $\lambda \rightarrow 0$ $(\infty)$ with fixed $\alpha$ all spins point up (down) along the $z$-axis, which means a perfect polarization of the particle number such as $N \rightarrow N_{1}$ $(N_{2})$. Alternately, for fixed $\lambda$, the limit $N \rightarrow N_{1}$ $(N_{2})$ corresponds to $\alpha \rightarrow + \infty$ $(- \infty)$.  Because the $\psi_{1}$ component has one singly-quantized vortex at the center, the total phase $\Theta$ in Eq. (\ref{sigmamoddimless}) is  
\begin{equation}
\Theta = \tan^{-1} \frac{y}{x}. \label{skyrmtotpha}
\end{equation}
Substituting Eqs. (\ref{skyrmansatz}) and (\ref{skyrmtotpha}) into Eq. (\ref{sigmamoddimless}) yields 
\begin{eqnarray}
E=\int d^{2} r \biggl\{ \frac{1}{2} (\nabla \sqrt{\rho_{\rm T}})^{2} + V_{\rm eff} \rho_{\rm T} 
+\frac{\rho_{\rm T}^{2}}{2} \biggl[ c_{0} + c_{1} \frac{r^{2}-4\lambda^{2} e^{-\alpha r^{2}}}{r^{2}+4\lambda^{2} e^{-\alpha r^{2}}} \nonumber \\
+ c_{2} \biggl( \frac{r^{2}-4\lambda^{2} e^{-\alpha r^{2}}}{r^{2}+4\lambda^{2} e^{-\alpha r^{2}}} \biggr)^{2} \biggr] \biggr\}. \label{symmskyreneg}
\end{eqnarray}
Thus, there are now fewer degrees of freedom in the energy functional: the total density $\rho_{\rm T}$ and the two variational parameters $\lambda$ and $\alpha$. In the above equation, we introduced an effective, radially symmetric, confining potential 
\begin{equation}
V_{\rm eff} \equiv \frac{r^{2}+4 \lambda^{2} e^{-\alpha r^{2}} \left[ (\alpha r^{2} +1)^{2} +1 \right] }{2 (r^{2}+4\lambda^{2} e^{-\alpha r^{2}})^{2}}
- \frac{\Omega r^{2}}{r^{2}+4\lambda^{2} e^{-\alpha r^{2}}} + \frac{r^{2}}{2}, 
\label{skyrmeffpot}
\end{equation} 
which determines the shape of the total density $\rho_{\rm T}$. The total density $\rho_{\rm T}$ can be calculated by solving the equation
\begin{equation}
-\frac{ \left( \nabla^{2} \sqrt{\rho_{\rm T}} \right) }{2 \sqrt{\rho_{\rm T}}} + V_{\rm eff} + \rho_{\rm T} \biggl[ c_{0} +  c_{1} \frac{r^{2}-4\lambda^{2} e^{-\alpha r^{2}}}{r^{2}+4\lambda^{2} e^{-\alpha r^{2}}} + c_{2} \biggl( \frac{r^{2}-4\lambda^{2} e^{-\alpha r^{2}}}{r^{2}+4\lambda^{2} e^{-\alpha r^{2}}} \biggr)^{2} \biggr] = \mu,
\label{totalden1}
\end{equation} 
where the chemical potential $\mu$ is fixed by the normalization condition $\int d^{2} r \rho_{\rm T} = 1$. 

Using the condition $\int d^{2} r \rho_{\rm T}=1$, we numerically calculate the chemical potential $\mu$ in Eq. (\ref{totalden1}) with the Thomas-Fermi approximation, where $\rho_{\rm T}$ is assumed to be zero in the region in which $\rho_{\rm T}$ is negative. Then, the total energy $E$ becomes a function of $(\lambda,\alpha)$ by using the Thomas-Fermi expression for $\rho_{\rm T}$. When the size of the skyrmion is large enough\cite{Kasamatsuskyrm}, the Thomas-Fermi approximation can be used to evaluate $\rho_{\rm T}$, and then the energy can be minimized with respect to only two variational parameters $\lambda$ and $\alpha$. Otherwise a density singularity associated with the vortex core appears in $\rho_{\rm T}$ so that the quantum pressure term cannot be ignored. 

\begin{figure}[btp]
\begin{center}
\includegraphics[height=0.28\textheight]{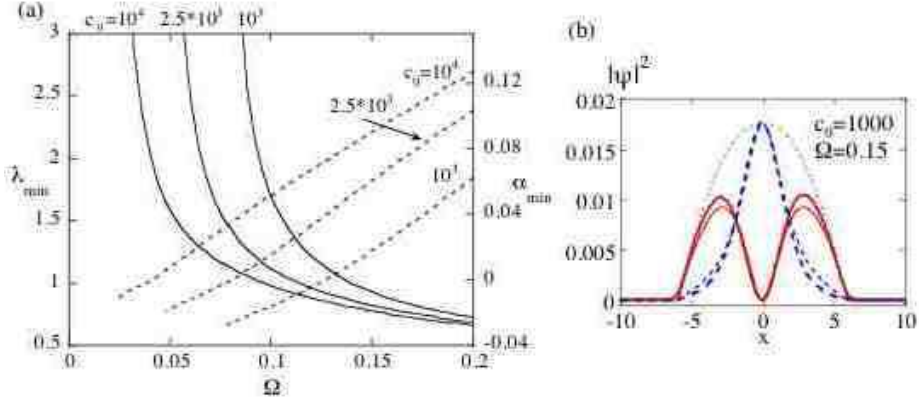}
\end{center}
\caption{(a) Optimized values of the variational parameters $\lambda$ (solid curve) and $\alpha$ (dashed curve) as a fuction of rotation frequency $\Omega$ and for $c_{0}=1000$, 2500, 10000 with $c_{1}=c_{2}=0$. (b) The cross-sections of $|\psi_{1}|^{2}$ (solid curve), $|\psi_{2}|^{2}$ (dashed curve) and the total density $\rho_{\rm T}$ (dotted curve) along the $x$-axis at $y=0$, where bold and thin curves represent the results obtained from the numerical and variational calculations, respectively.} 
\label{optimizedsize}
\end{figure}

For the SU(2) symmetric case without rotation ($\Omega=0$), the minimum energy occurs when $\lambda \rightarrow \infty$ and $\alpha \rightarrow - \infty$, which implies that $N=N_{2}$, i.e., complete polarization of the particle number. It follows that the global minimum for $\Omega=0$ is a nonvortex state for any value of $N_{2}$ for a fixed total particle number $N$. This is because the interaction energy terms now satisfy the SU(2) symmetry so that the energy of the nonvortex state is degenerate under the change of the ratio $N_{1}/N_{2}$ with fixed $N$. However, above a certain critical value of $\Omega$, there appears an energy minimum for particular finite values of $\lambda$ and $\alpha$. Since the minimized energy is lower than that of the nonvortex state with $\lambda \rightarrow +\infty$, an appropriate external rotation can provide {\it global stability} for the skyrmion. Figure \ref{optimizedsize}(a) shows the values of $\lambda$ and $\alpha$ that give the minimum of the total energy for $c_{0}=1000$, 2500, and 10000. The size of the skyrmion decreases with increasing $\Omega$, as revealed by a decrease in $\lambda_{\rm min}$. The same effect occurs for an increase in $\alpha$. The energy minimum appears at a certain critical frequency of $\Omega$ where $\lambda_{\rm min}$ ($\alpha_{\rm min}$) diverges to $+\infty$ ($-\infty$) and the critical frequency decreases as $c_{0}$ increases. Therefore, the condensates with larger $c_{0}$ can have a stable skyrmion at a lower rotation frequency. The comparison with the numerical solution in Fig. \ref{optimizedsize}(b) shows that the optimized variational functions given in Eqs. (\ref{skyrmansatz}) and (\ref{skyrmtotpha}) nearly reproduce the exact numerical solution. Hence, our approach here greatly improves upon the analytic treatment for the vortex states beyond the usual Thomas-Fermi approximation in Refs. \refcite{Ho2,Chui,Jezek}. However, full numerical calculations show that for $c_{0}=1000$, 2500, 10000 additional vortices are nucleated above $\Omega > 0.17$, 0.11, 0.05 and the ansatz (\ref{skyrmansatz}) for a single skyrmion is no longer valid.

The parameters $c_{1}$ and $c_{2}$ in Eq. (\ref{symmskyreneg}), the values of which can be controlled through the scattering lengths, predominantly determine the size of the skyrmion. The coefficient $c_{1}$, which is proportional to the difference between the intracomponent interactions $u_{1}$ and $u_{2}$, may be regarded as a (pseudo)magnetic field that tends to cause the spin to align along the $z$-axis; the positive (negative) $c_{1}$ arranges the spins downward (upward). For the negative $c_{1}$ ($u_{1}<u_{2}$) the stable size of the skyrmion shrinks as $|c_{1}|$ increases, which implies an increase in the fraction of the rotating $\psi_{1}$ component and a decrease in the fraction of the $\psi_{2}$ component at the vortex core, (i.e., the spins align upward). For the positive $c_{1}$ ($u_{1}>u_{2}$) the value of $\lambda$ increases rapidly and eventually goes to infinity (concurrently, $\alpha$ goes to $-\infty$), corresponding to the perfect ``spin-down" alignment (occupation of the nonrotating $\psi_{2}$ component only). 

The sign of $c_{2}$ determines the nature of the spin-spin interaction associated with $S_{z}$. For positive $c_{2}$ ($u_{1}+u_{2}>2u_{12}$) the interaction is antiferromagnetic, that is, a spatial mixture of the spin-up and spin-down components. The presence of a vortex causes an effective phase separation. Then, there appears to be no significant change of the spin profile compared to that of $c_{2}=0$ in Fig. \ref{axisymvor}(a). In contrast, for negative $c_{2}$ ($u_{1}+u_{2}<2u_{12}$), the system enters the ferromagnetic phase, where the spin domains are spontaneously formed. When there is a skyrmion in the $\psi_{1}$ component, its size shrinks with $|c_{2}|$ in a way similar to the $|c_{1}|$-dependence. This is because each particle number is not conserved in the present case; the energetically favorable configuration in a ferromagnetic phase tends to an overall spin polarization, that is, a perfect polarization of the particle number. In this regime, depending on the rotation frequency, there are two energy minima, one of which leads to a perfect polarization of the $\psi_{1}$ vortex state, but the other leads to the $\psi_{2}$ nonvortex state, which is stable. 

\begin{figure}[btp]
\begin{center}
\includegraphics[height=0.50\textheight]{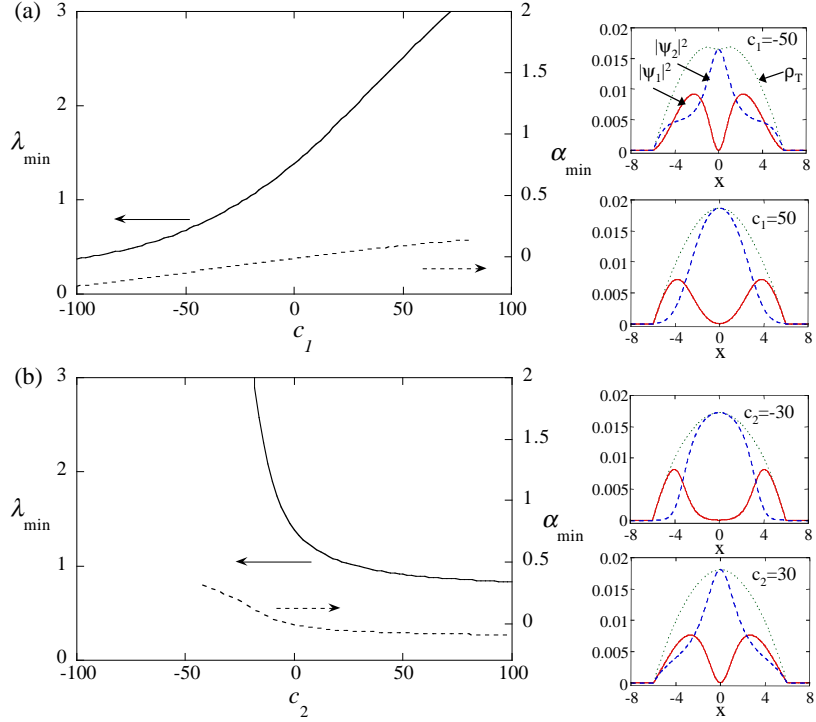}
\end{center}
\caption{Optimized values of the variational parameters $\lambda$ (solid curve) and $\alpha$ (dashed curve) as a function of (a) $c_{1}$ and (b) $c_{2}$ under the condition of fixed particle number of each component $N_{1}=N_{2}$ for $c_{0}=10^{3}$, $\Omega=0.15$. We fix $c_{2}=0$ for (a) and $c_{1}=0$ for (b). The right panels show the cross sections of the optimized variational functions $|\psi_{1}|^{2}$ (solid curve), $|\psi_{2}|^{2}$ (dashed curve), and the total density $\rho_{\rm T}$ (dotted curve) along the $y=0$ line ((a) top: $c_{1}=-50$, bottom: $c_{1}=50$ and (b) top: $c_{2}=-30$, bottom: $c_{2}=30$).} 
\label{optimfixNcase}
\end{figure}

So far, we have allowed a change in the particle number of each component to facilitate the calculation of the minimum skyrmion size. However, the actual experiments on two-component BECs have been done under a condition in which each particle number is fixed. This restriction can be taken into account by imposing the relation
\begin{equation}
\frac{N_{1}}{N_{2}} = \frac{\int d^{2} r \rho_{T} (1 + S_{z})}{\int d^{2} r \rho_{T} (1 - S_{z})}.
\label{ratioparti}
\end{equation}
For a given $\lambda$ the value of $\alpha$ is uniquely determined by Eq. (\ref{ratioparti}), where both $\rho_{\rm T}$ and $S_{z}$ are functions of two variational parameters. Therefore, the energy minimization can be done with respect to one variational parameter, which we choose to be $\lambda$. Let us here consider the case $N_{1}/N_{2}=1$, and investigate the stable size of a skyrmion. In this case, we find that the stable size is not affected by the change in rotation frequency $\Omega$. Figure \ref{optimfixNcase} shows the optimized values of $\lambda$ and $\alpha$ as a function of $c_{1}$ and $c_{2}$. No perfect polarization of the particle number occurs, so that the skyrmion may exist for all values of $c_{1}$ and $c_{2}$; moreover, in this case, even if the optimal variational parameter exist, they do not ensure a local minimum of the energy. In Fig. \ref{optimfixNcase}(a), for positive $c_{1}$, to enlarge the ``spin-down" domain containing the $\psi_{2}$ component, one should increase both $\lambda_{\rm min}$ and $\alpha_{\rm min}$. When this domain increases, so does the size of the vortex core of the $\psi_{1}$ component. For negative $c_{1}$, the domain with the rotating $\psi_{1}$ component tends to increase. As a result, the vortex core of the $\psi_{1}$ component shrinks and the excess $\psi_{2}$ component, which fills the vortex core, protrudes into the outside of the $\psi_{1}$ component. In the $c_{2}$ dependence, although there are no drastic change in the antiferromagnetic $c_{2}>0$ regime, for $c_{2}<0$ a rapid increase occurs in $\lambda_{\rm min}$ and $\alpha_{\rm min}$ when $|c_{2}|$ increases [Fig. \ref{optimfixNcase}(b)]. This means that spin-up or spin-down domains grow and their boundary becomes sharper for larger $|c_{2}|$. These results are consistent with the numerical ones shown in Fig. \ref{JILAvortexdensity}.

\subsection{Structure of nonaxisymmetric vortex states: a pair of merons}\label{meronsect}
\subsubsection{A meron pair: Numerical analysis}
In this section, we discuss a nonaxisymmetric vortex state. The spin texture of this state involves a pair of ``merons"\cite{Girvin} or ``Mermin-Ho vortices"\cite{VLeonhardt,Ho3,Mueller} (``meron ($\mu \epsilon \rho o \sigma$)" means ``fraction" in Latin). Our previous study revealed that a meron pair (a vortex molecule) can be stabilized thermodynamically in a two-component BEC if it is rotated and has an {\it internal coherent coupling}\cite{Kasamatsupre}. The two components interact not only through their mean-field interactions but also through the relative phase of the order parameters. Combination of these two effects enables us to explore a new regime of rich vortex structures which are absent in the conventional binary system\cite{Ho2,Chui,Jezek,Ripoll4,Mueller2,Kasamatsu3}. If the strength of the coupling drive is increased gradually from zero and its frequency is adiabatically tuned to be resonant, one can obtain a stationary state with a nearly equal-weight superposition of the two states\cite{Matthews3}. The internal coherent coupling can be included in the formulation through Eq. (\ref{intjosenerg}), and we obtain the time-independent, coupled GP equations by substituting $\Psi_{i}({\bf r},t)=\Psi_{i} ({\bf r}) e^{-i \mu t/\hbar}$ into Eqs. (\ref{2TDCGP1}) and (\ref{2TDCGP2}). Both components are characterized by the common chemical potential $\mu$ because, in the presence of the coherent coupling, the conserved quantity is the total particle number rather than the individual particle numbers. For simplicity, the detuning parameter $\Delta$ is assumed to be zero. In terms of the pseudospin picture, the Rabi term given in Eq. (\ref{intjosenerg}) is represented as $- \omega_{\rm R} \rho_{\rm T} S_{x}$, which plays a role of a ``transverse magnetic field" along the $x$-axis.

\begin{figure}[btp]
\begin{center}
\includegraphics[height=0.48\textheight]{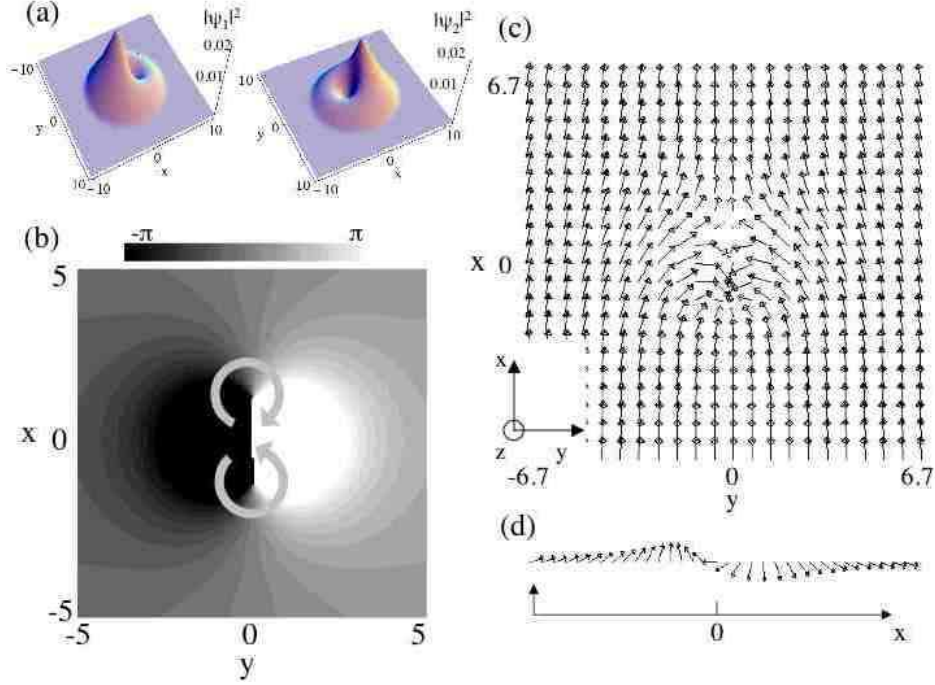}
\end{center}
\caption{A simulated meron pair. The values of the parameters are $u_{1}=u_{2}=u_{12}=1000$ ($c_{0}=1000$, $c_{1}=c_{2}=0$), $\Omega=0.15$ and $\omega_{\rm R} = 0.05$. (a) Density profile of the condensates $|\psi_{1}|^{2}$ and $|\psi_{2}|^{2}$ for $c_{0}=10^{3}$, $c_{1}=c_{2}=0$, $\Omega=0.15$ and $\omega_{\rm R}=0.05$. The two components have the same particle number. (b) Gray scale plot of the relative phase $\phi=\theta_{1}-\theta_{2}={\rm arg} \psi_{1} - {\rm arg} \psi_{2}$. Arrows show the direction of the circulation in relative-phase space. (c) Vectorial representation of the pseudospin texture of the vortex state (a), projected onto the $x$-$y$ plane. (d) Cross section of the spin texture (c) along the $x$ axis at $y=0$. } 
\label{nonaxivor}
\end{figure}
\begin{figure}[btp]
\begin{center}
\includegraphics[height=0.30\textheight]{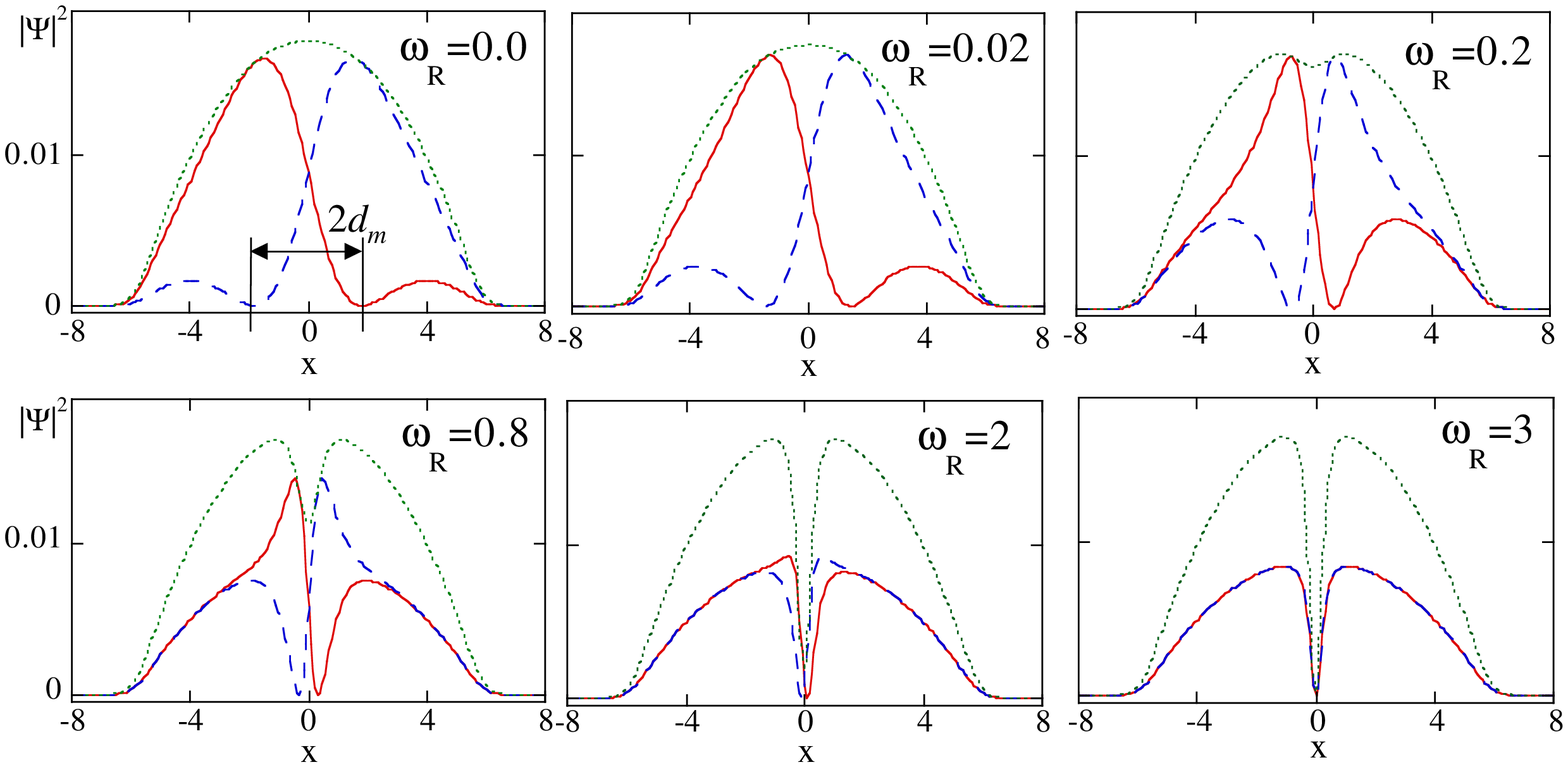}
\end{center}
\caption{Cross sections of a meron pair ($|\psi_{1}|^{2}$: solid-curve, $|\psi_{2}|^{2}$: dashed-curve, and $\rho_{\rm T}$: dotted-curve) along the $x$-axis at $y=0$ for $c_{0}=10^{3}$, $c_{1}=c_{2}=0$, $\Omega=0.15$ and $\omega_{\rm R}=$0, 0.02, 0.2, 0.8, 2, 3.} 
\label{meronfig}
\end{figure}

A typical solution of the vortex state with coherent coupling is shown in Fig. \ref{nonaxivor}. Each component has one off-axis vortex and the density peak of one component is located at the vortex core of the other component. This results in a coreless vortex in which the total density has no singularity. Because of the coherent coupling, the profile of the relative phase $\phi({\bf r})=\theta_{1}-\theta_{2}$ plays an important role in optimizing the structure, which is shown in Fig. \ref{nonaxivor}(b). The relative phase shows that the central region is characterized as a vortex-antivortex pair. In other words, the two vortices are connected by a branch cut of the relative phase with the $2 \pi$ difference, which is a characteristic domain wall structure in two-component BECs with internal coherent coupling\cite{Son}. Then, the vortex in one component attracts that in the other component, due to their being bound by the domain wall, thus forming a ``vortex molecule"\cite{Kasamatsupre}. As $\omega_{\rm R}$ increases, the size of the pair decreases as seen in Fig. \ref{meronfig}. Beyond $\omega_{\rm R} \simeq 3.0$, the two vortices completely overlap despite the intercomponent repulsive interaction.   

The corresponding spin texture is shown in Fig. \ref{nonaxivor}(c) and (d). Because of the transverse ``magnetic field", the spins are oriented in the $x$ direction everywhere except in the central domain-wall region where they tumble rapidly by $2\pi$. There exist two points, corresponding to the vortex centers, at which ${\bf S}$ is parallel to the $z$-axis. In Fig. \ref{nonaxivor}(c), the spin around the singularity with ${\bf S} = +{\bf \hat{z}}$ has a radial distribution, characterized by $(S_{x},S_{y}) \propto (-x,-y)$, and rolls by 90$^{\circ}$ as it goes outward, becoming perpendicular to the $x$-axis. For the spin around ${\bf S} = -{\bf \hat{z}}$, the distribution is instead hyperbolic $(S_{x},S_{y}) \propto (x,-y)$. This texture is known as a ``radial-hyperbolic" pair of merons, which has been discussed in the study of topological defects in superfluid $^{3}$He (Ref. \refcite{Salomaa}) and a double-layer quantum Hall system\cite{Girvin}. Since the trapping potential is axisymmetric, the energy is degenerate with respect to the orientation of the molecular polarization. If the molecule is polarized along the $y$-axis, the texture features a ``circular-hyperbolic" pair, but it can continuously transform into the radial-hyperbolic pair. 

We find that, in the SU(2) symmetric case, i.e., $u_{1}=u_{2}=u_{12}$ ($c_{1}=c_{2}=0$), both the topological charge density $q({\bf r})$ and the effective velocity ${\bf v}_{\rm eff}$ exhibit a radially isotropic profiles like that in Fig. \ref{skyrmtopological}, even though each component forms a nonaxisymmetric vortex configuration. This is due to the fact that the meron pair and the axisymmetric skyrmion are equivalent as a topological excitation for the case of $u_{1}=u_{2}=u_{12}$; they can transform to each other via an overall rotation of the pseudospin. Then, giving an infinitesimal value to $\omega_{\rm R}$ (i.e. applying a transverse magnetic field) is sufficient to stabilize a nonaxisymmetric configuration. We thus obtain a further insight of the nonaxisymmetric vortex by rotating the basis of the spinor so that the internal coupling becomes simpler. The Rabi term plays the role of a transverse ``magnetic field", making the $x$-axis as a preferred axis. Actually, by rotating the spinors as $\psi_{\pm}=(\psi_{1} \pm \psi_{2})/\sqrt{2}$ (the basis along with pseudo-$x$ axis), the coupled GP equations become 
\begin{eqnarray}
\biggl[ \frac{1}{2} \biggl( \frac{\nabla}{i} -{\bf \Omega} \times {\bf r} \biggr)^{2} + \tilde{V} + c_{0} \rho_{\rm T} \biggr] \psi_{+} 
= (\mu + \omega_{\rm R}) \psi_{+} , \\
\biggl[ \frac{1}{2} \biggl( \frac{\nabla}{i} -{\bf \Omega} \times {\bf r} \biggr)^{2} + \tilde{V} + c_{0} \rho_{\rm T} \biggr]  \psi_{-} 
 = (\mu - \omega_{\rm R}) \psi_{-} . \label{timeindGPeqhenkan} 
\end{eqnarray}
Here, the internal coupling is just the chemical potential difference between the ``$+$" and ``$-$" components. Then, the nonaxisymmetric structure in Fig. \ref{nonaxivor} is transformed to the axisymmetric vortex state in Fig. \ref{axisymvor}, that is, a skyrmion, where, the vortex core of the ``$+$" component is filled with the nonrotating ``$-$" component.  As one increases $\omega_{\rm R}$ the number of the ``$-$" particle decreases due to a decrease in the chemical potential of the $\psi_{-}$ component, and thus the vortex cores eventually become empty. This corresponds to the overlap of the vortex cores as shown in Fig. \ref{meronfig}. Therefore, the Rabi term alone does not break the axisymmetry of the topological excitation. However, inclusion of both the Rabi term and unequal coupling constants $u_{1} \neq u_{2} \neq u_{12}$ induces an anisotropy of the meron pair as will be discussed later.  

\subsubsection{A meron pair: Variational analysis}
In the case of $u_{1}=u_{2}=u_{12}$ $(c_{1}=c_{2}=0)$, the spin profile of the meron pair in Fig. \ref{nonaxivor} may be parametrized as\cite{Kasamatsupre,Kasamatsuskyrm}
\begin{equation}
S_{x}=\frac{r^{2}-4\lambda^{2} e^{-\alpha r^{2}}}{r^{2}+4\lambda^{2} e^{-\alpha r^{2}}}, \hspace{3mm} 
S_{y}=\frac{-4 \lambda y e^{-\alpha r^{2}/2}}{r^{2}+4 \lambda^{2} e^{-\alpha r^{2}}}, \hspace{3mm}
S_{z}=\frac{-4 \lambda x e^{-\alpha r^{2}/2}}{r^{2}+4 \lambda^{2} e^{-\alpha r^{2}}}. \label{meronprofiloe} \\
\end{equation}
Here, we assume that the meron pair is polarized along the $x$-axis. The only difference between this case and that for the skyrmion in Eq. (\ref{skyrmansatz}) is that the forms of $S_{x}$ and $S_{z}$ have been exchanged. In this case, the ratio of the particle number given in Eq.  (\ref{ratioparti}) is always unity, because $S_{z}$ is an odd function. The locations of the vortex cores are determined by two extremes of $S_{z}$, and are given by  
\begin{equation}
x^{2} = 4 \lambda^{2} e^{-\alpha x^{2}}. 
\label{moldistance}
\end{equation}
Then, it is natural to use the following form for the total phase $\Theta$:
\begin{equation}
\Theta = \tan^{-1} \frac{y}{x-2 \lambda e^{-\alpha r^{2}/2}} + \tan^{-1} \frac{y}{x+2 \lambda e^{-\alpha r^{2}/2}}.
\label{merontotpha}
\end{equation}
Substituting Eq. (\ref{meronprofiloe}) and Eq. (\ref{merontotpha}) into Eq. (\ref{sigmamoddimless}) with the Rabi term $- \omega_{\rm R} \rho_{\rm T} S_{x}$, we obtain the following total energy:  
\begin{eqnarray}
E=\int d^{2} r \biggl[ \frac{1}{2} (\nabla \sqrt{\rho_{\rm T}})^{2} + V_{\rm eff} \rho_{\rm T} -\omega_{\rm R} \rho_{T} \frac{r^{2}-4\lambda^{2} e^{-\alpha r^{2}}}{r^{2}+4\lambda^{2} e^{-\alpha r^{2}}} +c_{0} \frac{\rho_{\rm T}^{2}}{2} \biggr], \label{meroneneg}
\end{eqnarray}
where the effective confining potential $V_{\rm eff}$ is the same as that in Eq. (\ref{skyrmeffpot}). Because for any values of $\lambda$ and $\alpha$ Eq. (\ref{meroneneg}) gives the same energy as that of the axisymmetric skyrmion given by Eq. (\ref{symmskyreneg}), the skyrmion and the meron pair have the same optimized values of $\lambda$ and $\alpha$ if $\omega_{\rm R}=0$ (i.e., for the SU(2) symmetric case), and their energies are degenerate. 

\begin{figure}[btp]
\begin{center}
\includegraphics[height=0.21\textheight]{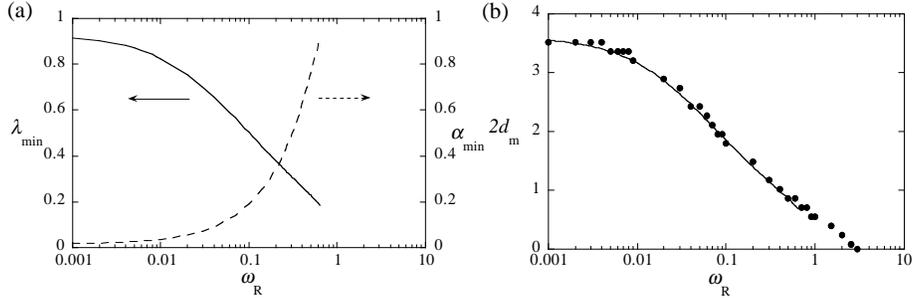}
\end{center}
\caption{Properties of a meron pair. (a) Optimized values of the variational parameters $\lambda$ (solid curve) and $\alpha$ (dashed curve) as a function of $\omega_{\rm R}$ for $c_{0}=10^{3}$, $\Omega=0.15$, $c_{1}=0$, and $c_{2}=0$. (b) The corresponding size $2d_{\rm m}$ of the meron pair (see Fig. \ref{meronfig} for definition) as a function of $\omega_{\rm R}$, calculated from Eq. (\ref{moldistance}) with $c_{0}=10^{3}$. Solid circles are the result obtained by numerical simulations.}  
\label{optimRabicase}
\end{figure}

Just as the $c_{1}$-term may be regarded as a magnetic field along the $z$-axis, the Rabi frequency $\omega_{\rm R}$ may be regarded as a magnetic field along the $x$-axis. Thus, the magnitude of $\omega_{\rm R}$ also influcences the stable size of the meron pair. The difference from the $c_{1}$-term is that the Rabi term is proportional to $\rho_{\rm T}$ instead of $\rho_{\rm T}^{2}$. We calculated the minimum values of $\lambda$ and $\alpha$ as a function of $\omega_{\rm R}$ under a slow rotation. The result is shown in Fig. \ref{optimRabicase}(a). As $\omega_{\rm R}$ increases, the minimized values of $\lambda$ ($\alpha$) becomes smaller (larger), and eventually vanishes (diverges). This behavior corresponds to a decrease in the size $d_{\rm m}$ of the meron pair (see in Fig. \ref{meronfig} for the definition) as shown in Fig. \ref{optimRabicase}(b), where $d_{\rm m}$ is calculated from Eq. (\ref{moldistance}). This indicates that the binding of the meron pair becomes stronger with increasing $\omega_{\rm R}$. The variational result agrees well with that obtained with direct numerical simulations. 

The meron pair arises from competion between repulsive and attractive interactions\cite{Kasamatsupre}. For well-separated merons, the repulsive interaction between them originates from the second and third terms of Eq. (\ref{sigmamoddimless}), which are the gradient energy of the pseudospin and the hydrodynamic kinetic energy of the ${\bf v}_{\rm eff}$-field. The contribution of the other terms are nearly independent of meron separation except for small $d_{\rm m}$ ($<0.30$ for $c_{0}=10^{3}$). Then, the Thomas-Fermi approximation cannot apply to the evaluation of the total energy; $\lambda_{\rm min}$ drops suddenly to zero with increasing $\omega_{\rm R}$ because there is no energy barrier associated with the quantum pressure of $\rho_{\rm T}$. On the other hand, the attractive force between the merons arises from a tension $T_{\rm d}$ in the domain wall of the relative phase, which is estimated to be $\sim T_{\rm d} d_{\rm m}$; for a homogeneous system $T_{\rm d} = 8 |\psi_{1}|^{2} |\psi_{2}|^{2} k / \rho_{\rm T}$ with the characteristic domain size $k^{-1}= (|\psi_{1}||\psi_{2}|/2\omega_{\rm R} \rho_{\rm T})^{1/2}$ (Ref. \refcite{Son}), and thus $T_{\rm d} \propto \sqrt{\omega_{\rm R}}$. Thus, the competition between the repulsive force and the attractive force creates an energy minimum and the two vortices can form a bound pair. 

\subsubsection{Effect of symmetry breaking terms}
\begin{figure}[btp]
\begin{center}
\includegraphics[height=0.52\textheight]{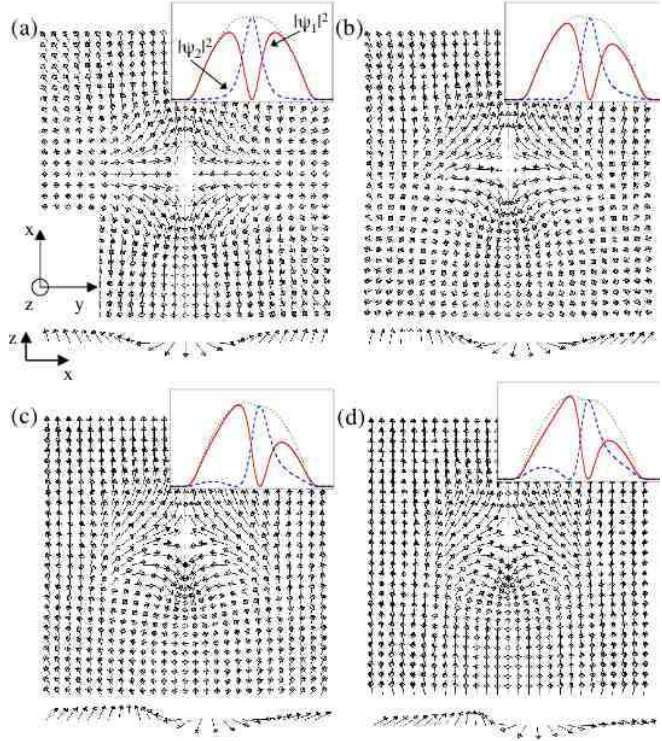}
\end{center}
\caption{Equilibrium structures of the spin texture in the presence of a longitudinal magnetic field ($c_{1}$) and a transverse magnetic field ($\omega_{\rm R}$). The range of the plot is $[-4 \leq x,y \leq +4]$. The values of the parameters are $\Omega=0.15$, $c_{0}=1000$, $c_{1}=-10$, $c_{2}=0$, and (a) $\omega_{\rm R}=0$, (b) $\omega_{\rm R}=0.02$, (c) $\omega_{\rm R}=0.05$, (d) $\omega_{\rm R}=0.10$. The insets show the cross sections of $|\psi_{1}|^{2}$ (solid curve), $|\psi_{2}|^{2}$ (dashed curve), and the total density $\rho_{\rm T}$ (dotted curve) along the $y=0$ line within the range $[-8 \leq x \leq +8]$.} 
\label{texturechange}
\end{figure}

Let us consider the effect of the symmetry breaking terms on a skyrmion and a meron pair. When $u_{1} \neq u_{2} \neq u_{12}$ ($c_{1}$, $c_{2} \neq 0$), the axisymmetry of the topological excitation is broken. First, we neglect the spin-spin interaction term ($c_{2}$ term) and investigate the dependence of the stable structure on $c_{1}$ and $\omega_{\rm R}$. This is equivalent to applying a longitudinal and transverse (pseudo)magnetic field. We prepare a stable axisymmetric skyrmion by applying the longitudinal magnetic field $c_{1}=-10$, and then turn on the transverse magnetic field $\omega_{\rm R}$. Figures \ref{texturechange}(a)-(d) show variations of the spin texture as well as those of the cross-section of the condensate density along the $y=0$ line when the Rabi frequency $\omega_{\rm R}$ is increased. The transverse magnetic field shifts the skyrmion off-center and converts it into a meron. At the same time, another meron enters from outside and forms an ``asymmetric" meron pair shown in Figs. \ref{texturechange}(c) and (d). This is a second-order phase transition because there is no energy barrier to destabilize the axisymmetric skyrmion with respect to the transverse magnetic field. In this case, the distribution of the topological charge $q({\bf r})$ moves from the center as $\omega_{\rm R}$ increases, and there is no dramatic change in the global shape from the isotropic skyrmion of Fig. \ref{skyrmtopological}. As $\omega_{\rm R}$ increases further, the peak of $q({\bf r})$ moves back to the center and its value goes to infinity when the two merons merge.

\begin{figure}[btp]
\begin{center}
\includegraphics[height=0.48\textheight]{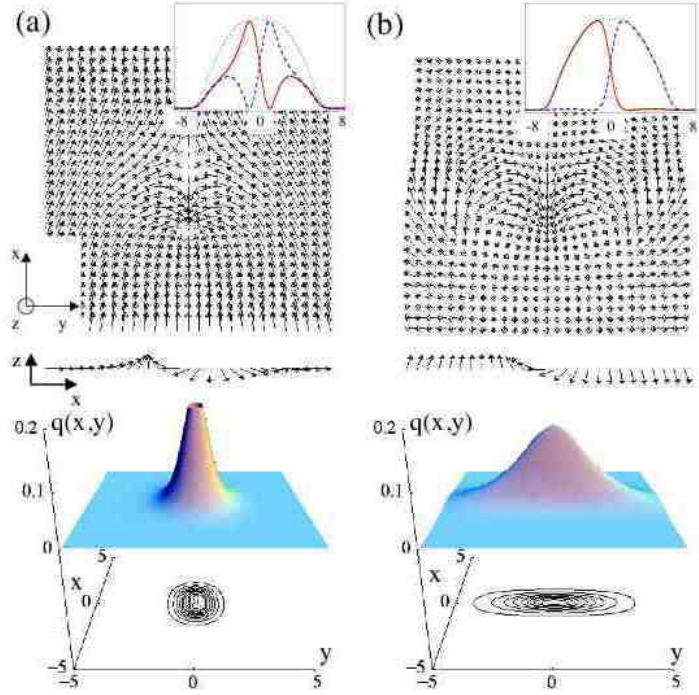}
\end{center}
\caption{Equilibrium structures of the spin texture in the presence of a spin-spin interaction $c_{2}$ and a transverse magnetic field ($\omega_{\rm R}$). The values of the parameters are $\Omega=0.15$, $c_{0}=1000$, $c_{1}=0$, $\omega_{\rm R}=0.05$, and (a) $c_{2}=20$, (b) $c_{2}=-20$. (top) The spin textures are displayed in the region $[-4 \leq x,y \leq 4]$. The insets show the cross sections of $|\psi_{1}|^{2}$ (solid curve), $|\psi_{2}|^{2}$ (dashed curve), and the total density $\rho_{\rm T}$ (dotted curve) along the $y=0$ line within the range $[-8 \leq x \leq +8]$. The surface plots at the bottom are the spatial distributions of the topological charge density $q({\bf r})$.} 
\label{c2ferroantiferro}
\end{figure}
Next, we study the effects of the spin-spin interaction term ($c_{2}$ term) and the dependence of the stable structure on $\omega_{\rm R}$ ($c_{1}$ is fixed at zero). First, we show that, in contrast to the case of $c_{1}$, the structure of the meron pair is sensitive to $c_{2}$. Figure \ref{c2ferroantiferro} shows the equilibrium structure of the condensate density, the spin texture and the topological charge density $q({\bf r})$ for both the antiferromagnetic case ($c_{2}=-20$) and the ferromagnetic case ($c_{2}=20$). For the antiferromagnetic case there is no significant difference in the density and spin profile, compared with the solution of $c_{2}=0$ in Figs. \ref{nonaxivor} and \ref{meronfig}. However, the topological charge density (i.e., vorticity) is distributed anisotropically in such a way that its distribution is elongated along the direction of polarization of the meron pair. For the ferromagnetic case, spin domains are formed, which gives rise to a considerable change in the density and spin profile as seen in Fig. \ref{c2ferroantiferro}(b). Most spins align up or down along the $z$-axis on either side, sandwiching the domain wall across which the spin flips rapidly. If $\omega_{\rm R}$ is increased further, the spins on both sides lay along the $x$-axis, forming the well-defined meron pair as in Fig. \ref{meronfig}. Then, the topological charge density is distributed with the direction perpendicular to the direction of polarization of the meron pair, being concentrated on the domain-wall region. This anisotropy of the meron pair gives an interesting situation when the condensates undergo a rapid rotation; the anisotropic interaction between the meron pairs generates a distorted lattice of ``vortex molecules"\cite{Kasamatsupre}, which will be discussed in Sec. \ref{Joseph}.

\subsection{Topologically nontrivial skyrmions in a 3-D configuration}\label{3Dskyrmion}
In contrast to the topologically {\it trivial} skyrmion discussed above, we review here the topologically {\it nontrivial} skyrmion excitations in multicomponent BECs\cite{Kawaja2}$^{-}$\cite{Savage}. These skyrmions have complicated 3-D structures and are characterized by spin inversion in a finite region of space and a typical topological charge $Q=1$. It is known that topological considerations allow for these excitations, but it is not known whether or not such a configuration is energetically favorable. 

The first example of a nontrivial skyrmion was discussed by Al Khawaja and Stoof\cite{Kawaja2}. They considered uniform two-component condensates subject to the SU(2) symmetry and the {\it radially symmetric} profile of the skyrmion by assuming that the normalized spinor is characterized by $\chi({\bf r})=\exp \{ - 2 i {\bf \Omega} ({\bf r}) \cdot {\bf S} \} \chi_{z}$, where ${\bf S}=\bm{\sigma}/2$ with the Pauli matrix $\bm{\sigma}$, $\chi_{z}=(1,0)^{T}$, and ${\bf \Omega} ({\bf r}) = \omega(r) {\bf r}/r$ which determines a specific spin texture of the skyrmion. The physical meaning of this formula is that the average spin at a position ${\bf r}$ is rotated by an angle 2$\Omega({\bf r})$ from its initial orientation with the axis of rotation ${\bf \Omega} ({\bf r})/\Omega({\bf r})$. The boundary conditions for $\omega(r)$ that represents the spin distribution for the skyrmion are the following. First, at $r \rightarrow \infty$ all spins must be oriented as in the ground state, that is, $\omega(r \rightarrow \infty)=0$. Along the $z$-axis the spins are also not rotated due to our assumption about ${\bf \Omega}$; thus, to have a nonsingular texture of the spinor with a non-zero winding number, we must require $\omega(0)=2\pi$. Finally, to avoid the singular behavior of ${\bf \Omega} ({\bf r})$ itself, we use only functions $\omega(r)$ with zero slope at the origin. Therefore, $\omega(r)$ is a monotonically decreasing function that starts from $2 \pi$ at the origin and reaches zero as $r \rightarrow \infty$. 

Al Khawaja and Stoof\cite{Kawaja2} used a variational ansatz $\omega(r)=4 \cot ^{-1} (r/\lambda)^{2}$ that satisfies the above boundary conditions, where $\lambda$ represents the size of the skyrmion. Substituting this ansatz into the energy functional in Eq. (\ref{energyfunctio2}) with $V_{i}=0$ and $\Omega=0$ gives the total energy as a function of the total density $\rho_{\rm T}$ and $\lambda$. For the given $\omega(r)$, they calculated $\rho_{\rm T}(r)$ exactly by solving numerically the differential equation for $\rho_{\rm T}(r)$ obtained by varying $E[\rho_{\rm T}(r),\lambda]$ with respect to $\rho_{\rm T}(r)$. By substituting this density profile back into the energy functional, the energy of the skyrmion is expressed as a function of only $\lambda$ and the equilibrium properties can be obtained by minimizing this energy with respect to $\lambda$. The resulting spin texture shows a complicated spin distribution; the spin $\langle S_{z} \rangle$ flips five times as it goes outward from the center in the $x$-$y$ plane, and $\langle S_{x} \rangle$ and $\langle S_{y} \rangle$ exhibit quadrupole profiles with one radial node in the $y$-$z$ plane\cite{Kawaja2}.

A crucial problem is the stability of the skyrmion. In equilibrium, the energy is minimized for the vanishing size of the skyrmion\cite{Kawaja2}. However, for sufficiently small skyrmion sizes, a nonequilibrium stability mechanism starts to work. A number of atoms in the center of the skyrmion, hereafter referred to as core atoms, will be trapped by an effective 3-D potential barrier, that is, a repulsive shell with a finite radius that is created by the gradient term of the spin texture $\hbar^{2} |\nabla \chi ({\bf r})|^{2}/2m$. As the skyrmion shrinks in size, the barrier height of the repulsive shell increases and the radius decreases. This leads to a squeezing of the core, thereby increasing its energy, and ultimately stabilizing the skyrmion. In this case, it is not an equilibrium state of the condensate because the core atoms will tunnel through the barrier and give the skyrmion a finite lifetime. Detailed calculations of this tunneling rate\cite{Kawaja2,Zhang} showed that only a few atoms in the core of the skyrmion are needed to stabilize it and to give it a sufficiently long lifetime. 

\begin{figure}
\begin{center}
\begin{minipage}{3.9cm}
\includegraphics[height=0.125\textheight]{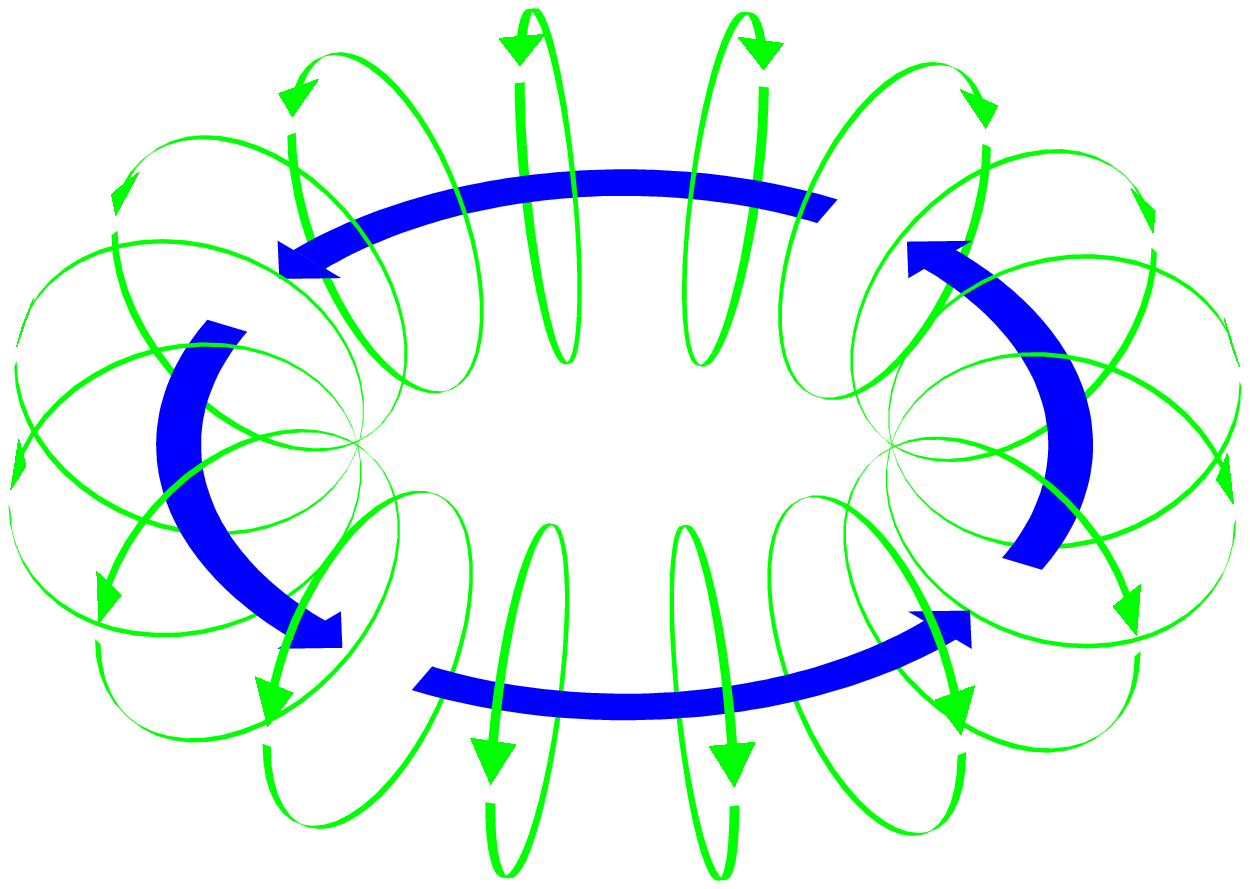}
\end{minipage}
\begin{minipage}{4.3cm}
\includegraphics[height=0.30\textheight]{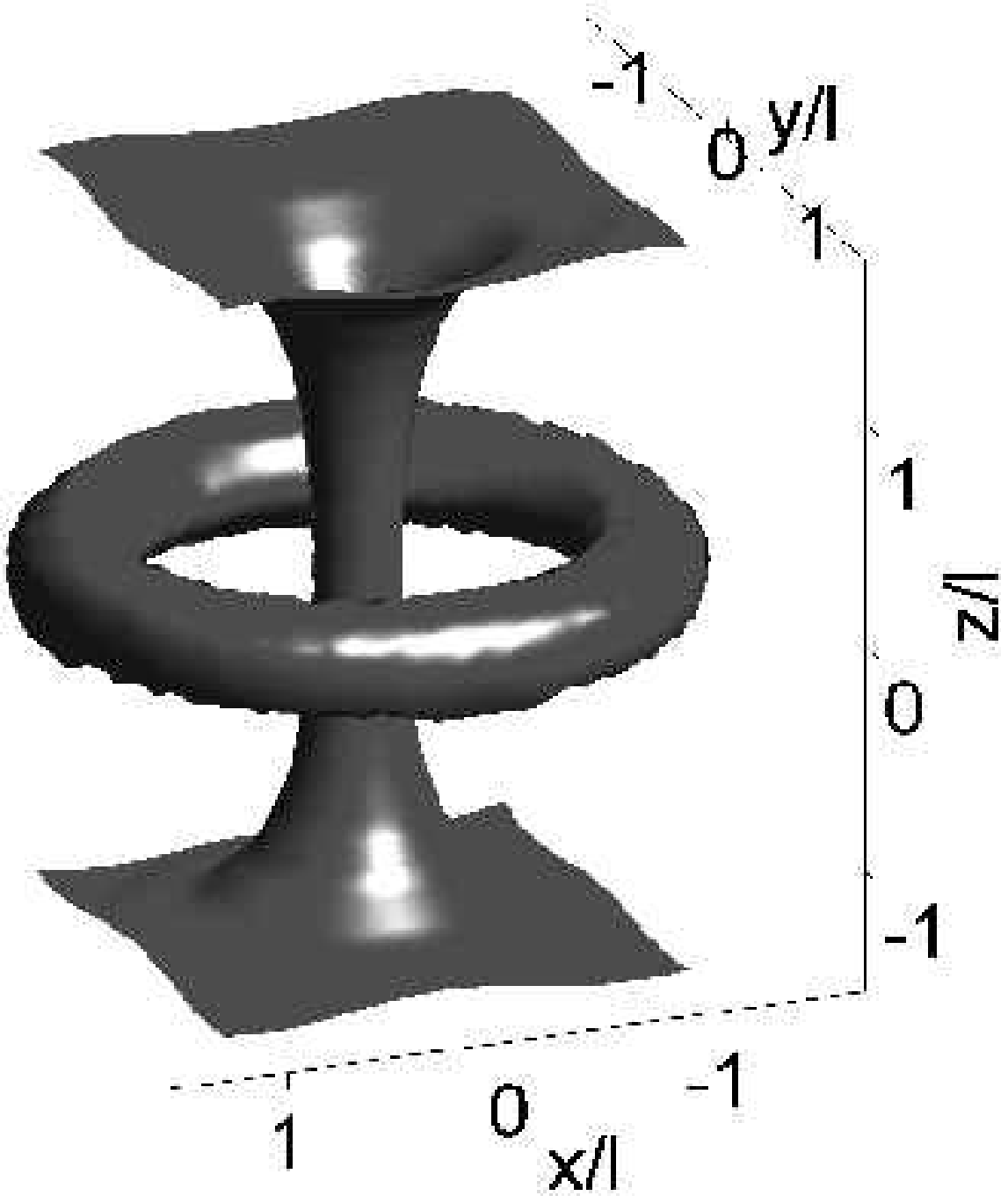}
\end{minipage}
\end{center}
\caption{A 3-D skyrmion (left), as a vortex ring supporting a superflow, and a 3-D skyrmion (right) with a constant surface density, which was obtained by numerical simulations. The vortex core and the ring are shown as the surface of constant atomic density. [Ruostekoski and Anglin, Phys. Rev. Lett. {\bf 86} 3934 (2001), reproduced with permission. Copyright (2001) by the American Physical Society.]} \label{f1}
\end{figure}

Other interesting structures of nontrivial 3-D skyrmion were investigated in Refs. \refcite{Ruostekoski,Battye,Savage}. The authors in Refs. \refcite{Ruostekoski,Battye,Savage} also considered the rotational operator of the spinor as $\chi({\bf r})=\exp \{ i \lambda({\bf r}) \bm{\sigma} \cdot ({\bf r}/r) \} \chi_{z}$, where $\lambda$ is a monotonic function imposed by the boundary condition for $\lambda$ as $\lambda(0)=0$ and $\lambda=\pi$ at the boundary of the condensate; for this boundary condition the topological charge becomes $Q=1$. The corresponding condensate wave functions are
\begin{eqnarray}\label{skyrmsolsr}
\left(
\begin{array}{r}
\psi_1({\bf r}) \\
\psi_2({\bf r})
\end{array} \right)
= \sqrt{\rho_{\rm T}} \left(
\begin{array}{r}
\cos[\lambda({\bf r})]-i\sin[\lambda({\bf r})]\cos\theta \\
-i\sin[\lambda({\bf r})]\sin\theta\exp(i\phi)
\end{array} \right).
\end{eqnarray}
Here, $(\lambda, \theta, \phi)$ can be understood as the spherical angles of the 3-sphere (in four-dimensional parameter space). Because of the topological stability of the skyrmion, any continuous deformation of Eq. (\ref{skyrmsolsr}), without altering the asymptotic boundary values, is still a skyrmion with $Q=1$. With this realization of the skyrmion, $|\psi_2|^2$ vanishes both on the $z$-axis and at infinity, and is concentrated in a toroidal region. On the other hand, $|\psi_1|^2$ vanishes on the circle $\theta=\pi/2$ and $r=\lambda^{-1}(\pi/2)$, a region where $|\psi_2|^2$ is concentrated. Expanding $\psi_1$ about any point on this circle, we find $\psi_1 \propto \delta r + i\delta\theta$, indicating that the nodal circle of $\psi_1$ represents a vortex line. Hence the $\psi_1$ component forms a vortex ring, within whose core $\psi_2$ resides and flows azimuthally with the winding number of one. The skyrmion can be viewed as a quantized vortex ring in one component, which is filled with the other component that also carries the quantized circulation along the ring. This is shown in Fig. \ref{f1}. This configuration closely resembles the ``cosmic vortons" which might have been formed in the early universe from superconducting cosmic strings\cite{Davis}. 

Ruostekoski and Anglin\cite{Ruostekoski} proposed a carefully designed sequence of Rabi transitions that create the skyrmion configuration of Eq. (\ref{skyrmsolsr}). They argued that the subsequent dissipative time evolution exhibited unstable dynamics that caused the vortex ring component to shrink to zero radius. However, since the general multicomponent system is characterized by several kinds of two-body scattering lengths, whether a stable skyrmion exists or not is a highly nontrivial problem. For homogeneous two-component condensates, Battye {\it et al.} found that the energetically stable skyrmion can exist under the condition that phase separation occurs, that is, if the condition (\ref{sepcond}) holds\cite{Battye}. In this case, the two components can strongly repel each other and the toroidal filling due to the line component can prevent the vortex ring from shrinking to nothing. Consequently, a filled vortex ring, as opposed to an empty vortex ring, can be more stable against collapse. Moreover, in the skyrmion the filling has one unit of angular momentum about the $z$-axis, resulting in a $1/r^{2}$ centrifugal barrier that further hinders the ring from shrinking. 

Savage and Ruostekoski extended the analysis to a harmonically trapped system and found that there are additional instabilities that will play a crucial role in experiments\cite{Savage}. For sufficiently large total number of atoms in the vortex-line component, the nonlinear repulsion between the two components is strong enough to inhibit the collapse of the vortex ring, stabilizing the skyrmion against shrinking for cylindrically symmetric initial states. However, due to the inhomogeneous potential, the skyrmion is still unstable with respect to the drift of the vortex line towards the edge of the BEC. Once the vortex line drifts to a low-density region, the nonlinear repulsion is no longer strong enough to prevent the shrinking of the skyrmion and the collapse occurs as described above. The drift reduces the total angular momentum of the atoms, indicating an energetic instability. Physically, this results from dissipation (e.g., from thermal atoms). Thus, additional physical mechanism, such as rotation or optical pinning potentials, will be required for stabilization. 

Rotation of only the vortex-line component gives the simplest example for stabilization, but this is difficult. Moreover, rotation of both components introduces a new instability mechanism. For sufficiently rapid rotations the line component again reaches the boundary, resulting in $Q<1$. At the same time the ring vortex accommodated the rotation by twisting and finally breaking at the condensate boundary. Nevertheless, for a small range of rotation frequency near $0.085 \omega_{\perp}$, the skyrmion was stable against both shrinkage of a vortex ring and drift of a vortex line. Then, the stable skyrmion configuration requires a shift of the rotation axis to off-center . The breaking of cylindrical symmetry implies a family of degenerate skyrmions parametrized by the azimuthal angle. Another method for stabilizing the skyrmion is to inhibit the drift towards low-density regions by creating a positive density gradient around the vortex line. This can be implemented with a blue-detuned Gaussian laser beam along the $z$ axis, providing a cylindrically symmetric repulsive Gaussian dipole potential perpendicular to the $z$ axis. Under these conditions, we expect that atomic BECs, which have been imprinted with a topological configuration, will relax via dissipation to a stable skyrmion configuration. 

\subsection{Dynamic properties of vortices}\label{stabdyn}
The dynamic properties  of vortices in two-component BECs are also expected to be extremely rich because of the various kinds of interactions between the two superfluids. A remarkable feature of this system arises from ``buoyancy"\cite{Anderson2,McGee,Ohberg}, which refers to a net mean-field force on a constituent of a multicomponent condensate, the force being dependent on the various intraspecies and interspecies interaction parameters. 

As shown in Sec. \ref{experim}, the unstable dynamics of the vortex was observed, depending on which component has a vortex in the $^{87}$Rb condensate\cite{Matthews}. This phenomenon is understood from linear stability analysis of the vortex state based on the mean-field approach\cite{Ripoll3,Ripoll31,Skrybin}. We use the following notation for the states: $(1,0)$ for the state with the vortex in $|1\rangle $ and $(0,1)$ for the state with the vortex in $|2\rangle $. In the JILA experiment, two components had the same number of particles ($N_1=N_2=N$) but, in general, one could consider any ratio between the populations of the different hyperfine states. For $^{87}$Rb atoms, the relation of the coupling constants $g_{1}>g_{12}>g_{2}$ ($a_{1}:a_{12}:a_{2}=1.03:1:0.97$) favors a configuration that has its first component spread over the largest part of the space, as shown in the previous discussion of equilibrium. Numerical simulations show that for equal populations $N_{1}=N_{2}=N$, and with the above coupling constants, the stationary states $(1,0)$ is stable while the other state $(0,1)$ is unstable\cite{Ripoll3,Ripoll31,Skrybin}.

The origin of the instability of the state $(0,1)$ is purely dynamical, which can be understood within the framework of the mean-field theory for the two-component system without dissipation. Actually, the instability does not lead to expulsion of the vortex from the condensate, but instead to periodic transfer of the vortex from one component to the other. To study the vortex stability, one starts from the time-dependent, coupled Gross-Pitaevskii equations for the condensate wave functions of each species: 
\begin{eqnarray}
i\hbar \frac{\partial \Psi_{1}}{\partial t}=\left[ -\frac{\hbar^{2}\nabla^{2}}{2m}+V_{1}+g_{1}|\Psi_{1}|^{2}+g_{12}|\Psi_{2}|^{2} \right] \Psi _{1}, \label{m6}  \\
i\hbar \frac{\partial \Psi_{2}}{\partial t}=\left[ -\frac{\hbar^{2}\nabla^{2}}{2m}+V_{2}+g_{2}|\Psi_{2}|^{2}+g_{12}|\Psi_{1}|^{2} \right] \Psi_{2}.  \label{m7}
\end{eqnarray}
These equations are the particular case of Eqs. (\ref{2TDCGP1}) and (\ref{2TDCGP2}) in which the drive is turned off ($\hat H_1$, $\omega_{\rm R}$, $\Delta=0$). Equations (\ref{m6}) and (\ref{m7}) conserve the number of particles in each hyperfine state. However, the angular momentum of each component is no longer a conserved quantity. Instead, the conserved quantity is the total angular momentum $\langle L_{z} \rangle= \int d^3r ( {\Psi}_{1}^{*} L_{z} \Psi_{1} + {\Psi}_{2}^{*} L_{z} \Psi_{2})$. As in the JILA experiments, both potentials are assumed to be spherically symmetric and have the form $V_1({\bf r})=V_2({\bf r})=\frac 12 m \omega_\perp^2 (r^2+z^2)$. For stationary configurations in which each component has a well-defined value of the angular momentum, the time and angular dependences of the wave functions are factored out as
\begin{equation}
\Psi _i(r,z,\phi )=e^{-i\mu _it/\hbar }e^{i \gamma_i \theta }\psi_i(r,z)  \hspace{4mm} (i=1,2). \label{m9}
\end{equation}
Here, we focus on the two axisymmetric configurations with vorticity $(\gamma_1,\gamma_2)=(1,0)$ and $(0,1)$, which correspond to the single vortex states for the $|1\rangle $ and $|2\rangle $ components shown in Fig. \ref{JILAvortexdensity}.

Linear stability analysis of the two axisymmetric vortex states leads to the following results. Among the normal modes of the $(1,0)$ case, there is a negative eigenvalue, which means that there is a path in the configuration space along which the energy decreases. This path, which is associated with the mode of the dipole type, belongs to a perturbation that causes the initial displacement of the vortex from the trap center. However, this drift instability of the vortex is affected by dissipation; in particular, without dissipation, the configuration is dynamically stable. But, it can be energetically stabilized by turning on rotation as discussed in Sec. \ref{axisym}. In the $(0,1)$ case, on the other hand, there are normal modes with complex frequencies. The unstable modes have a similar shape of the negative-eigenvalue modes of the $(1,0)$ case, which means that the perturbations also displace the vortex from the center. The imaginary part of the eigenvalues implies that vortices with unit charge in $|2\rangle $ are dynamically unstable under a generic perturbation of the initial configuration and this instability grows without dissipation. This result is consistent with the JILA experiments in Fig. \ref{JILAwatch}, in which a vortex in the $|2\rangle $ component was unstable and collapsed.

Numerical simulations of the vortex dynamics for large perturbations were done by Garc\'{i}a-Ripoll and P\'{e}rez-Garc\'{i}a\cite{Ripoll3}. The results showed that the linearly stable state $(1,0)$ is robust and survives under a wide range of perturbations; the vortex only shows a precession around the center. In contrast, the unstable configuration $(0,1)$ gives rise to recurrent dynamics: the first component and the vortex oscillate synchronously, where the vortex core in $|2\rangle $ pins the density peak of $|1\rangle $. These oscillations grow in amplitude until the vortex spirals out. The displaced vortex carries less than the total angular momentum, and so the vortex precesses around the center. Because of this and the conservation of the total angular momentum, the vortex is transferred from $|2\rangle $ to $|1\rangle $. Although the dynamics is not completely periodic, this mechanism exhibits some recurrence and the vortex eventually returns to $|2\rangle $. The preceding behavior persists even for strong perturbations in a 2-D condensate. However, for large perturbations of a 3-D condensate, the dynamics may be chaotic.

The preceding results are valid when $N_{1}=N_{2}$. For an arbitrary ratio of the populations $N_1/N_2$ and arbitrary values of the coupling constants $g_{1}$, $g_{2}$ and $g_{12}$, the dynamical stability conditions can be estimated analytically\cite{Ripoll31,Skrybin}. Under the two-mode approximation\cite{Ripoll31}, where the wave function is approximated as $\Psi_{i} \simeq a_{i}(t) \Psi_{g}({\bf r}) + b_{i}(t) \Psi_{e}({\bf r})$ with the eigenfunctions of the $d$-dimensional harmonic oscillator $\Psi_{g}({\bf r}) = (1/\pi)^{d/2} e^{-r^{2}/2}$ and $\Psi_{e}({\bf r}) = (2/d\pi)^{d/2} r e^{-r^{2}/2} e^{i \theta}$, the configuration $(1,0)$ is stable if
\begin{equation}
\left( \sqrt{\frac{N_1}{N_2}}-1\right) ^2>1-\frac{a_{1}}{a_{12}}. \label{m10}
\end{equation}
For $^{87}$Rb, the inequality (\ref{m10}) is always satisfied because the right-hand side is negative, which proves that the configuration with a vortex in $|1\rangle $ is always linearly stable. Note that the stability properties do not depend on the total number of particles but only on the ratio between the populations. The stability condition of the configuration $(0,1)$ can be obtained in a similar manner as
\begin{equation}
\left( \sqrt{\frac{N_2}{N_1}}-1\right) ^2>1-\frac{a_{2}}{a_{12}}. \label{m11}
\end{equation}
This inequality fails for a certain range of $N_1/N_2$. For the case of $^{87}$Rb the unstable range is $0.68<N_{2}/N_{1}<1.38$, which means that certain choices of the population imbalance allow stabilization of the vortex in $|2\rangle$. More detailed calculations were done by Skyrbin\cite{Skrybin} who derived the stability condition as
\begin{equation}
\frac{a_{2}}{a_{12}} > \frac{N_{1}}{N_{2}} \left( 2 \sqrt{2 \left( \frac{N_{2}}{N_{1}} +1 \right)} -3 \right),  \label{m12}
\end{equation}
which gives an unstable range of $0.63<N_{2}/N_{1}<1.69$, which is slightly different from Eq. (\ref{m11}). These results predict the possibility to dynamically stabilize the axisymmetric vortex states for various multiple-condensate systems. 

The effect of the intercomponent interaction was also demonstrated in a precession of an off-centered vortex core\cite{Anderson2}. This is a consequence of a Magnus effect; that is, an applied force on a rotating cylinder in a fluid causes pressure imbalances at the cylinder surface that makes the cylinder drift orthogonal to the force. The experiment observed that the precession frequency of the filled core is slower than that of the empty core\cite{Anderson2}. The slower precession of filled cores can be understood in terms of the buoyancy effect. Because of its slightly smaller scattering length, the $| 2 \rangle$ component has negative buoyancy with respect to the $| 1 \rangle$ component, and consequently tends to sink inward towards the center of the condensate. With increasing amounts of the $|2\rangle$ component in the core, the inward force on the core begins to counteract the outward buoyancy of the vortex velocity field, resulting in a reduced precession velocity. It is predicted that with a filling component of sufficiently negative buoyancy in the core, the core precession may stop or even precess in the direction opposite to the direction of the fluid flow\cite{McGee}. 
 
\subsection{Quantized vortices in spin-1 Bose-Einstein condensates}\label{spin1vor}
Another multicomponent Bose-condensed system is a ``spinor BEC" characterized by the hyperfine spin $F=1$. An optical trap removes the restriction of the confinable hyperfine spin states, realizing new multicomponent superfluids with internal degrees of freedom\cite{Stenger}$^{-}$\cite{Kuwamoto}. In this section, we review briefly the study of quantized vortices in such a BEC with $|{\bf F}|=1$ hyperfine spin states $|F=1, m_{F}=\pm1,0 \rangle$. 

To describe the $F=1$ spinor BEC, a generalized GP mean-field model was introduced by Ohmi and Machida\cite{Ohmi} and Ho\cite{Ho3}. With the mean-field approximation, the total Hamiltonian of the system reduces to an energy functional similar to Eq. (\ref{energyfunctio2}):
\begin{eqnarray}
E [\Psi_{i}, \Psi_{i}^{\ast} ] = \int d {\bf r} \biggl[ \sum_{i} \Psi_{i}^{\ast} \left( -\frac{\hbar^{2} \nabla^{2}}{2m} + V({\bf r}) - \Omega L_{z} \right) \Psi_{i} + \frac{g_n}{2} \sum_{i,j} \Psi_{i}^{\ast} \Psi_{j}^{\ast} \Psi_{j} \Psi_{i} \nonumber \\ 
+ \frac{g_{s}}{2} \sum_{\alpha} \sum_{ijkl} \Psi_{i}^{\ast} \Psi_{j}^{\ast} (\hat{F}_{\alpha})_{ik} (\hat{F}_{\alpha})_{jl} \Psi_{k} \Psi_{l} \biggr]. 
\label{energyfunspin1}
\end{eqnarray}
Here we consider the system in a rotating frame with ${\bf \Omega}=\Omega \hat{\bf z}$ and the subscripts represent $\alpha = (x,y,z)$ and $i,j,k,l = (0, \pm1)$. The condenate wave functions are characterized by the hyperfine sublevels $m_{F} = 1, 0, -1$ as $\Psi = (\Psi_{1}, \Psi_{0}, \Psi_{-1})$. If one of the component $\Psi_{i}$ is fixed at zero, Eq. (\ref{energyfunspin1}) reduces to the energy functional of the two-component case Eq. (\ref{energyfunctio2}); the crucial difference from the two-component problem is the presence of the $g_{s}$-term. The coupling constants $g_{n}$ and $g_{s}$ are written as   
\begin{equation}
g_{n}=\frac{4 \pi \hbar^{2}}{m} \frac{2a_{2}+a_{0}}{3}, \hspace{5mm}
g_{s}=\frac{4 \pi \hbar^{2}}{m} \frac{a_{2}-a_{0}}{3},
\end{equation}
where $a_{0}$ and $a_{2}$ are the scattering lengths for two colliding atoms with total spin angular momentum $F_{\rm tot}=0$ and $2$. The spin operators $\hat{F_{\alpha}} (\alpha = x, y, z)$ can be expressed in matrix form as 
\begin{eqnarray}
 F_x = \frac{1}{\sqrt{2}}
       \left( 
              \begin{array}{ccc}
              0 & 1 & 0\\
              1 & 0 & 1\\
              0 & 1 & 0
              \end{array}
       \right),  \hspace{2mm}
 F_y = \frac{i}{\sqrt{2}}
       \left(
              \begin{array}{ccc}
              0 & -1 &  0\\
              1 &  0 & -1\\
              0 &  1 &  0
              \end{array}
       \right),  \hspace{2mm}
 F_z =
       \left( \;
              \begin{array}{ccc}
              1 &  0 &  0\\
              0 &  0 &  0\\
              0 &  0 & -1
              \end{array}
       \right).
\end{eqnarray}
Note that the term proportional to $g_{s}$ in Eq. (\ref{energyfunspin1}) describes the spin exchange interaction, which is ferromagnetic for $g_{s}<0$ and antiferromagnetic for $g_{s}>0$. The spinor BEC of $F=1$ $^{23}$Na atoms studied by the MIT group was demonstrated to have an antiferromagnetic interaction\cite{Stenger}, while the $F=1$ $^{87}$Rb condensate has a ferromagnetic one\cite{Chang,Klausen}. The sign of $g_{s}$ plays a crucial role in determining the stable vortex structure. The time-independent coupled GP equations are derived by the requirement that the energy functional (\ref{energyfunspin1}) be extremal:
\begin{eqnarray}
\left(-\frac{\hbar^{2} \nabla^{2}}{2m} + V({\bf r}) - \Omega L_{z} \right) \Psi_{1} + (g_{n}+g_{s})(\sum_{i} \Psi_{i}^{\ast} \Psi_{i})\Psi_{1} - 2g_{s} \Psi_{-1}^{\ast} \Psi_{-1} \Psi_{1} \nonumber \\ 
+ g_{s} \Psi_{-1}^{\ast} (\Psi_{0})^{2} = \mu_{1} \Psi_{1}, \label{3gpe1} \\
\left(-\frac{\hbar^{2} \nabla^{2}}{2m} + V({\bf r}) - \Omega L_{z} \right) \Psi_{0} + (g_{n}+g_{s})(\sum_{i} \Psi_{i}^{\ast} \Psi_{i})\Psi_{0} -g_{s} \Psi_{0}^{\ast} \Psi_{0} \Psi_{0} \nonumber \\ 
+ 2g_{s} \Psi_{0}^{\ast} \Psi_{1} \Psi_{-1} = \mu_{0} \Psi_{0}, \label{3gpe2} \\
\left(-\frac{\hbar^{2} \nabla^{2}}{2m} + V({\bf r}) - \Omega L_{z} \right) \Psi_{-1} + (g_{n}+g_{s})(\sum_{i} \Psi_{i}^{\ast} \Psi_{i})\Psi_{-1} - 2g_{s} \Psi_{1}^{\ast} \Psi_{1} \Psi_{-1} \nonumber \\ 
+ g_{s} \Psi_{1}^{\ast} (\Psi_{0})^{2} = \mu_{-1} \Psi_{-1}, \label{3gpe3}
\end{eqnarray}
where the three chemical potentials $\mu_{i}$ ($i=0, \pm 1$) play the role of Lagrange multipliers. They are determined so as to conserve the total particle number $N= \int d {\bf r} (| \Psi_1 ({\bf r}) |^2 + | \Psi_0 ({\bf r}) |^2 + | \Psi_{-1} ({\bf r}) |^2) $ and magnetization $M = \int d {\bf r} (| \Psi_{1} ({\bf r}) |^2 - | \Psi_{-1} ({\bf r}) |^2)$. The angular momentum $\langle L_{z} \rangle = \sum_{i} \int d {\bf r} \Psi_{i} L_{z} \Psi_{i}$ should also be considered as a conserved quantity. 

\subsubsection{Coreless vortex states}
First consider the structure of axisymmetric vortex states. The detailed analysis was made by Isoshima and Machida\cite{Isoshima}. In this structure, the vortex configuration is classified by the combination of the winding number $\gamma_{i}$ of the condensate wave function $\Psi_{j}=\sqrt{\rho_{j}(r)} \exp [i (\alpha_{j} + \gamma_{j} \theta)]$ with the cylindrical coordinate ($r,\theta,z$), where the homogenity of the wave functions along the $z$-axis is assumed, and $\alpha_{j}$ is the overall phase of the $j$-th component. The phase factors of $\Psi_{j}$ are determined such that the energy of the $g_{s}$ term in the integrand of Eq. (\ref{energyfunspin1}) is minimized; this term reads
\begin{eqnarray}
E_{s} &=& \frac{g_{s}}{2} \sum_{\alpha} \sum_{ijkl} \Psi_{i}^{\ast} \Psi_{j}^{\ast} (\hat{F}_{\alpha})_{ik} (\hat{F}_{\alpha})_{jl} \Psi_{k} \Psi_{l} \nonumber \\ 
&=& \frac{g_{s}}{2} \biggl\{ 2 \rho_{0} \left[ \rho_{1} + \rho_{-1} + \sqrt{\rho_{1} \rho_{-1}} (e^{i (\alpha_{1}+\alpha_{-1}-2\alpha_{0}) + i (\gamma_{1}+\gamma_{-1}-2\gamma_{0}) \theta} + {\rm c.c.}) \right] \nonumber \\
& & + (\rho_{1} - \rho_{-1})^{2}   \biggr\}. 
\label{spinexint}
\end{eqnarray}
To minimize $E_{s}$, $\alpha_{j}$ and $\gamma_{j}$ should satisfy 
\begin{eqnarray}
2 \alpha_{0} &=& \alpha_{1} + \alpha_{-1} + n \pi,  \\
2 \gamma_{0} &=& \gamma_{1} + \gamma_{-1}, \label{betacond}
\end{eqnarray}
where $n$ is an integer, and the odd (even) $n$ corresponds to the antiferromagnetic (ferromagnetic) interaction. The global phase $\alpha_{j}$ has no effect on the vortex structure, and will therefore be set to zero. The possible combination of $\gamma_{j}$ satisfying Eq. (\ref{betacond}) gives this system a characteristic vortex structure. If the values of $\gamma_{j}$ are restricted to being $\gamma_{j} \leq 1$, we have $(\gamma_{1},\gamma_{0},\gamma_{-1})=(1,0,-1)$, $(1,1,1)$ and $(1,1/2,0)$. In the case of $(1,1/2,0)$, the value $1/2$ is not allowed in the axisymmetric system, so that the $\Psi_{0}$-component vanishes; this vortex state is called the ``Alice state" (half-quantum vortex state)\cite{VLeonhardt,Isoshima}. Therefore, there is one-to-one correspondence between the Alice state and the axisymmetric vortex state in two-component BECs discussed in Sec. \ref{axisym} (i.e., $\Psi_{1} (\Psi_{-1}) \rightarrow \psi_{1} (\psi_{2})$ in Fig. \ref{axisymvor}). The structure and the stability of these vortex states for various $\Omega$ and the total magnetization $M$ were thoroughly investigated in Ref. \refcite{Isoshima}.

\begin{figure}
\begin{center}
\begin{tabular}{cc}
\includegraphics[width=6cm]{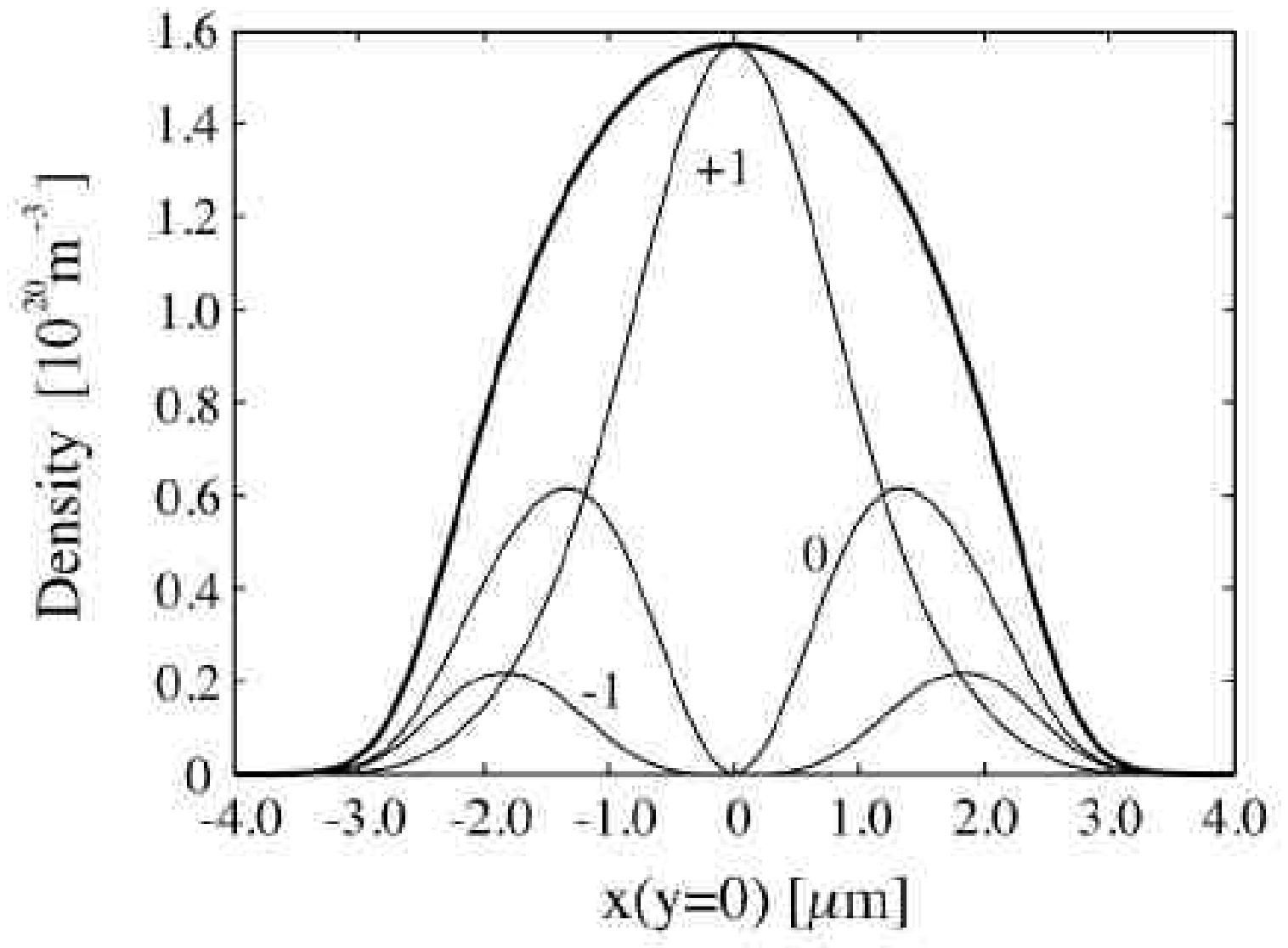}
&
\includegraphics[width=4.2cm]{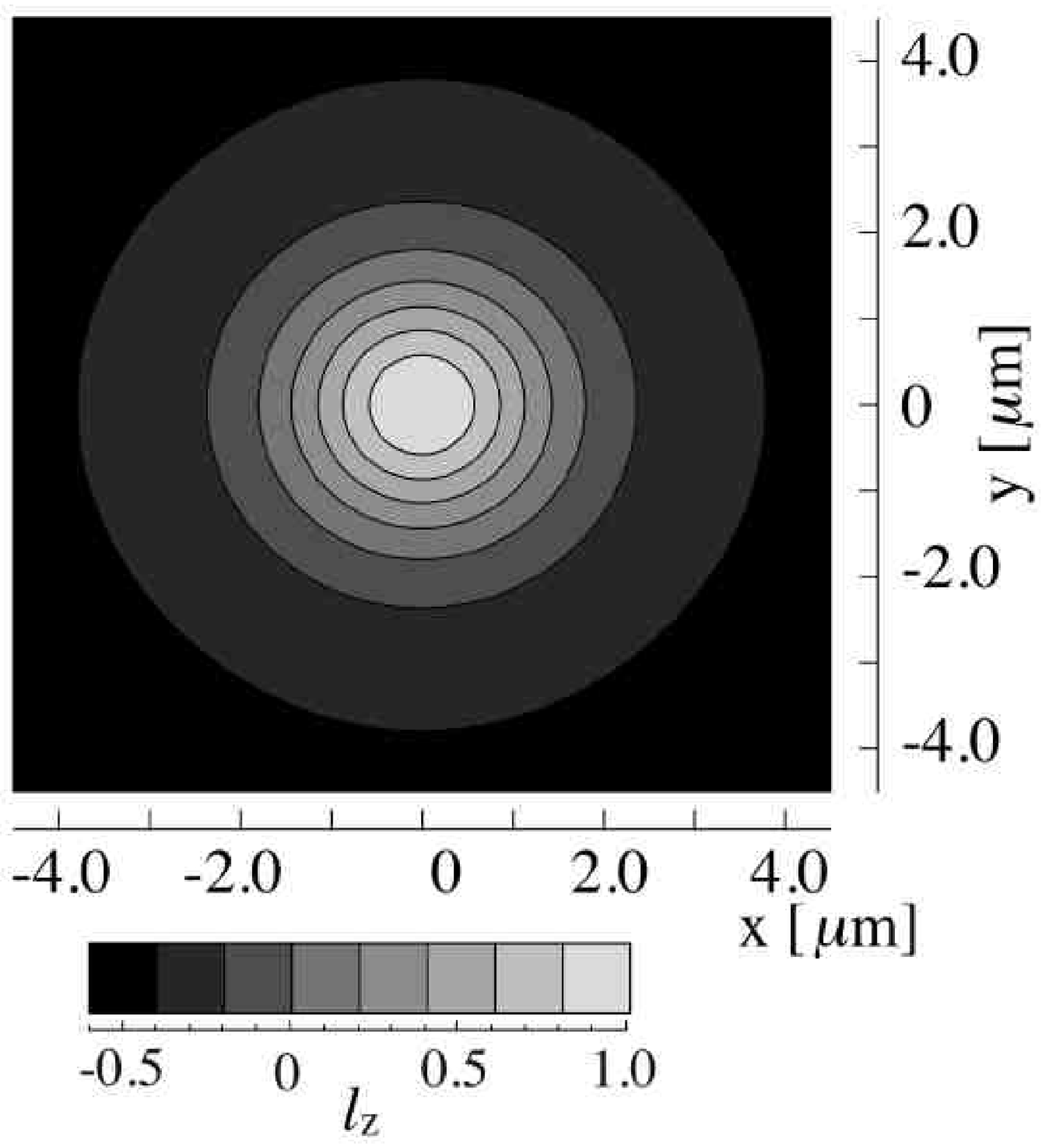}\\
(a)
 & 
(b) 
\end{tabular}
\end{center}
\caption{\label{Misusi}A $(0,1,2)$ vortex in a ferromagnetic spinor BEC. The density profile of the condensates (a) and the density map of the $l_z$-vector (b) for the $(0,1,2)$ vortex at $\Omega/\omega_{\perp}=0.35$ and M/N=0.21. In (a), the bold curve is the total density $\rho_{\rm T}=\sum_j | \Psi_j |^2$ and the thin curves show the density of each component $|\Psi_j|^2$ ($j=1,0,-1$). [Mizushima, Machida, and Kita, Phys. Rev. Lett. {\bf 89} 030401-1 (2002), reproduced with permission. Copyright (2002) by the American Physical Society.]}
\end{figure}

Also, the winding number $\gamma_{j}$ can exceed unity. Mizushima {\it et al.} discussed the $(0,1,2)$ vortex shown in Fig. \ref{Misusi} and showed that the ferromagnetic interaction supports the thermodynamic stability of this structure\cite{Mizushima3}. The central region of the harmonic trap is occupied by the non-rotating $\Psi_{1}$ component. The $\Psi_{0}$ component with $\gamma_{0}=1$ is pushed outward, while the $\Psi_{-1}$ component with $\gamma_{-1}=2$ occupies the outermost region. The resulting total density is nonsingular. This configuration is favorable for the ferromagnetic case, because it is favorable to spatial phase separation; the presence of vortices with different $\gamma_{j}$ effectively cause phase separation in the radial direction as shown in Fig. \ref{Misusi}(a)
 
Putting this coreless vortex into the pseudospin representation reveals that it has the structure of a spin texture. The wave functions of Fig. \ref{Misusi}(a) can be parametrized as
\begin{eqnarray}\label{MHansatz}
\left(
\begin{array}{c}
\Psi_{1} \\
\Psi_{0} \\
\Psi_{-1}
\end{array} \right)
= \sqrt{\rho_{\rm T} (r)} \left(
\begin{array}{c}
\cos^{2} \frac{\beta(r)}{2} \\
e^{i \theta} \sqrt{2} \sin \frac{\beta(r)}{2} \cos \frac{\beta(r)}{2} \\
e^{2 i \theta} \sin^{2} \frac{\beta(r)}{2}
\end{array} \right),
\end{eqnarray}
where the bending angle $\beta(r)$ has the range $0<\beta(r)<\pi$. The spin direction is denoted by the $\hat{l}$-vector and is given as $\hat{l}(r)=\hat{z} \cos \beta(r) + \sin \beta(r) (\cos \theta \hat{x} + \sin \theta \hat{y})$ [the contour density of $l_{z}$ is shown in Fig. \ref{Misusi}(b)], where $\beta$ varies from $\beta (0) =0$ to $\beta (R)=\pi/2$ ($=\pi$) for a Mermin-Ho (Anderson-Toulouse) vortex ($R$ is the outer boundary of the cloud). In other words, $l_{z}=\cos \beta(r)$ varies from $l_{z}(0)=1$ to $l_{z}(R)=0$ $(-1)$, where $\hat{l}$-vector points up at the center and points outward (downward) at the circumference for the Mermin-Ho (Anderson-Toulouse) vortex. In the superfluid $^{3}$He-A phase, the stability of these vortices is guaranteed by the constraint that the $\hat{l}$-vector is perpendicular to the vessel wall\cite{Salomaa}. There is no such boundary condition in the trapped BEC system. In this case, the $(0,1,2)$ vortex of Fig. \ref{Misusi} can be interpreted as a vortex with an intermediate boundary condition ($\pi/2 < \beta(R) < \pi$); we can control the value of $\beta(R)$ from the Mermin-Ho condition to the Anderson-Toulouse one by merely changing the total magnetization $M$\cite{Mizushima3}. In another numerical study, Martikainen {\it et al.} analyzed the coreless vortex state as a function of rotation frequency without fixing the total magnetization, and found that $\beta(R)$ increases with $\Omega$ and the upper value of $\beta(R)$ is $3\pi/4$, above which additional vortices nucleate. This implies that a Anderson-Toulouse vortex can never be the ground state of the system. 

Full 2-D numerical simulations of the coupled GP equations (\ref{3gpe1})-(\ref{3gpe3}) also reveal non-axisymmetric structures of vortices, whose stability depends sensitively on the spin-exchange interaction $g_{s}$, the external rotation $\Omega$ and the total magnetization\cite{Yip,Mizushima3,Martikainen}. For example, the axisymmetric $(1,1,1)$ vortex is always unstable, leading to the displacement of the vortex cores from the center. In the antiferromagnetic regime, the vortex cores are displaced such that the condensates has two singularities, through an overlap of the $\Psi_{1}$ and $\Psi_{-1}$ components or it has three singularities to form a triangle configuration. The breaking of the axisymmetry mainly leads to a decrease in the overlapping area of the condensate wave functions. In comparison with axisymmetric states, these non-axisymmetric vortices have the advantage that they can easily adjust to changes in $\Omega$. As $\Omega$ increases, the two or three separate singularities adjust their interdistances from the center and change the value of $L_z$ to gain energy due to the term $-\Omega L_z$. In this sense, the non-axisymmetric vortices are more flexible for a change in $\Omega$ compared with the axisymmetric ones. 
 
\subsubsection{Monopoles}
Another interesting topological excitation that occurs in the spin-1 BEC system is a singular pointlike defect, called a ``monopole" due to its similarity to magnetic monopoles in particle physics\cite{Hooft}. A monopole is characterized by a unit vector that is radial with respect to  a unique central point (i.e., the ``hedgehog" structure). Stoof {\it et al.}\cite{Stoofspin} proposed monopole excitations in an antiferromagnetic spin-1 BEC. The monopole is characterized by the spinor 
\begin{eqnarray}\label{monopleansatz}
\left(
\begin{array}{c}
\Psi_{1} \\
\Psi_{0} \\
\Psi_{-1}
\end{array} \right)
= \sqrt{\frac{\rho_{\rm T} ({\bf r})}{2}} e^{i \theta({\bf r})} \left(
\begin{array}{c}
-m_{x}({\bf r}) + i m_{y}({\bf r}) \\
\sqrt{2} m_{z}({\bf r})\\
m_{x}({\bf r}) + i m_{y}({\bf r})
\end{array} \right),
\end{eqnarray}
where $\theta$ is the overall phase. The form of Eq. (\ref{monopleansatz}) originates from the fact the antiferromagnetic interaction implies zero average spin $|\langle {\bf F} \rangle | \equiv \bar{\Psi} {\bf F} \Psi/\rho_{\rm T} = 0$. The spherically symmetric monopole is described by the radial unit vector ${\bf m}({\bf r}) = {\bf r}/r$. Minimization of the energy of the symmetric configuration yields $\theta=0$ and the density singularity at the origin $\rho_{\rm T}(0)=0$. Then, the spinor component $\Psi_0$ resembles a dark soliton and $\Psi_{\pm 1}$ form perfectly overlapping, straight singly-quantized vortex lines with opposite circulation, the latter being perpendicular to the phase kink plane. Stoof {\it et al.} demonstrated that this particular spin texture is a consequence of the unit winding number and the minimization of the gradient energy\cite{Stoofspin}. This configuration could be created through the phase imprinting technique achieved by the carefully designed sequences of laser pulses\cite{Changspin}. 

\begin{figure}
\begin{center}
\begin{minipage}{5.5cm}
\includegraphics[height=0.22\textheight]{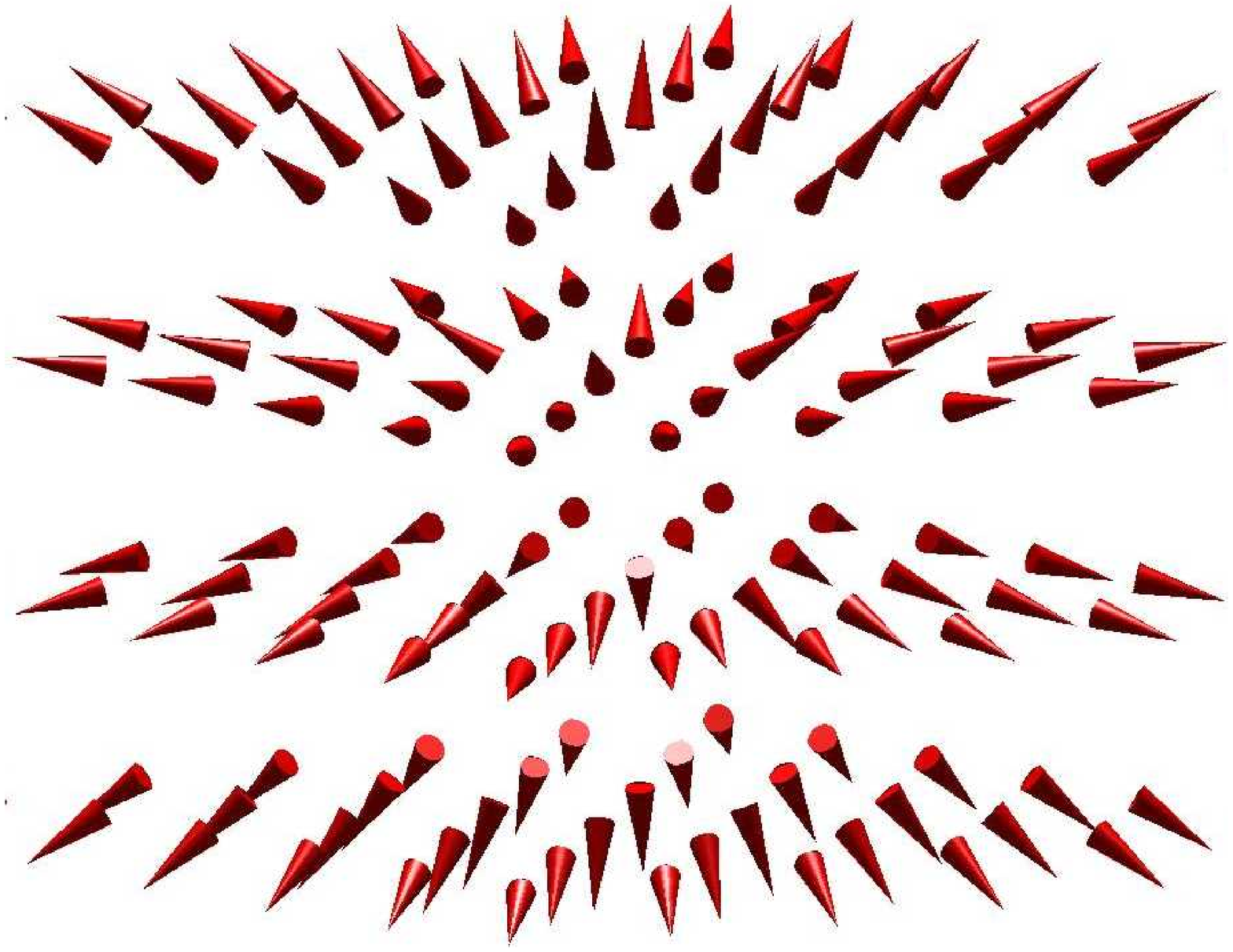}
\end{minipage}
\begin{minipage}{5.5cm}
\includegraphics[height=0.22\textheight]{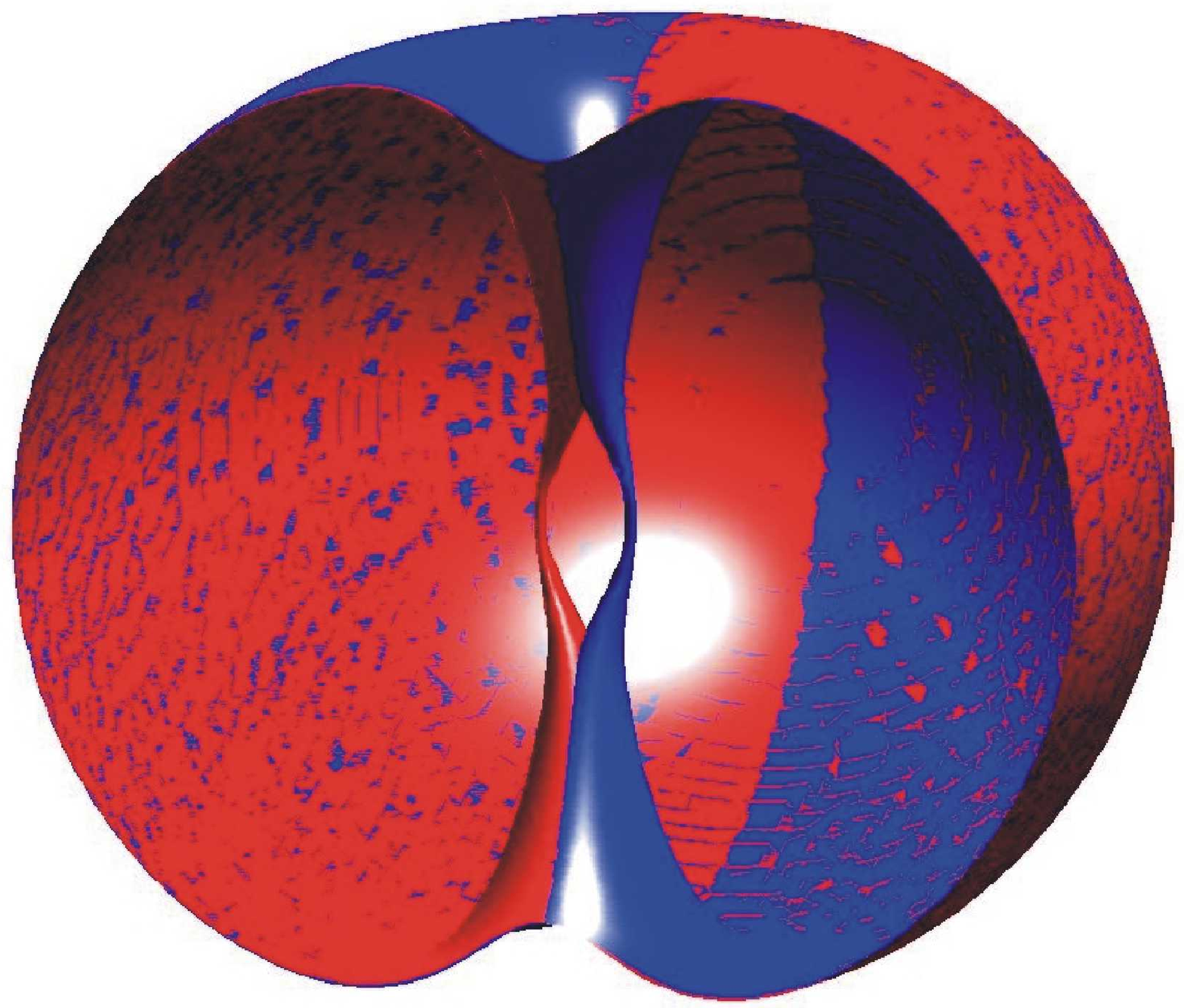}
\end{minipage}
\end{center}
\caption{A stable half-quantum vortex ring (Alice ring). To form it, the energy of a spherically symmetric monopole was minimized by continuously deforming the field configuration. The calculation was done for a condensate trapped in a spherically symmetric trap with the frequency $\omega=2\pi \times 10$ Hz and the total particle number $N=4 \times 10^{6}$ in a weakly antiferromagnetic regime $g_{s}/g_{n}=0.04$. The left figure shows the asymptotic distribution of the spin quantization axis ${\bf m}({\bf r})$; for $r \gg \xi _{a}$ it forms the radial hedgehog. For visualization purposes, the unoriented ${\bf m}({\bf r})$ field is drawn by cones. The right figure shows the constant surface density for $|\Psi _{1}({\bf r})|^{2}$ (light grey) and for $|\Psi _{-1}({\bf r})|^{2}$ (dark grey), where the monopole core is deformed for $r \leq \xi _{a}$ with the two line vortices separating. [Ruostekoski and Anglin, Phys. Rev. Lett. {\bf 91} 190402-1 (2003), reproduced with permission. Copyright (2003) by the American Physical Society.]} \label{f22}
\end{figure}
The theoretical study by Ruostekoski and Anglin\cite{Ruostekoskispin} demonstrated that the spherically symmetric monopole is stable only in the strongly antiferromagnetic regime. This is due to the following two length scales 
\begin{equation}
\xi_{s} \equiv \sqrt{\frac{3}{8 \pi \rho_{\rm T} (2a_{2}+a_{0})}}, \hspace{5mm} \xi_{a} \equiv \sqrt{\frac{3}{8 \pi \rho_{\rm T} (a_{2}-a_{0})}}.
\end{equation}
They represent the length scales over which $\rho_{\rm T}$ and $|\langle {\bf F} \rangle|$, respectively, tend to their bulk value when subjected to localized perturbations. In the weakly antiferromagnetic regime, the wavelength ($\sim \xi_{a}$) at which the antiferromagnetic constraint may be violated is much larger than that ($\sim \xi_{s}$) at which the total density constraint fails. Then, the monopole core extends to a larger size, which is favorable for having nonzero average spin instead of a density zero. The monopole point defect may be continuously deformed into a circular line defect without changing the topological invariant. This is done by punching a hole in the spherically symmetric core. To keep $\Psi$ single-valued on the disc bounded by the ring, the macroscopic phase $\theta$ must change by $\pi$ around any loop that links the defect circle, while there is also a $\pi$-disclination in ${\bf m}$ on the disc. This structure is identified as a half-quantum vortex line, forming a closed circular ring, called an ``Alice ring" [see Fig. \ref{f22}]. To keep $\Psi$ single-valued on the ring itself, one can either have $\rho_{\rm T}$ vanish there, or have $|\langle {\bf F} \rangle |= 1$ on the ring instead. By decreasing the ratio $g_s/g_n$, the spherically symmetric monopole core with $\rho_{\rm T}(0)=0$ becomes unstable against formation of an energetically stable Alice ring with $|\langle {\bf F} \rangle|= 1$ core. According to a linear stability analysis of this case, the radially symmetric monopole is unstable in a homogeneous condensate for $g_s/g_n \lesssim 0.17$, while according to the numerical simulation for a trapped condensate the monopole is unstable for $g_s/g_n\lesssim 0.16$, using an experimentally feasible set of parameters for $^{23}$Na ($\omega_{\perp}=\omega_{z}=2 \pi \times 10$ Hz and $N=4 \times 10^{6}$).

The creation of the monopole is not limited to the antiferromagnetic spinor condensates. Martikainen {\it et al.} studied in detail the two-component monopole characterized as 
\begin{eqnarray}\label{monopleferroansatz}
\left(
\begin{array}{c}
\Psi_{1} \\
\Psi_{0} \\
\Psi_{-1}
\end{array} \right)
= \sqrt{\frac{\rho_{\rm T} ({\bf r})}{2}} e^{i \theta({\bf r})} \left(
\begin{array}{c}
-m_{x}({\bf r}) + i m_{y}({\bf r}) \\
\sqrt{2} m_{z}({\bf r})\\
0
\end{array} \right).
\end{eqnarray}
To ensure the stability of this texture against phase separation, the spin-1 condensate must have ferromagnetic interactions, which makes the $^{87}$Rb spinor condensate a potential candidate. In other words, contrary to a previous expectation, the preparation of a monopole is not restricted to the antiferromagnetic $^{23}$Na condensate because one can use spinors that do not have the order-parameter space of the ground state. 

\section{Vortex states in rapidly rotating two-component BECs}\label{fast}
Rapid rotation of a BEC generates a large number of quantized vortices, with the result that the condensate can mimic a rigid body rotation despite the irrotational nature of superfluidity\cite{Doneley}. It is well known that the most stable configuration of an assembly of quantized vortices in a rotating {\it single-component} superfluid ($^{4}$He-II or atomic gas BECs) is a triangular lattice; also it is known that this is analogous to an Abrikosov lattice of magnetic fluxes in a type-II superconductor. At present, most experiments on vortex lattices in atomic-gas BECs have been done in single-component systems. Thus, a natural question arises: what happens in rapidly rotating two-component BECs? The vortex lattices in such systems are more complicated than those in single component condensates because of the intercomponent interactions. This section reviews the studies of vortex states in two-component BECs with a large number of vortices. We also describe a recent experiment\cite{Volker} that found interlaced square vortex lattices in two-component BECs. 

First, we show results obtained through analysis based on the ``mean-field quantum Hall regime''\cite{Mueller2}. This regime will be valid when the BEC in a harmonic trapping potential is rotated with angular frequency $\Omega$ close to the radial trapping frequency $\omega_{\perp}$\cite{Schweikhard,Ho4}. This method has the great advantage that one can analytically describe the vortex structure, although the method is effective only near the limit $\Omega \simeq \omega_{\perp}$. Next, we show the results of the direct numerical calculation of the coupled GP equations (\ref{bingp1}) and (\ref{bingp2})\cite{Kasamatsu3}. They reveal a rich variety of vortex structures that have eluded analytic treatment. Depending on the strength of the intercomponent interaction and the rotation frequency, vortex lattices can have the form of triangular lattices, square lattices, double-core vortex lattices, and ``serpentine" vortex sheets. These properties can be understood by using the concept of ``pseudospin", in which the vortex lattices are interpreted as {\it a lattice of skyrmion}. Finally, we discuss the effect of internal coherent coupling on the lattice structure, caused by the anisotropy of the resulting vortex molecules (meron pairs). 

Similar discussions can also be done for spin-1 BECs. Kita {\it et al.} studied vortex-lattice structures of antiferromagnetic spin-1 BECs using the phenomenological Ginzburg-Landau equation\cite{Kita}, which is formally similar to the mean-field quantum Hall approach. Their study revealed that the conventional Abrikosov lattice with hard vortex cores is unstable and the vortex cores shift their locations. The system has many metastable configuration of vortices, depending sensitively on the ratio $g_{2}/g_{0}$ in Eq. (\ref{energyfunspin1}). The vortices are characterized by distribution of magnetization and the difference in the number of circulation of quanta per unit cell, all of which makes the characterization of the vortex lattice structure complicated. Further discussions on the phase diagram of the ground state in rotating spin-1 bosons are reported in Reijnders {\it et al.}\cite{Reijnders}. This paper also describes a study of vortex-lattice structures with the mean-field quantum Hall approach and the exact diagonalization study of the quantum liquid phase. In addition, Mizushima {\it et al.} did numerical simulations of the GP equations (\ref{3gpe1})-(\ref{3gpe3}) for a ferromagnetic spinor BEC in a fast-rotating regime and also in an external magnetic field\cite{Mizushima4}. As a result, they could constuct a phase diagram of stable configurations on a plane of the rotation frequency and the total magnetization. 

\subsection{Mean-field quantum Hall approach}
The ``mean-field quantum Hall regime'' is the regime obtained in a rapidly rotating limit, where mean-field theory remains valid. Thus, each component can still be characterized by a condensate wave function $\Psi_{i}$ ($i=1,2$). The angular momentum of the system is so high that $\Psi_{i}$ is made up of the orbitals in the lowest Landau level (LLL) in the plane perpendicular to the rotation axis. Ho showed that this regime will emerge in a 3-D Bose gas at sufficiently high angular momenta\cite{Ho4}, and recently the JILA group has achieved this regime experimentally\cite{Schweikhard}. (See also the recent detailed arguments in Ref.\refcite{Baym2}-\refcite{AftalionLLL} for the use of LLL approximation and the structure of a vortex lattice.) In this regime, the wave function acquires an analytic structure that allows exact evaluation of the energy of a vortex lattice. As a result, it is possible to evaluate a wide range of lattice structures in rapidly rotating condensates, a feat that would be impractical for numerical calculations because of the time and the accuracy required. 

An extension of this argument to the two-component BECs was made by Mueller and Ho\cite{Mueller2} for the following situation. If each component has an equal number of bosons, and if the trapping potentials of the two components are identical, the two components will have the same size and the same density of vortices. In this case, one expects that each component will contain identical vortex lattices, with one lattice displaced relative to the other because of the presence of the intercomponent repulsive interaction. We shall refer to such a configuration of vortex lattices as staggered vortex lattices. While in the current experiment, the coupling constants satisfy $g_{1}\sim g_{2}\sim g_{12}$, considerable insights are gained by studying vortex lattices for a wider range of coupling constants. Mueller and Ho considered the case $g_{1}\sim g_{2} \neq g_{12}$ and investigated a wide range of vortex lattice structures as a function of $\delta = g_{12}/\sqrt{g_{1}g_{2}}$, and we summarize their result here.  

The condensate wave functions $\Psi_{1}$ and $\Psi_{2}$ of a rotating two-component BEC are determined by minimizing the free energy $F = E  -\mu_{1} N_{1} -\mu_{2}N_{2}$, where $E$ is the total energy functional (\ref{energyfunctio2}), $\mu_{i}$ ($i=1,2$) are the chemical potentials that determine the number of bosons $N_{i}$ in each component.  For simplicity, we assume the same cigar-shaped trapping potential $V=m(\omega_{\perp}^{2} r^{2}+ \omega_{z}^{2} z^{2})/2$ with $\omega_{\perp} \gg \omega_{z}$ for each component. As discuss in Ref.\refcite{Ho4}, a slow variation of the trapping potential along the $z$-axis allows one to apply a Thomas-Fermi approximation for the $z$ dependence of $\Psi_{i}$ and write $F$ as 
\begin{eqnarray}\label{genk}
F = \int d^{3} r \biggl\{ \sum_{i=1,2} \Psi^{\ast}_{i} \biggl[ \frac{1}{2m} \left(\frac{\hbar}{i} {\bf \nabla} - m \omega_{\perp} \hat{\bf z} \times {\bf r}\right)^2 + (\omega_{\perp}-\Omega) L_{z} - \mu_{i}(z) \biggr] \Psi_{i} \nonumber  \\
+  \frac{1}{2}g_{1}|\Psi_{1}|^{4} + \frac{1}{2}g_{2} |\Psi_{2}|^{4} + g_{12}|\Psi_{1}|^2|\Psi_{2}|^{2} \biggr\}
\end{eqnarray}
with $\mu_{i}(z)=\mu_{i}- m \omega_{z}^{2}z^2 / 2$. As $z$ is treated as a parameter, it is convenient to write $\Psi_{i}=\sqrt{\rho_{i}(z)}\psi_{i}({\bf r},z)$ with $\int |\psi_{i}({\bf r}, z)|^2\, d^{2}r =1$; hence, the normalization condition $\int  d^{3} r |\Psi_{i}|^2 = N_{i}$ is now written as $\int dz \rho_{i}(z) = N_{i}$. 

The first term of the integrand in Eq. (\ref{genk}) $h_{\rm L}=(-i \hbar \nabla -m \omega_{\perp} \hat{\bf z} \times {\bf r})^{2}/2m$ is identical to the Hamiltonian that describes the motion of an electron in a magnetic field $B \hat{\bf z}$ in the symmetric gauge (i.e., $(-i \hbar \nabla - e {\bf A}/c)^{2}/2m$ with a vector potential ${\bf A}=B \hat{\bf z} \times {\bf r}/2$). The eigenstates of $h_{\rm L}$ are referred to as the Landau levels and described in polar coordinates $(r,\theta)$ as 
\begin{equation}
u_{n,m}({\bf r}) = C_{nm} e^{i (n-m) \theta -r^{2}/4b_{\rm ho}^{2}} \biggl( \frac{r}{b_{\rm ho}} \biggr)^{|m-n|} L^{|m-n|}_{(m+n-|m-n|)/2} \biggl( \frac{r^{2}}{2b_{\rm ho}^{2}} \biggr), \label{LLorbital}
\end{equation}
where $b_{\rm ho}=\sqrt{\hbar/m\omega_{\perp}}$, $C_{nm}$ is the normalization constant, and $L^{n}_{m}(x)$ the associated Laguerre polynomial. Since $u_{n,m}$ are also eigenstates of $L_{z}$ with the eigenvalue $\hbar(m-n)$, the eigenenergy of $h_{\rm L} - (\omega_{\perp} - \Omega) L_{z}$ reads
\begin{equation}
\epsilon_{nm} = \hbar (\omega_{\perp} + \Omega) n + \hbar (\omega_{\perp} - \Omega) m + \hbar \omega_{\perp},
\end{equation}
where $n$ is the index of the Landau level, and $m$ is the index that classifies the degenerate states in the $n$-th Landau level. This degeneracy with respect to $m$ is lifted by the interaction. When $\Omega$ approaches $\omega_{\perp}$, the condensate density becomes so dilute because of the centrifugal effect that the mean-field interaction energy decreases. If the interaction energy becomes less than the energy spacing of the Landau level, then the wave functions $\psi_{i}$ can be written in terms of the orbitals $u_{0,m}({\bf r})$ in the LLL in the $xy$-plane, 
\begin{equation}
u_{0,m}({\bf r}) = \frac{1}{\sqrt{2 \pi m! }} \biggl[ \frac{x+iy}{b_{\rm ho}} \biggr]^{m} e^{-r^{2}/2b_{\rm ho}^{2}}. \label{LLLwave1}
\end{equation}
The energy functional then becomes
\begin{eqnarray}
F = \int dz \biggl\{ \sum_{i=1,2} \left[ \hbar(\omega_{\perp} - \Omega) \frac{\langle r^2 \rangle_{i} }{b_{\rm ho}^2} - \mu_{i}(z) + \hbar\omega_{\perp} \right] \rho_{i}(z) \nonumber \\
+ \int {\rm d}^2 r \biggl[ \frac{1}{2} \sum_{i=1,2} g_{i} \rho^{2}_{i} |\psi_{i}|^4  + g_{12} \rho_{1} \rho_{2} |\psi_{1}|^2 |\psi_{2}|^2  \biggr] \biggr\} \label{energy2ndstage} 
\end{eqnarray}
where $\langle r^2 \rangle_{i} = \int\! r^2 |\psi_{i}|^2 \, d^2 r$. 

We next introduce a variational wave function that represents a vortex lattice. Generally, wave functions in the LLL can be written as $\phi({\bf r}) = f(z) e^{-r^{2}/2b_{\rm ho}^{2}}$, where $f(z)$ is an arbitrary polynomial of $z=x+iy$ (but not of $z^{\ast}$). The form of a vortex lattice can be described by
\begin{equation}
\phi({\bf r}) = C \prod_{\alpha}(z - b_{\alpha}) e^{- r^2/2b_{\rm ho}^2},
\label{sigmatheta}
\end{equation}
where $C$ is a normalization constant and $\{ b_{\alpha} \}$ are the zeros of $\phi$, which are the positions of vortices. In Eq. (\ref{sigmatheta}), we assume perfect periodicity of the lattice. If the zeros form an infinite lattice with unit cell size $v_{c}$, the density $|\phi|^2$ can then be written as a product of a Gaussian and a function which is periodic under lattice translation:
\begin{equation}
|\phi|^2 =  e^{-r^2/\sigma^2} g({\bf r}),  \,\,\,\,\,\, g({\bf r}) = g({\bf r} + {\bf R}),
\label{new} 
\end{equation}
where ${\bf R} = n_{1}{\bf B}_{1} + n_{2}{\bf B}_{2}$, $n_{i}$ are integers, and ${\bf B}_{1}$, ${\bf B}_{2}$ are the basis vectors of the lattice. The width $\sigma$ includes the contribution of the number of vortices of the system, which is determined below. 

It is convenient to introduce the complex basis vectors $(b_{1},b_{2})$ defined by $b_{i}=(\hat{\bf x} + i \hat{\bf y}) \cdot {\bf B}_{i}$; then the unit cell size is $v_{c}=i(b_{1}b_{2}^{\ast} - b_{1}^{\ast} b_{2}) / 2$. If the lattice is oriented so that $b_{1}$ is real, we can parametrize ${\bf B}_{1} = b_{1} \hat{\bf x}$ and ${\bf B}_{2} = b_{1} (u \hat{\bf x} +v \hat{\bf y})$, and obtain $b_{2}=b_{1} (u+iv)$ and $v_{c}=b_{1}^{2} v$. Except for the normalization constant, the form of Eq. (\ref{sigmatheta}) is similar to the Weierstrass's $\sigma$-function\cite{elliptic}
\begin{equation}
\sigma(z)=z \prod_{m,n=-\infty}^{\infty} \biggl[ \biggl( 1 - \frac{z}{\Omega_{mn}} \biggr) \exp \biggl( \frac{z}{\Omega_{mn}} + \frac{z^{2}}{2 \Omega_{mn}^{2}} \biggr)  \biggr],  
\end{equation}
where the term with $m=n=0$ is omitted from the product and $\Omega_{mn}=2m\omega_{1}+2n\omega_{2}$ with the half-periods $\omega_{1}$ and $\omega_{2}$. By using a relation between the Weierstrass's $\sigma$-function and the Jacobi $\theta$-function\cite{elliptic} 
\begin{equation}
\theta(\zeta,\tau) = \frac{1}{i} \sum^{+\infty}_{n=-\infty} (-1)^{n} e^{-i \pi \tau (n+1/2)^{2} + 2 \pi i \zeta (n+1/2)},
\end{equation}
we can write $f(z) = \theta(\zeta, \tau) h(\zeta)$, where $\zeta = (x+iy)/b_1 = \bar x+i\bar y$, $\tau=u+iv=b_{2}/b_{1}$, and $h(\zeta)$ is an entire function without zeros. To ensure the normalization of $\phi$, the function $h(\zeta)$ must be of the form $h(\zeta) = \exp(c_{1}\zeta + c_{2}\zeta^2)$. 

The density of the system is then $|\phi({\bf r})|^2 = |\theta(\zeta,\tau)|^2 |e^{c_{1}\zeta + c_{2}\zeta^{2}}|^2 e^{-r^2/b_{\rm ho}^2}$. For a vortex lattice with inversion symmetry about the origin ${\bf r}=0$, we have $c_{1}=0$. In addition, if the shape of the condensate is cylindrically symmetric, we have $c_2=\pi/(2 v_c)$, which gives $|\phi({\bf r})|^2 = |\theta(\zeta,\tau)|^2 e^{-r^{2}/\sigma^{2}}$ with
\begin{equation}
\frac{1}{\sigma^{2}} = \frac{1}{b_{\rm ho}^{2}} - \frac{\pi}{v_{c}}.  \label{sigma}
\end{equation}
Following Eq. (\ref{new}), we can write $g({\bf r})$ as
\begin{equation}
g({\bf r}) = \left| \theta(\zeta, \tau) e^{-\pi y^{2}/v_{c}} \right|^{2}.
\end{equation}
It follows from the quasi-periodic properties of the Jacobi theta-function  
\begin{eqnarray}
\theta(\zeta +1 , \tau) &=& \theta(\zeta, \tau),
\\
\theta(\zeta + \tau, \tau)
&=& - e^{-i\pi(\tau + 2\zeta)}\theta(\zeta, \tau)
\end{eqnarray}
that $g({\bf r})= g({\bf r}+{\bf R})$. The periodicity of $g({\bf r})$ implies $g({\bf r}) = v_{c}^{-1} \sum_{\bf K} g_{\bf K} e^{i{\bf K}\cdot {\bf r}}$, where $\{ {\bf K} \}$ are the reciprocal lattice vectors. ${\bf K}=m_{1}{\bf K}_{1}+ m_{2}{\bf K}_{2}$, and ${\bf K}_{i}$ are the basis vector of the reciprocal lattice, ${\bf K}_{1} = (2\pi/v_{c}){\bf B}_{2}\times \hat{\bf z}$, ${\bf K}_{2} =
(2\pi/v_{c})\hat{\bf z} \times {\bf B}_{1}$, and
\begin{equation}
v_{c} {\bf K}^2 = \frac{(2\pi)^2}{v}\left( (vm_{1})^2 + (m_{2} - um_{1})^2\right).
\end{equation}

The explicit form of $g_{\bf K}$ is calculated in a straightforward way. We obtain 
\begin{eqnarray}
|\theta(\zeta, \tau)|^2 &=& \sum_{m_{1},m_{2}} e^{i \pi u [m_{2}(m_{2}+1)-m_{1}(m_{1}+1)]} e^{- \pi v [m_{2}(m_{2}+1)-m_{1}(m_{1}+1) +1/2]} \nonumber \\ 
& & \times e^{2 i \pi \bar{x} (m_{2}-m_{1})} e^{- 2 \pi \bar{y} (m_{1}+m_{2}+1)} \nonumber \\
&=& \sum_{m}(-1)^{m} e^{2\pi i m \bar x} e^{-\pi v m^2/2} L_{m}, \label{thetasqure} 
\end{eqnarray}
where 
\begin{equation}
L_{m} = \frac{1}{2} \sum_{m'} \left( 1 - e^{i\pi (m+m')} \right) e^{\left(i\pi u m - 2\pi \bar y-\pi vm'/2\right)m^\prime} .
\label{Lm} 
\end{equation}
The factor $( 1 - e^{i\pi (m+m')} )/2$ is added in Eq. (\ref{Lm}) because $m+m'$ must be odd. By applying the Poisson summation formula 
\begin{equation}
\sum_{n=-\infty}^{+\infty} f(n) = \sum_{k=-\infty}^{+\infty} \int_{-\infty}^{+\infty} dx f(x) e^{2 \pi i k x}
\end{equation}
to Eq. (\ref{Lm}), we have
\begin{eqnarray}
L_{m} &=& \frac{1}{\sqrt{2v}} \sum_{k} \left[ e^{-\pi(2k + um + 2i\bar y)^2/2 v } + (-1)^{m+1} e^{-\pi(2k +1 + um + 2i\bar y)^2/2 v } \right] \nonumber \\
&=& \frac{1}{\sqrt{2v}} \sum_{k} (-1)^{(m+1)k} e^{-\pi(k + um + 2i\bar y)^2/2 v}. 
\label{sumLm} 
\end{eqnarray}
We thus have
\begin{eqnarray}
|\theta(\zeta, \tau)|^2 &=& \frac{1}{v} \sum_{m,n} (-1)^{m+n+mn} \sqrt{2 v} e^{-v_{c} |{\bf K}|^{2} /8\pi} e^{i {\bf K} \cdot {\bf r}} e^{2\pi y^{2}/v_{c}} \nonumber \\
&\equiv& \left[\frac{1}{v_{c}} \sum_{\bf K} g_{\bf K} e^{i{\bf K}\cdot {\bf r}} \right] e^{2\pi y^2/v_{c}},
\label{finalthetat2} 
\end{eqnarray}
where ${\bf r} = x \hat{\bf x} + y \hat{\bf y}$, ${\bf K} = (2\pi m \hat{\bf x} - 2\pi (n+um)/v \hat{\bf y})/b_1$, and 
\begin{equation}
g_{\bf K}= (-1)^{m+n + mn} e^{-v_{c}|{\bf K}|^2/8\pi} \sqrt{\frac{v_{c}}{2}}.
\label{gK2}
\end{equation}

If $g_{1}=g_{2}$, the two components with equal particle numbers and trapping potentials have identical vortex lattices. A sufficiently small difference in $g_{1}-g_{2}$ should not change this structure. If the intercomponent interaction is repulsive $g_{12}>0$, the mean-field interaction energy is minimized by interlacing the two lattices; if the vortex lattice of one component is deformed, the other has to follow to keep the interaction energy minimal. Under these assumptions, the normalized condensates $\psi_{1}$ and $\psi_{2}$ in Eq. (\ref{energy2ndstage}) would be represented by densities of the form
\begin{eqnarray}
|\psi_{1}|^{2} &=& \frac{1}{\pi \sigma^2} \sum_{{\bf K}} \tilde{g}_{\bf K} e^{i{\bf K}\cdot {\bf r}} e^{-r^2/\sigma^2}, \label{Phi2}\\
|\psi_{2}|^{2} &=& \frac{1}{\pi \sigma^2} \sum_{{\bf K}} \tilde{g}_{\bf K} e^{i{\bf K}\cdot ({\bf r}-{\bf r}_{0})} e^{-r^2/\sigma^2}, \label{Phi4}\\
\tilde{g}_{\bf K} &=& \frac{{g}_{\bf K}}{\sum_{{\bf K'}} g_{\bf K'} e^{-\sigma^2 {\bf K'}^2/4}}.
\label{gK} 
\end{eqnarray}
The wave function is described by variational parameters $\rho_{i}(z)$, $\sigma^2$, the basis vectors ${\bf B}_{i}$ (which determine the unit cell size $v_{c}$), and the relative displacement ${\bf r}_{0}$ between the two lattice.

By integrating Eqs. (\ref{Phi2}) and (\ref{Phi4}), one would see that terms up to those of relative order $v_c/\sigma^2$, the radius of the condensate is $\langle r^{2} \rangle_{1}= \langle r^{2} \rangle_{2}= \sigma^2$.  Defining the quantities $I$ and $I_{12}$ as $\int d^2 r |\Phi_{i}|^4 \equiv I/ (\pi \sigma^2)$ and $\int d^2 r |\Phi_{1}|^2 |\Phi_{2}|^2 = I_{12}/(\pi\sigma^2)$, we have
\begin{eqnarray}
I &=& \frac{1}{2} \sum_{{\bf K}, {\bf K'}} \tilde{g}_{\bf K} \tilde{g}_{\bf K'} e^{-\sigma^2 |{\bf K} + {\bf K'}|^2/8}, \label{I} \\
I_{12} &=& \sum_{\bf K}\tilde{g}_{\bf K}  \tilde{g}_{\bf K'}  e^{-i{\bf K'}\cdot {\bf r}_{0}} e^{-\sigma^2 |{\bf K} + {\bf K'}|^2/8}, \label{I12} 
\end{eqnarray}
and the energy functional $F$ takes the form
\begin{eqnarray}
F =  \int dz \biggl\{ -\left[ \mu(z) -\hbar\omega_{\perp} - \hbar(\omega_{\perp}-\Omega) \frac{\sigma^2}{b_{\rm ho}^{2}} \right] (n_{1}+n_{2})
\nonumber \\
 + \frac{1}{2 \pi \sigma^{2}} \left( n_{1}^2 g_{1}I +  n_{2}^2 g_{2}I +
2n _{1}n_{2}g_{12}I_{12} \right) \biggr\}.
\label{newK} \end{eqnarray}
Since we assume here a large number of vortices, the size of the condensate is much larger than the unit cell and thus $\pi \sigma^{2}/v_{c} \gg 1$. We can therefore ignore all ${\bf K}+{\bf K'}\neq 0$ terms in Eqs. (\ref{I}) and (\ref{I12}), since $\sigma^2 {\bf K}^2 > \pi \sigma^2/v_{c}$. We then have
\begin{equation}
I = \sum_{\bf K} \left|\frac{g_{\bf K}}{g_{\bf 0}}\right|^2,
\,\,\,\,\,\,
I_{12} = \sum_{\bf K} \left|\frac{g_{\bf K}}{g_{\bf 0}}\right|^2
{\rm cos}{\bf K}\cdot {\bf r}_{0},
\label{newI} \end{equation}
where $g_{\bf K}$ is given by Eq. (\ref{gK2}) and the sum of ${\bf K}$ is over the integers $m_{1}$ and $m_{2}$.  Since the expressions of $I$ and $I_{12}$ in Eq. (\ref{newI}) are independent of $\sigma^2$, the minimization of $F$ in Eq. (\ref{newK}) with respect to $\sigma^2$ and $\rho_{i}$ is greatly simplified. The optimized $\sigma^2$ and $\bm{\rho}=(\rho_1,\rho_2)$ are given by
\begin{eqnarray}\label{opt1}
\sigma^2 &=& b_{\rm ho}^2 \frac{\mu(z)-\hbar\omega_\perp}{3\hbar (\omega_{\perp} - \Omega)}, \\
\bm{\rho}(z) &=& \frac{2 \sigma^2}{3 b_{\rm ho}^2} \left[ \mu(z)-\hbar\omega_\perp \right] {\bf G}^{-1}\cdot{\bf 1}, \label{result1} 
\end{eqnarray}
where
\begin{eqnarray}
{\bf G} &=& \left(
\begin{array}{cc} g_{1} I & g_{12}I_{12}\\ g_{12}I_{12} & g_{2} I
\end{array}\right),\,\,\,\,\,
{\bf 1} =
\left(
\begin{array}{c}
1\\
1\\
\end{array}
\right),
\end{eqnarray}
and the corresponding free energy $F$ is given by 
\begin{eqnarray}
F &=& - \int dz \frac{1}{3} \left[ \mu(z)-\hbar\omega_\perp \right] {\bf 1}\cdot \bm{\rho}(z), 
\label{result2} 
\end{eqnarray}
It is clear from Eqs. (\ref{opt1}) through (\ref{result2}) that the solution for the case where $g_{1}-g_{2} \ll |g_{1}+g_{2}|$ is very close to that of $g_{1}=g_{2}$. The lattice shape (parameterized by ${\bf r}_0$, $u$,
and $v$) enters the grand potential only through the factor ${\bf 1\cdot G^{-1}\cdot 1}$.  When $g_1=g_2$ this term is inversely proportional to $J=I+\delta I_{12}$ ($\delta = g_{12}/g_{1}$), and the most favorable lattice is the one that minimizes $J$.

\begin{figure}[btp]
\begin{center}
\includegraphics[height=0.30\textheight]{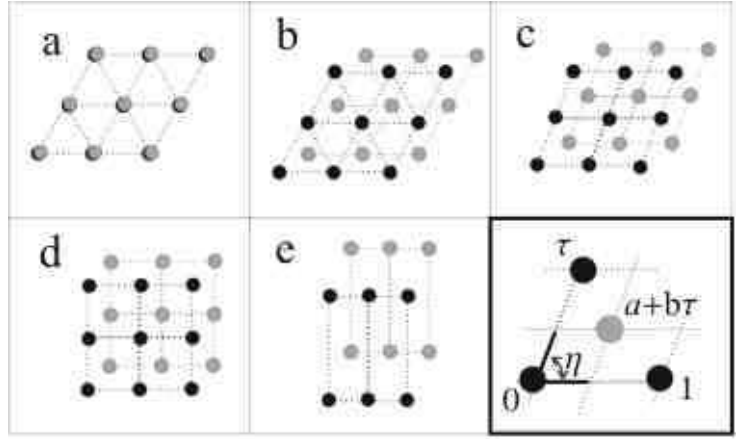}
\end{center}
\caption{Vortex lattice phases in rotating two-component BECs. Black and grey dots represent vortices of each of the two BECs. The panels (a) through (e) show the vortex structure in each of the phases described in the text. The final panel depicts the lattice types; the black and grey dots, respectively, occupy positions in the complex plane $\{ m + n \tau \}$ and $\{ (a+m) + (b+n) \tau \}$, where $m$ and $n$ are integers. All minimal-energy configurations have $a=b$. [Mueller and Ho, Phys. Rev. Lett. {\bf 88} 180403-3 (2002), reproduced with permission. Copyright (2002) by the American Physical Society.] }
\label{muellerandhoresult}
\end{figure}
In the minimization it is convenient to measure lengths in unit of the the basis vector ${\bf B}_{1} = b_{1}\hat{\bf x}$ and also to write the complex representation of ${\bf B}_{2}$ and ${\bf r}_{0}$ as $\tau= u+iv=|\tau|e^{i\eta}$ and $r_{0} \equiv a+b\tau$, respectively.  The phase diagram of the vortex lattice as a function of the ratio $\delta= g_{12}/g$ is shown in Figs. \ref{muellerandhoresult}. The main features are
\begin{itemize}
\item (a) $\delta < 0$: In this region the vortices of the two components coincide with each other ($a=b=0$) to form a triangular lattice ($\tau=e^{i \pi/3}$).
\item (b) $0 < \delta < 0.172$: The vortex lattice in each component remains triangular. However, the relative displacement $r_{0}$ between the two lattices undergoes a first-order change so that one lattice is displaced from the center of the triangle of the other component ($a=b=1/3$).  The lattice type (characterized by $\tau=e^{i\pi/3}$) remains constant within this interval.
\item (c) $0.172 < \delta < 0.373$: At $\delta= 0.172$, ${\bf r}_{0}$ moves from the center of the triangle (i.e. half of the unit cell) to the center of the (rhombic) unit cell ($a=b=1/2$).  The angle $\eta$ abruptly changes from $60^{\circ}$ to $67.95^{\circ}$ at $\delta= 0.172$, and increases continuously to $90^{\circ}$ as $\delta$ increases to $0.372$. As a result, the lattice type is no longer fixed and the unit cell is a rhombus.  The modulus of $\tau$, however, remains fixed across this region.
\item (d) $0.373<\delta<0.926$: The two lattices are ``mode-locked'' into a centered square structure throughout the entire interval $(\tau=i,a=b=1/2)$.
\item (e) $0.926<\delta$: The lattice type again varies continuously with interaction $\delta$. Each component's vortex lattice has a rectangular unit cell ($\eta=\pi/2$) whose aspect ratio $|\tau|$ increases with $\delta$. Both $^{87}$Rb and $^{23}$Na have interaction parameters within this range.  At $\delta=1$, ($g_{1}=g_{2}=g_{12}=g$), the aspect ratio is $\sqrt{3}$. For indistiguishable components, the combined lattice would then appear triangular.
\end{itemize}

As described in Sec. \ref{basic}, the condition of phase separation for nonrotating condensates is given by Eq. (\ref{sepcond}) (i.e., $\delta >1$). The presence of a vortex lattice naturally modulates the density of each component, with the high density regions of one component overlapping with the low density regions of the other.  Thus, the system is effectively phase-separated whenever staggered vortex lattices are present, even for $\delta<1$.  In particular, the vortex lattice near $\delta=1$ (Fig. \ref{muellerandhoresult}(e)) is made up of alternating rows of vortices of each component, and the system therefore contains stripes in which one component has a high density and the other component has a very low density. 

\subsection{Numerical simulations for rotating two-component BECs in the Thomas-Fermi regime}
This section addresses a full numerical study of the GP equations on the vortex lattice structure in two-component BECs\cite{Kasamatsu3}. The assumptions in the LLL calculation shown in the last subsection are that (i) the rotation frequency is assumed to be very close to the radial trapping frequency and that (ii) a vortex lattice is assumed to be perfectly regular\cite{Mueller}. Therefore, a vast experimentally accessible parameter regime of the lattice configuration remain to be explored. The full numerical simulations of Eq. (\ref{nondimgpeq1}) and (\ref{nondimgpeq2}) reveal a rich variety of vortex structures which have eluded the analytic treatment. Here, we show some detailed results of our simulations. 

\begin{figure}[btp]
\begin{center}
\includegraphics[height=0.28\textheight]{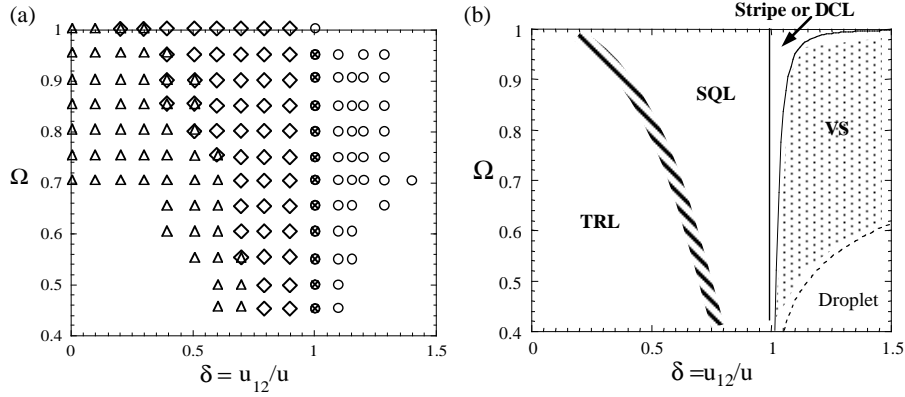}
\end{center}
\caption{$\Omega$-$\delta$ phase diagram for vortex states in rotating two-component BECs. (a) Table of data for the phase diagram in (b). $\triangle$: triangular lattice, $\diamond$: square lattice, $\otimes$: stripe or double-core vortex lattice, $\circ$: vortex sheet. Because of the continuous change between triangular and square lattices, their boundary is shown by a combination of $\triangle$ and $\diamond$. The plots at $\Omega=1$ show the results obtained by Mueller and Ho [Phys. Rev. Lett. {\bf 88}, 180403 (2002)] based on the LLL approximation. (b) Vortex phase diagram. The boundaries of the vortex sheet phase are given by the condition $b=\Lambda_{p}$ (solid curve) and $2b=R_{\rm TF}$ (dashed curve) (see text).}
\label{phasedia}
\end{figure}

We numerically calculated the equilibrium solutions of Eq. (\ref{nondimgpeq1}) and (\ref{nondimgpeq2}) through the conjugate gradient method\cite{Modugno,Crecipe}. Following Mueller and Ho, we also assume an identical trapping potential for both components ($V_{1}=V_{2}=V$) and the identical intracomponent interactions $g_{1}=g_{2}$. We confine ourselves to the 2-D problem. The dimensionless coupling constants (defined in Eqs. (\ref{nondimgpeq1}) and (\ref{nondimgpeq2})) are $u_{1}=u_{2}=u=4000$ under the assumption $N_{1}=N_{2}$; the normalization of each wave function is taken as $\int d^{2}r |\psi_{i}|^{2} = 1/2$. The intercomponent interaction $u_{12}$ and the rotation frequency $\Omega$ are treated as variable parameters. Figure \ref{phasedia} shows the numerically obtained phase diagram of vortex states in the $(\delta \equiv u_{12}/u, \Omega)$ plane. The upper limit of the rotation frequency is $\Omega = 1$, because for $\Omega > 1$ the centrifugal potential dominates the harmonic trapping potential and therefore a BEC cannot be trapped. Also, we do not discuss the slow rotation regime below $\Omega=0.40$. The criterion of phase separation $\delta=1$ is reflected on the structure of vortex states, as explained below. 

\begin{figure}[btp]
\begin{center}
\includegraphics[height=0.38\textheight]{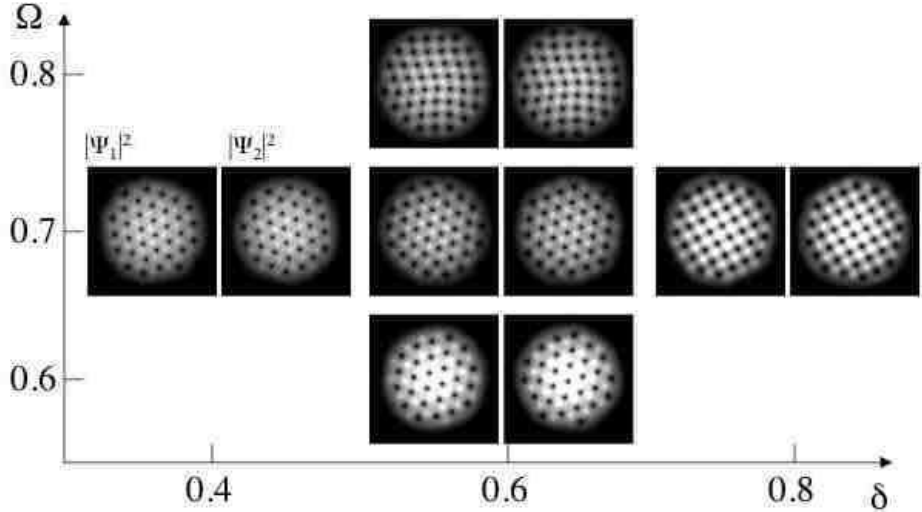}
\caption{Two-dimensional density profile of the condensates $\psi_{1}$ and $\psi_{2}$ with $u=4000$ in the overlapping region $\delta<1$. We choose typical values of rotation frequency $\Omega=0.6$, 0.7, 0.8, and those of the intercomponent interaction strength $\delta=0.4$, 0.6, 0.8. The plot range of the density profile is $[-10 < x,y < +10]$ in units of $b_{\rm ho}$.}
\label{vortexlattice}
\end{center}
\end{figure}

In the overlapping region $\delta<1$, two types of regular vortex lattices are obtained as the equilibrium state. For $\delta=0$, the two components do not interact; therefore, triangular vortex lattices form as that in a single component BEC. As $\delta$ increases, the positions of vortex cores in one component gradually shift from those of the other component and the triangular lattices become distorted. Eventually, the vortices in each component form a square lattice rather than a triangular one, which is consistent with the result of the LLL calculation done by Mueller and Ho\cite{Mueller2}. However, Fig. \ref{phasedia} shows that the stable region of the square lattice depends not only on $\delta$ but also on $\Omega$. Figure \ref{vortexlattice} shows the structural transition in closer detail. While an increase in $\delta$ indeed causes the deformation of the lattices from triangular to square, the transition occurs at a significantly higher value of $\delta$ than that of the LLL result ($\delta=0.373$). For example, for $\Omega=0.7$ the transition occurs at $\delta \simeq 0.65$. This implies that an increase of rotation frequency also causes the transition from triangular to square lattices. From Fig. \ref{vortexlattice}, the lattice changes from triangular to square around $\Omega=0.7-0.8$ for $\delta=0.6$. We find that the two vortex lattices are interlaced in such a manner that a peak in the density of one component is located at the density minimum of the other. As a result, the total density $\rho_{\rm T}= |\psi_{1}|^{2} + |\psi_{2}|^{2}$ obeys the Thomas-Fermi distribution applied to the overlapping two-component BECs with rigid body rotation $\rho_{\rm T}(r) = \sqrt{2 \alpha/\pi} -\alpha r^{2}$ with $\alpha=(1-\Omega^{2})/u(1+\delta)$. We have confirmed that this generic feature holds for other parameter regimes. 

\begin{figure}[btp]
\begin{center}
\includegraphics[height=0.3\textheight]{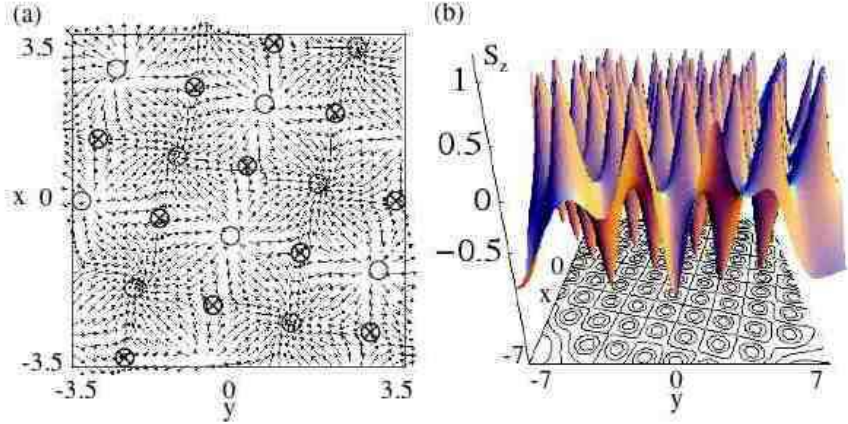}
\end{center}
\caption{Spin texture for the square lattice solution $\delta=0.8$ and $\Omega=0.7$ (cf. Fig. \ref{vortexlattice}). (a) The spin texture profile projected onto the $x$-$y$ plane. The positions of the defects with $+ \hat{\bf z}$ and $- \hat{\bf z}$ are marked by $\bigcirc$ and $\bigotimes$, respectively. (b) The spatial distribution of $S_{z}$.} 
\label{skyrmlatidfce}
\end{figure}

The pseudospin and the NL$\sigma$M provide insights for understanding why the square lattice is stabilized in two-component BECs. According to the energy functional of Eq. (\ref{energyfunctio2}), two components interact via the intercomponent interaction $g_{12} |\psi_{1}|^{2} |\psi_{2}|^{2}$; however, the velocity field in one component is independent of that of the other. Therefore, the intercomponent relation is determined only by the density distribution of the condensates that minimizes the interaction energy. According to the NL$\sigma$M (\ref{sigmamoddimless}), we can rewrite the interaction energy in terms of the total density $\rho_{\rm T}$ and the $z$-component of pseudospin $S_{z}=|\chi_{1}|^{2}-|\chi_{2}|^{2}$. In this representation, the spin-up component of $S_{z}$ corresponds to the density peak of $\psi_{1}$ at the vortex core of $\psi_{2}$, and vice versa for spin-down components. Now, we have assumed $u_{1}=u_{2}=u$, so that $c_{1}=0$ and the interaction energy reads 
\begin{eqnarray}
E_{\rm int} &=& \int d^{2} r \frac{\rho_{\rm T}^{2}}{2} \biggl( c_{0} + c_{2} S_{z}^{2} \biggr) \nonumber \\
&=& \int d^{2} r \frac{u \rho_{\rm T}^{2}}{4} \biggl[ (1+\delta)  + (1-\delta) S_{z}^{2} \biggr]. 
\label{intenergy}
\end{eqnarray} 
This expression is similar to that of a binary alloy, in which $N/2$ ``A" atoms and $N/2$ ``B" atoms are distributed on $N$ lattice sites \cite{Kasamatsu3}. A larger $\delta$ makes a smoother total density $\rho_{\rm T}$ that is more favorable for reducing the first term of Eq. (\ref{intenergy}). This results in a shift of the positions of vortex cores of each component. Therefore, the spin-spin interaction $c_{2}$ in the second term plays a crucial role in stabilizing the square lattice. When the coefficient $1 - \delta$ is positive, $c_{2}>0$ and the interaction between spins is anti-ferromagnetic, which favors a square lattice because the triangular lattice can become frustrated. This can be understood by representing the vortex lattice state with the pseudospin. The vortex lattices of two-component BECs can be regarded as a square lattice of skyrmions, in which skyrmions with a radial-disgyration texture ($+\hat{\bf z}$) alternate with those having a hyperbolic-disgyration texture ($-\hat{\bf z}$). The profile of $S_{z}$ in Fig. \ref{skyrmlatidfce}(b) shows a square lattice structure, which is equivalent to the ``continuous" Ising model of spins; actually, the second term of Eq. (\ref{intenergy}) is equal to the first term of the Ising Hamiltonian in a continuous limit $H_{\rm spin}=J\sum_{i,j} S_{z}({\bf r}_{i}) [S_{z}({\bf r}_{i}) + ({\bf r}_{j}-{\bf r}_{i}) \cdot \nabla S_{z}({\bf r}_{i})+\cdot \cdot \cdot]$. As both $\delta$ and $\Omega$ increase, the interlocking of vortex lattices becomes stronger and the anti-ferromagnetic nature is made more pronounced, which lead to a nontrivial boundary between triangular lattices and square lattices in the phase diagram of Fig. \ref{phasedia}. 

\begin{figure}[btp]
\begin{center}
\includegraphics[height=0.17\textheight]{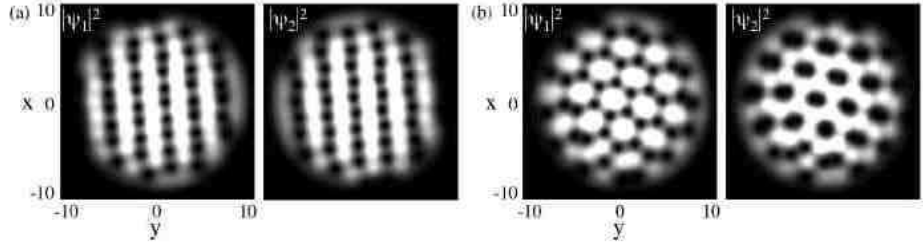}
\end{center}
\caption{The density profiles of the condensates $\psi_{1}$ (left) and $\psi_{2}$ (right) for $\Omega=0.7$ and $\delta=1$, i.e., $u_{1}=u_{2}=u_{12}$. The profiles (a) and (b) were obtained by numerical simulation starting from different trial wave functions. The energy deference between these states is 
$\Delta E \sim 10^{-5} \hbar \omega_{\perp} N$.}
\label{vordouble}
\end{figure}
As $\delta$ exceeds unity, $c_{2}$ becomes negative and the system enters a ferromagnetic phase. Then, the condensates undergo phase separation to spontaneously form domains having the same spin component. Concurrently, vortices begin to overlap at $\delta \sim 1$. For $\delta=1$, an SU(2) symmetric case, interesting structures appear as shown in Fig. \ref{vordouble}. In Fig. \ref{vordouble}(a), each condensate density forms a stripe pattern such that the lines of vortices for component one overlap those for component two. Mueller and Ho analytically derived the lattice structure in Fig. \ref{vordouble}(a), with the assumption of a perfect lattice\cite{Mueller}. However, our numerical treatment shows that the lattice is significantly different when the energy is only slightly different [Fig. \ref{vordouble}(b)]. For the same parameters, we can obtain another equilibrium state called a ``double-core vortex lattice" in Fig. \ref{phasedia}, where a vortex lattice of component 2 is made up of pairs of vortices with the same circulation and the vortices in component 1 surround those pairs. Therefore, various metastable structures will be observed in this parameter region.

\begin{figure}[btp]
\begin{center}
\includegraphics[height=0.45\textheight]{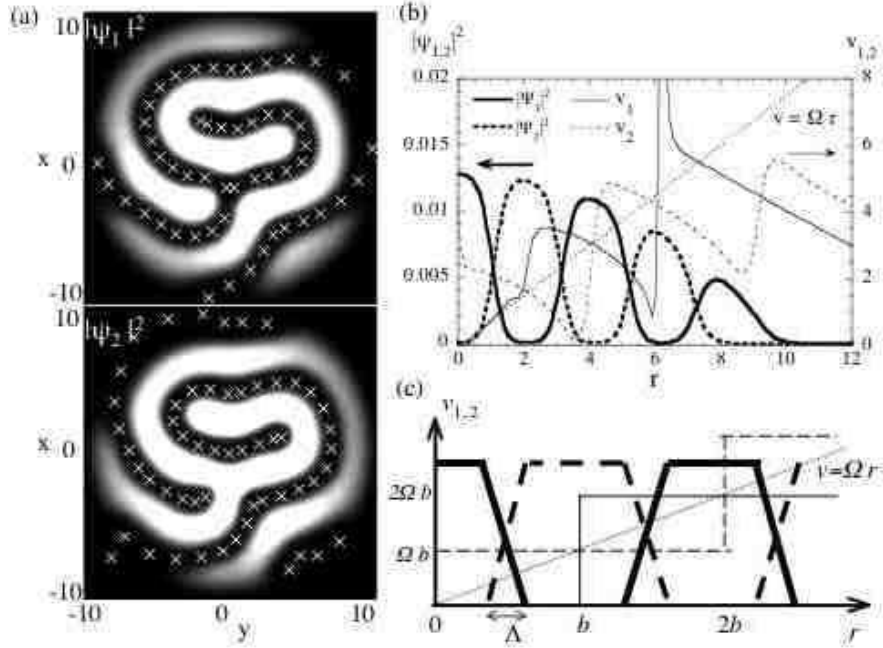}
\caption{Vortex sheets. (a) The density profiles of the condensates for $\Omega=0.75$ and $\delta=1.1$. The vortex sheets are made up of chains of singly quantized vortices whose positions are marked by $\times$. (b) Bold curves show the density profiles of $\Psi_{1}$ and $\Psi_{2}$ in the radial component, and thin curves the corresponding velocity profiles $|{\bf v}_{i}|=|\nabla {\rm arg}\psi_{i}|$. The velocity of rigid body rotation $v_{\rm s}=\Omega r$ is shown by a dotted line for comparison. (c) Model of the vortex sheet state. See the text for details.}
\label{vorsheet}
\end{center}
\end{figure}

For strongly phase-separated region $\delta>1$ ($c_{2}<0$), the domains of the same spin component, at which the other-component vortices are located, merge further, resulting in the formation of ``serpentine" vortex sheets. A typical example is shown in Fig. \ref{vorsheet}(a). Singly quantized vortices line up in sheets, and the sheets of component 1 and 2 are interwoven. Figure \ref{vorsheet}(b) shows that each superfluid velocity $v_{i}$ $(i=1,2)$ jumps at the vortex sheet, following approximately the velocity of solid-body rotation. In the region $\delta>1$, even though the value of $\delta$ is changed, the total density is fixed by the Thomas-Fermi distribution with $\delta = 1$.

A stationary vortex sheet has been observed in rotating superfluid $^{3}$He-A\cite{sheethe}. In that system, the vortices are bound to a topologically stable, domain-wall soliton across which the direction of the orbital angular momentum of the Cooper-pairs have opposite orientations. Moreover, experiments with rotating superfluid $^{3}$He-A show that the equilibrium separation between the sheets is determined by the competition between the surface tension $\sigma$ of the domain wall and the kinetic energy of the superflow\cite{sheethe}. 

\begin{figure}
\begin{center}
\includegraphics[height=0.25\textheight]{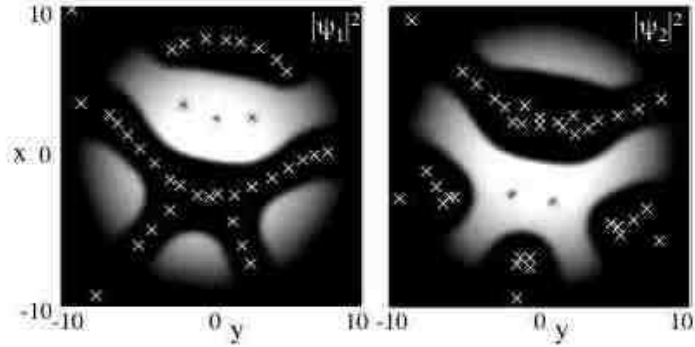}
\end{center}
\caption{Density profiles of the condensates for $\Omega=0.55$ and $\delta=1.5$. Positions of vortices are marked by $\times$.}
\label{rotatingdrop}
\end{figure}

How do these properties differ for two-component BECs? To estimate the sheet separation in two-component BECs, we consider a simple model shown in Fig. \ref{vorsheet}(c). In this model, each velocity is assumed to be uniform between the vortex sheets of the {\it same} component; specifically, the value of $v_{i}$ increases by $2 \Omega b$ across every sheet. The total density $\rho_{\rm T}=|\psi_{1}|^{2}+|\psi_{2}|^{2}$ is constant, and the domain boundary with the penetration depth $\Lambda$ is approximated by the linear profile. We calculate the free energy $F=E-\Omega L_{z}$ in the range $0<r<2b$ with the sheet distance $b$ being determined so as to minimize $F$ per unit area. Although the condensate density is lower near the edge because of the trapping potential, this model should be valid in the central region. We first estimate the optimized penetration depth $\Lambda_{\rm p}$ by minimizing the surface tension $\sigma$ of a single domain wall with respect to $\Lambda$\cite{Tim}, which is the sum of the quantum pressure energy $E_{\rm qp}/2 \pi b \approx (\rho_{\rm T}/2 \Lambda)$ and the interaction energy in the overlapping region $E_{\rm int}/2 \pi b =u (\delta-1) \Lambda \rho_{\rm T}^{2}/12$. Minimization of the surface tension with respect to $\Lambda$ yields $\Lambda_{\rm p} = \sqrt{6/u (\delta-1)\rho_{\rm T}}$. Then, the corresponding surface tension is written as 
\begin{equation}
\sigma = \rho_{\rm T}^{3/2} \sqrt{\frac{u (\delta-1)}{6}}.
\end{equation}
Next, the flow energy per area in rotating frame is 
\begin{equation}
\frac{1}{4\pi b^{2}} \int_{0}^{2b} d^{2}r \sum_{i} \rho _{i}  \frac{(v_{i}-\Omega r)^{2}}{2} = 
\frac{29 \rho_{\rm T} \Omega^{2} b^{2}}{768}, 
\end{equation}
where $b \gg \Lambda_{\rm p}$ is assumed. Thus, the free energy per unit area is 
\begin{equation}
\frac{F}{4 \pi b^{2}} = \frac{29 \rho_{\rm T} \Omega^{2} b^{2}}{768}
+ \frac{2 \sigma}{b}+\frac{u \rho_{\rm T}^{2}}{2}. 
\end{equation}
Minimizing this energy with respect to $b$, one obtains 
\begin{equation}
b=\biggl( \frac{768 \sigma}{29 \Omega^{2} \rho_{\rm T}} \biggr)^{1/3}=\biggl( \frac{768}{29 \Omega^{2} \Lambda_{\rm p}} \biggr)^{1/3}. \label{sheetsizeb}
\end{equation}
By using the Thomas-Fermi density at $r=0$ with $\delta=1$ as the value of $\rho_{T}$ in the denominator of Eq. (\ref{sheetsizeb}), the sheet spacing $b \propto (\delta-1)^{1/6} (1-\Omega^{2})^{1/12}/\Omega^{2/3}$ is consistent with what we found numerically; for example, for parameters used in Fig. \ref{vorsheet} we obtain $b=3.09b_{\rm ho}$. The vortex sheet is expected in the region $b>\Lambda_{\rm p}$ and $2b<R_{\rm TF}=\sqrt{2/\sqrt{\pi \alpha}}$, as shown in the shaded region in Fig. \ref{phasedia}(b). When $2b>R_{\rm TF}$ (i.e., when $b$ becomes comparable with the size of the condensate), a strong ferromagnetic interaction reduces the size of the domain wall region and the serpentine structure disappears as shown in Fig. \ref{rotatingdrop}, realizing rotating ``droplets" in which most vortices are located in the low density region. 

What structures can be observed experimentally? The two-component BECs realized in JILA are a mixture of the states $|1,-1 \rangle$ and $|2,1 \rangle$ of $^{87}$Rb. This mixture has scattering lengths in the ratio of $a_{11} : a_{22} : a_{12} = 1:0.94:0.97$ and so $\delta \sim 1$ for equal particle numbers. Therefore, lattices with partially overlapping vortices, as shown in Fig. \ref{vordouble}, are expected to be observed. $^{85}$Rb atoms in the $|2,2 \rangle$ state make an interesting system because both the intercomponent scattering length and the scattering length of $^{85}$Rb $|2,2 \rangle$ can be controlled by using the Feshbach resonance. The MIT group has observed the phase separation of a mixture of $^{23}$Na BECs with $|1,0\rangle$ and $|1,1\rangle$ state\cite{Miesner}, which has $\delta \geq 1$ ($a_{1}:a_{2}:a_{12}=1:1.035:1.035$). A mixture of $^{41}$K and $^{87}$Rb BECs which is reported recently \cite{Modugno} lies deeply in a phase-separated region. The vortex sheets should therefore be observed at high rotation frequencies. 

\subsection{Effect of internal coherent coupling on lattice structure}\label{Joseph}
As discussed in Sec. \ref{meronsect}, an external coherent-coupling field makes bound vortex pairs in a two-component system, inducing an effective attractive interaction between them. A question then naturally arises: what happens when an external field is applied to rapidly a rotating two-component BEC? Since two components interact with each other through their relative phase, the structure of the vortex states should be greatly affected by this effect. Zhai {\it et al.} studied a similar situation that involves the structure of a vortex lattice in a condensate trapped in a double-layered potential\cite{Zhai2}. The interlayer coherence, which seems to be a small contribution in such a situation, was instead shown to have a significant effect on the structural transition of the lattices. Their treatment was in the limit $u_{12} \ll u$, and thus their analyses were simpler than the case discussed in this subsection. Here we present some numerical results about the structure of vortex lattices with a coupling field. 

\begin{figure}[btp]
\begin{center}
\includegraphics[height=0.48\textheight]{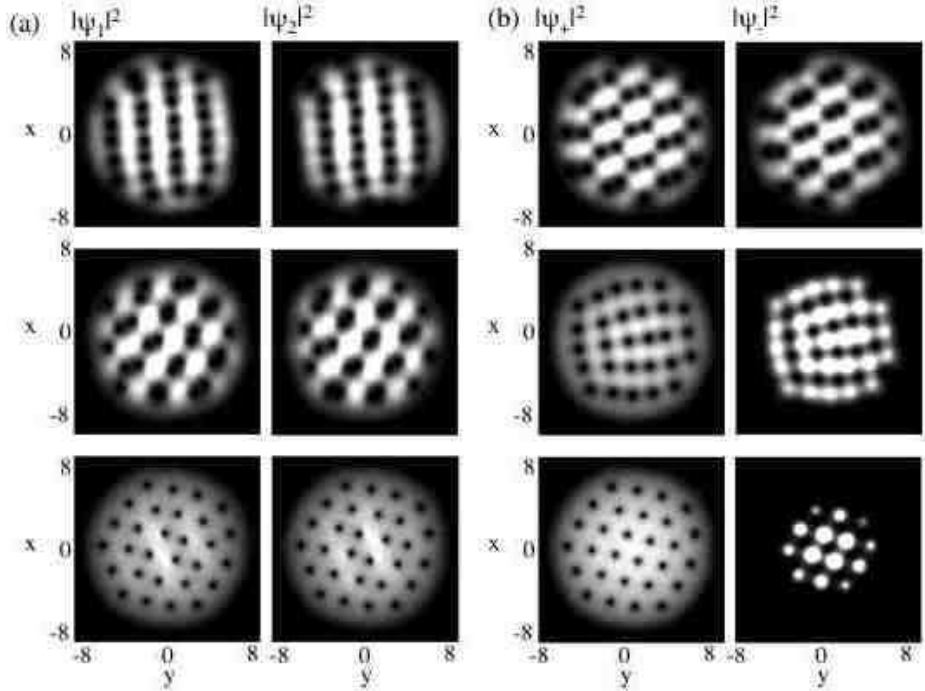}
\end{center}
\caption{(a) The density profiles of the equilibrium solutions $\psi_{1}$ and $\psi_{2}$ for $\Omega=0.8$ and $\omega_{\rm R}=0$, 0.2, 0.5 from top to bottom. (b) The corresponding profiles of $\psi_{+}$ and $\psi_{-}$ obtained by the basis transformation $\psi_{\pm} = (\psi_{1} \pm \psi_{2})/\sqrt{2}$.} 
\label{spintex1.5}
\end{figure}
First, we discuss the simple case $u_{1}=u_{2}=u_{12}$ ($c_{1}=c_{2}=0$) with the SU(2) symmetry. As the rotation frequency increases, a large number of vortices nucleate and form stripes or double-core lattices as shown in Fig. \ref{vordouble}. When $\omega_{\rm R}$ is turned on, the vortices begin to form pairs and then form a lattice of meron pairs. The resulting equilibrium structures for $\Omega=0.8$ and several values of $\omega_{\rm R}$ (0, 0.2, 0.5) are shown in Fig. \ref{spintex1.5}(a). The meron pairs are characterized by their axisymmetric topological charge and vorticity, yielding a their square lattice. This lattice of meron pairs is equivalently a lattice of skyrmions under the basis transformation $(\psi_{1},\psi_{2}) \rightarrow (\psi_{+},\psi_{-})$ as described in Eq. (\ref{timeindGPeqhenkan}) and shown in Fig. \ref{spintex1.5}(b). As $\omega_{\rm R}$ increases further, vortices of one component overlap completely with those of the other, forming a triangular lattice as in a single-component BEC. Viewed in terms of the $\psi_{\pm}$-basis, the decrease of the particle number in the $\psi_{-}$ component effectively weakens the mutual repulsion between the two components, thus giving rise to a transition from square lattices to triangular lattices. Therefore, application of an external coupling drive could be an excellent tool to explore a rich variety of vortex structures in two-component BECs. 

\begin{figure}[btp]
\begin{center}
\includegraphics[height=0.62\textheight]{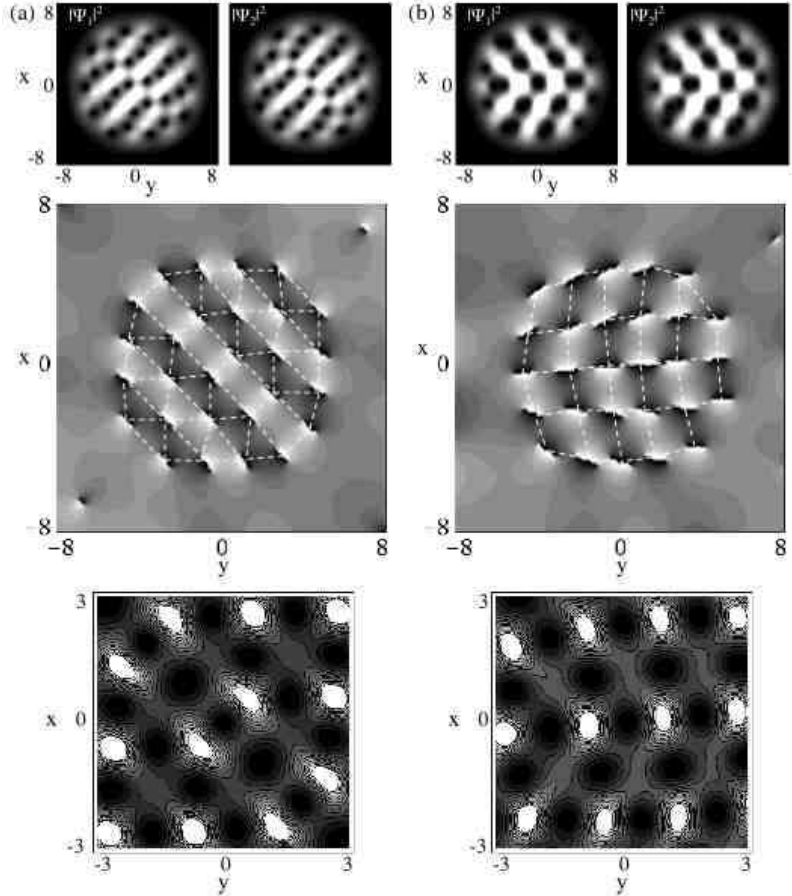}
\end{center}
\caption{Vortex lattices of SU(2)-symmetric BECs with internal coherent coupling. The upper panels are the profiles of the condensate density $|\psi_{1}|^{2}$ and $|\psi_{2}|^{2}$ and the lower, larger images are the relative phases for $\Omega=0.80$ and $\omega_{\rm R}=0.2$, the centers of mass of the molecules are linked by dashed lines. The bottom panels show the distributions of the topological charge $q({\bf r})$. (a) $c_{2}=20$ (antiferromagnetic). (b) $c_{2}=-20$ (ferromagnetic). } 
\label{spintex2}
\end{figure}
However, in the case that $u_{1} \neq u_{2} \neq u_{12}$, the meron pair is characterized by an anisotropic topological charge as shown in Fig. \ref{c2ferroantiferro}. A consequence of this anisotropy is that the resultant vortex state exhibits a distorted lattice structure. Typical examples are shown in Fig. \ref{spintex2}, where (a) and (b) correspond to the antiferromagnetic and ferromagnetic cases, respectively. From the bottom panels in Fig. \ref{spintex2}, the direction of the ordering depends strongly on the distribution of the topological charge in each meron pair. The collective mode of this distorted lattice is expected to exhibit anisotropic wave propagation, depending on the polarization of the meron pairs.

\subsection{Experimental observation of square lattices in two-component BECs}
\begin{figure}[btp]
\begin{center}
\includegraphics[height=0.60\textheight]{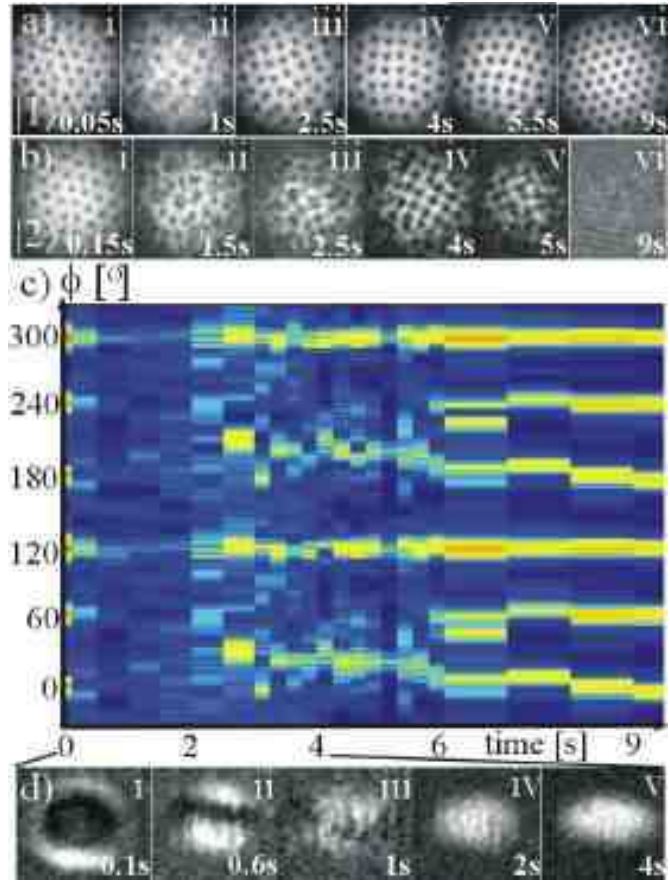}
\end{center}
\caption{Experimental observation of square lattices in two-component BECs. (a) Time sequence of images of state $|1\rangle$, after $\sim80\%$ population transfer to $|2\rangle$, showing the evolution from a hexagonal lattice over a turbulent stage to a square structure and back to a hexagonal lattice. (b) Similar to (a) except for state $|2\rangle$. In this case, a square lattice forms (iv) before the lattice decays. (c) Time evolution in reciprocal space. Intensity in an annulus along the $\varphi$ coordinate. The ordinate is the angle. The initial 6-peak structure of the hexagonal lattice vanishes quickly due to turbulence. The four peaks in $3-5.5\,\rm{s}$ indicate a square lattice. During $6-9\,\rm{s}$ a transition back to hexagonal lattices occurs. (d) Two-color edge-on images of the initial turbulent evolution. State $|1\rangle$ ($|2\rangle$) appears bright (dark) on gray background. The fine filament structures in (ii-v) are due to mutual filling of vortex cores. [Schweikhard {\it et al}., Phys. Rev. Lett. {\bf 93} 210403-2 (2004), reproduced with permission. Copyright (2004) by the American Physical Society.]} 
\label{TimeEvolutiuon}
\end{figure}

Schweikhard {\it et al.} observed well-separated vortices forming in two-component BECs that later formed ordered square lattices\cite{Volker}. As in Matthew {\it et al}'s experiment\cite{Matthews} mentioned in Sec. \ref{experim}, the two-component system consists of two hyperfine levels of the $^{87}$Rb atom: $|F=1,m_{F}=-1\rangle = |1\rangle$ and $|F=2,m_{F}=1\rangle = |2\rangle$. To study the rotational properties, they created a regular triangular vortex lattice in a BEC in the state $|1\rangle$ that contained $(3.5-4)\times 10^6$ atoms and rotated at a frequency $\Omega \approx 0.75\times\omega_{\perp}$ about the z-axis. Then, a short pulse of the coupling drive transfered a $80-85\%$ fraction of the population into state $|2\rangle$. After a variable wait time, they took two images of the condensates. One was a nondestructive phase-contrast image [Fig. \ref{TimeEvolutiuon}(d)], which was taken along an axis perpendicular to the rotation axis. The other was a destructive image to resolve the vortex core structures [Fig. \ref{TimeEvolutiuon}(a) and (b)] taken along the rotation axis after the free expansion. 

Figures \ref{TimeEvolutiuon} (a) and (b) show the formation and decay dynamics of the square lattice structure of the expanded $|1\rangle$ and $|2\rangle$ states. For the first period $0.1-0.25\,\rm{s}$, there is little dynamical behavoir in either component and certainly no structural transition in the vortex lattice. From $0.25 - 2\,\rm{s}$, turbulent behavior appears in both components in which vortex visibility degrades significantly, as shown in Figs. \ref{TimeEvolutiuon}[a(ii)] and \ref{TimeEvolutiuon}[b(ii)]. This turbulence is a direct consequence of the transition from overlapping hexagonal vortex lattices to interlaced square lattices. From $2 - 3\,\rm{s}$, square lattices emerge from the turbulent state (Fig. \ref{TimeEvolutiuon}[a(iii)]). From $3 - 5.5\,\rm{s}$ stable square lattices are observed in both components (Figs. \ref{TimeEvolutiuon}[a(iv)] and [b(iv)]). At this stage, around $4\,\rm{s}$, despite the large ($80-85\,\%$) initial population transfer to the state $|2\rangle$, the number of $|2\rangle$ atoms has decreased to only $1.5 \times10^5$, while the state $|1\rangle$ contains $(5-7)\times 10^5$ atoms. This is due to a larger inelastic scattering loss of the $|2\rangle$ state than that of the $|1\rangle$ state.  As the $|2\rangle$ state population continues to decay, the vortex lattice planes bend (Fig. \ref{TimeEvolutiuon}[a(v)]) and a transition back to a hexagonal lattice in the state $|1\rangle$ takes place (Fig. \ref{TimeEvolutiuon}[a(vi)]). No turbulence occurs during the transition from a square to a hexagonal lattice.

To discuss the structural transition more clearly, Fig. \ref{TimeEvolutiuon}(c) shows the time evolution of the intensity of the reciprocal lattice peaks of the vortex lattice. In this plot,  the triangular lattice has six peaks at $\phi=60^{\circ} \times n$ with integer $n$, whereas the square lattice has four peaks at $90^{\circ} \times n$. Initially six peaks are visible, which corresponds to the reciprocal lattice of a triangular vortex lattice. Because of subsequent turbulence these peaks vanish between $200\,\rm{ms}$ and $2\,\rm{s}$. After $2-3 \,\rm{s}$ a four-peak structure appears, which is a clear signature of a square lattice. At $5.5\,\rm{s}$ two additional peaks appear, which signals the onset of the transition back to the triangular lattice. This experiment thus demonstrated that, although the state is transient, the intercomponent interaction yields the square lattices of vortices as predicted in Refs. \refcite{Mueller2} and \refcite{Kasamatsu3}.

Figure \ref{TimeEvolutiuon}(d) shows a time sequence of side-view images, where the $|1\rangle$ ($|2\rangle$) state appears bright (dark). As can be seen in Fig. \ref{TimeEvolutiuon}[d(i)], a spatial phase separation from the initially overlapping two components to a ball-shell structure occurs within $50 - 100 \rm{ms}$. During the period of this axial separation, both the individual vortices and the overall vortex lattice remain remarkably stable, as viewed along the rotation axis (see Fig. \ref{TimeEvolutiuon}[a(i)] and Fig. \ref{TimeEvolutiuon}[b(i)]). Around $600\,\rm{ms}$, fine filament structures appear at the intercomponent boundary (Fig. \ref{TimeEvolutiuon}[d(ii)]), concurrently with the full development of vortex turbulence seen in Fig. \ref{TimeEvolutiuon}[a(ii)]. The filament structures distort and fill out the whole BEC as the vortex turbulence peaks at around $1\,\rm{s}$ (Fig. \ref{TimeEvolutiuon}[d(iii)]), and straightens up at around $2\,\rm{s}$ (Fig. \ref{TimeEvolutiuon}[d(iv)]), concurrently with the restoration of vortex visibility in top view images. Subsequently, the filaments become less visible as the state $|2\rangle$ decays (Fig. \ref{TimeEvolutiuon}[d(v)]). These structures are interpreted as vortex cores in either component being filled by fluid of the other component, forming a skyrmion lattice. In the experiment, interference techniques were used to investigate the detailed dynamics of the phase separation. Also, the stability of square-lattice structure was confirmed by observing the fast relaxation of the Tkachenko excitations, as in the single component BEC\cite{Coddington,Baym}$^{-}$\cite{Gifford}. 


\section{Conclusions}\label{conclusion}
We have reviewed the physics of quantized vortices in rotating multicomponent BECs, with an emphasis on unconventional vortex states in two-component system. The vector nature of the order parameter gives rise to intriguing structures and dynamics that are absent in a single component system. Because analogous physical phenomena and theoretical formulation are found in other physical systems which involve multicomponent order parameters, an indepth understanding of quantized vortices in multicomponent BECs may offer physical insight in other systems.

For the slowly rotating regimes, we discussed the structure of single vortex states in two-component BECs with and without internal coherent coupling by using both numerical simulations of the coupled GP equations and the variational analysis based on the nonlinear $\sigma$ model. The vortex structure is significantly altered from that of a single-component BEC because of the intercomponent interaction. We showed that a pseudospin is a useful concept to represent the vortex state, which reveals structures and energetics of the important topological excitations called skyrmions and merons. A vortex molecule (a meron pair) is predicted to be stabilized in the coherently coupled two-component BEC in a rotating potential, and the contributions that breaks SU(2) symmetry make the vortex molecule have anisotropic vorticity. We showed the theoretical studies on the 3-D nontrivial skyrmion, which is regarded as a composite vortex ring and is closely related to a proposed topological excitation in the early universe. Also, we gave an introdutory discussion on quantized vortices and monopoles in spin-1 BECs. Much remains to be investigated concerning the mutual interactions between exotic topological defects and their nontrivial dynamics.

We described how a fast rotation can stabilize a lattice of skyrmions and merons. Such a lattice consists of vortices in each component. The analytical and numerical studies revealed a rich variety of vortex phase diagrams in the variables of rotation frequency ($0.4<\Omega/\omega_{\perp}<1$) and an intercomponent interaction strength. The intercomponent repulsive interaction causes one lattice to displace relative to the other, stabilizing mainly the interlaced square lattices in an antiferromagnetic regime and the serpentine vortex sheets in a ferromagnetic regime. We proposed that an external driving field is a useful tool to control the strength of the interaction between the two components and, thereby, provides a unique means to study the rich variety of vortex phases. Much work is necessary to reveal further details on the lattices of skyrmions and merons in rapidly rotating two-component BECs; moreover there remain further interesting problems such as structural transitions and collective oscillations of the lattice of skyrmions and merons, and new phenomena related to double-layer quantum Hall physics\cite{Girvin}. 

Another type of two-component system of ultra-cold atoms is boson-fermion mixtures. Such studies, which includes $^{7}$Li-$^{6}$Li (Refs. \refcite{Truscott,Schreck}), $^{23}$Na-$^{6}$Li (Ref. \refcite{Hadzibabic}), and $^{87}$Rb-$^{40}$K (Refs. \refcite{Roatimix,Modugnomix,Goldwin}), are analogous to the system of $^{3}$He-$^{4}$He interpenetrating superfluids\cite{Edward}. In a spin-polarized fermionic system, $s$-wave fermion-fermion collisions are forbidded by the Pauli exclusion principle, and the Fermi pressure is the dominant contribution to the ground state properties. The boson-fermion mixture will provide another field to investigate rich vortex phenomena mediated by the boson-fermion interaction, which is repulsive\cite{Truscott,Schreck} or attractive\cite{Roatimix,Modugnomix,Goldwin}, and can be tuned through the Feshbach resonance\cite{StanFesh,InouyeFesh}. The stability of vortex states in a $^{87}$Rb-$^{40}$K mixture was studied theoretically in Ref.\refcite{Jezekmix}. Quatized vortices in spin-2 BECs are also an intriguing subject and will soon open up a new research field because spin-2 BECs have been successfully loaded in an optical trap\cite{Schmaljohann}$^{-}$\cite{Kuwamoto}. This system is even more attractive than the spin-1 system, because it is expected to have a rich variety of novel ground state phases, a magnetic response, and spin dynamics\cite{Koashi}$^{-}$\cite{Ueda}. A method of topological phase imprinting\cite{Leanhardt,Nakahara,Leanhardt2} is also a powerful tool to generate a quantized vortex\cite{Mottonen} and the experiment toward this direction is in progress\cite{Kawaguchi}.


\section*{Acknowledgements}

K.K. and M.T. acknowledge support by a Grant-in-Aid for Scientific Research (Grants No. 15$\cdot$5955 and No. 15341022) by the Japan Society for the Promotion of Science. 
M.U. acknowledges  support by a Grant-in-Aid for Scientific Research (Grant No.15340129) by the Ministry of Education, Culture, Sports, Science and Technology of Japan, and a CREST program by Japan Science and Technology Corporation (JST). 









\begin{thebibliography}{99}

\bibitem{Doneley} 
R. J. Donnelly, {\it Quantized Vortices in Helium II} (Cambridge University Press, Cambridge, 1991). 
\bibitem{Salomaa}
M.M. Salomaa and G.E. Volovik, {\it Rev. Mod. Phys.} {\bf 59}, 533 (1987); G.E. Volovik, {\it The universe in a helium droplet} (Oxford University Press, Oxford, 2003).
\bibitem{Joynt}
A. Knigavko and B. Rosenstein, {\it Phys. Rev. Lett.} {\bf 82}, 1261 (1999); B. Rosenstein, I. Shapiro, B. Ya. Shapiro, and G. Bel, {\it Phys. Rev.} {\bf B67}, 224507 (2003). 
\bibitem{Girvin}
S.M. Girvin, {\it The Quantum Hall effect: Novel Excitations and Broken Symmtries, Proceedings of the Les Houches Summer School of Theoretical Physics}, 1998 (Springer Verlag, Berlin, and Les Editions de Physique, Paris); for a brief review, S.M. Girvin, {\it Physics Today} {\bf 53}, 39 (2000). 
\bibitem{Chen}
Y.F. Chen, K.F. Huang, H.C. Lai, and Y.P. Lan, {\it Phys. Rev. Lett.} {\bf 90}, 053904 (2003).
\bibitem{Skyrmi}
T.H.R. Skyrme, {\it Proc. R. Soc. London}, Ser. A {\bf 260}, 127 (1961).
\bibitem{cosmo}
{\it Topological Defects and the Non-Equilibrium Dynamics of Symmetry Breaking 
Phase Transitions,} edited by Y.M. Bunkov and H. Godfrin (Kluwer Academic Publishers, Les Houches, 1999).
\bibitem{Pethickbook}
C.J. Pethick and H. Smith, {\it Bose-Einstein Condensation in Dilute Gases}, Cambridge University Press, Cambridge (2002).
\bibitem{Pitaevskiibook}
L. Pitaevskii and S. Stringari, {\it Bose-Einstein Condensation}, Oxford University Press, Oxford (2003).
\bibitem{Duine}
R.A. Duine and H.T.C. Stoof, {\it Phys. Rep.} {\bf 396}, 115 (2004).
\bibitem{Matthews}
M.R. Matthews, B.P. Anderson, P.C. Haljan, D.S. Hall, C.E. Wieman, and E.A. Cornell, {\it Phys. Rev. Lett.} {\bf 83}, 2498 (1999). 
\bibitem{Williams}
J.E. Williams and M.J. Holland, {\it Nature (London)} {\bf 401}, 568 (1999). 
\bibitem{Leanhardt}
A.E. Leanhardt, A. G\"{o}rlitz, A.P. Chikkatur, D. Kielpinski, Y. Shin, D.E. Pritchard, and W. Ketterle, {\it Phys. Rev. Lett.} {\bf 89}, 190403 (2002). 
\bibitem{Nakahara}
M. Nakahara, T. Isoshima, K. Machida, S.-I. Ogawa, and T. Ohmi, {\it Physica} {\bf B284-288}, 17 (2000); T. Isoshima, M. Nakahara, T. Ohmi, K. Machida, {\it Phys. Rev.} {\bf A61}, 063610 (2000); S.-I. Ogawa, M. M\"{o}tt\"{o}nen, M. Nakahara, T. Ohmi, H. Shimada, {\it Phys. Rev.} {\bf A66}, 013617 (2002). .
\bibitem{Madison} 
K.W. Madison, F. Chevy, W. Wohlleben, and J. Dalibard, {\it Phys. Rev. Lett.} {\bf 84}, 806 (2000).
\bibitem{Abo}
J.R. Abo-Shaeer, C. Raman, J. M. Vogels, and W. Ketterle, {\it Science} {\bf 292}, 476 (2001).
\bibitem{Haljan}
P.C. Haljan, I. Coddington, P. Engels, and E. A. Cornell, {\it Phys. Rev. Lett.} {\bf 87}, 210403 (2001).
\bibitem{Hodby}
E. Hodby, G. Hechenblaikner, S. A. Hopkins, O. M. Marag\'{o}, and C. J. Foot, {\it Phys. Rev. Lett.} {\bf 88}, 010405 (2002). 
\bibitem{Fetterrev}
 A.L. Fetter and A.A. Svidzinsky, {\it J. Phys. Condens. Matter}, {\bf 13}, R135 (2001).
\bibitem{Feder}
D.L. Feder, A.A. Svidzinsky, A.L. Fetter, and C.W. Clark, {\it Phys. Rev. Lett.} {\bf 86}, 564 (2001). 
\bibitem{Ripoll}
J.J. Garc\'{i}a-Ripoll and V.M. P\'{e}rez-Garc\'{i}a, {\it Phys. Rev.} {\bf A64}, 053611 (2001). 
\bibitem{Aftalion}
A. Aftalion and T. Riviere, {\it Phys. Rev.} {\bf A64}, 043611 (2001); A. Aftalion and I. Danaila, {\it Phys. Rev.} {\bf A68}, 023603 (2003). 
\bibitem{Modugno}
M. Modugno, L. Pricoupenko, and Y. Castin, {\it Eur. Phys. J.} {\bf D22}, 235 (2003). 
\bibitem{Rosenbusch}
P. Rosenbusch, V. Bretin, and J. Dalibard, {\it Phys. Rev. Lett.} {\bf 89}, 200403 (2002).
\bibitem{Ishosima}
T. Isoshima and K. Machida, {\it Phys. Rev.} {\bf A60}, 3313 (1999).
\bibitem{Dalfovo}
F. Dalfovo and S. Stringari, {\it Phys. Rev.} {\bf A63}, 011601 (2001).
\bibitem{Sinha}
S. Sinha and Y. Castin, {\it Phys. Rev. Lett.} {\bf 87}, 190402 (2001). 
\bibitem{Ripoll2}
J. J. Garc\'{i}a-Ripoll and V. M. P\'{e}rez-Garc\'{i}a, {\it Phys. Rev.} {\bf A63}, 041603(R) (2001). 
\bibitem{Anglin}
J. R. Anglin, {\it Phys. Rev. Lett.} {\bf 87}, 240401 (2001); {\it Phys. Rev.} {\bf A65}, 063611 (2002). 
\bibitem{Simula}
T. P. Simula, S. M. M. Virtanen, and M. M. Salomaa, {\it Phys. Rev.} {\bf A66}, 035601 (2002).
\bibitem{Williams3}
J. E. Williams, E. Zaremba, B. Jackson, T. Nikuni, and A. Griffin, {\it Phys. Rev. Lett.} {\bf 88}, 070401 (2002). 
\bibitem{Kramer}
M. Kraemer, L. Pitaevskii, S. Stringari, and F. Zambelli, {\it Laser Physics} {\bf 12}, 113 (2002). 
\bibitem{Kawaja1}
U. Al Khawaja, {\it Phys. Rev.} {\bf A68}, 063614 (2003). 
\bibitem{Madison2}
K. W. Madison, F. Chevy, V. Bretin, and J. Dalibard, {\it Phys. Rev. Lett.} {\bf 86}, 4443 (2001).
\bibitem{Tsubota}
M. Tsubota, K. Kasamatsu, and M. Ueda, {\it Phys. Rev.} {\bf A65}, 023603 (2002).
\bibitem{Penckwitt}
A.A. Penckwitt, R.J. Ballagh, and C.W. Gardiner, {\it Phys. Rev. Lett.} {\bf 89}, 260402 (2002). 
\bibitem{Kasamatsu}
K. Kasamatsu, M. Tsubota, and M. Ueda, {\it Phys. Rev.} {\bf A67}, 033610 (2003). 
\bibitem{Lundh}
E. Lundh, J.-P. Martikainen, and K.-A. Suominen, {\it Phys. Rev.} {\bf A67}, 063604 (2003). 
\bibitem{Lobo}
C. Lobo, A. Sinatra, Y. Castin, {\it Phys. Rev. Lett.} {\bf 92}, 020403 (2004). 
\bibitem{Engels}
P. Engels, I. Coddington, P.C. Haljan, and E.A. Cornell, {\it Phys. Rev. Lett.} {\bf 89}, 100403 (2002). 
\bibitem{Engels2}
P. Engels, I. Coddington, P.C. Haljan, V. Schweikhard, and E.A. Cornell, {\it Phys. Rev. Lett.} {\bf 90}, 170405 (2003). 
\bibitem{Coddington}
I. Coddington, P. Engels, V. Schweikhard, and E.A. Cornell, {\it Phys. Rev. Lett.} {\bf 91}, 100402 (2003). 
\bibitem{Schweikhard}
V. Schweikhard, I. Coddington, P. Engels, V.P. Mogendorff, and E.A. Cornell, {\it Phys. Rev. Lett.} {\bf 92}, 040404 (2004). 
\bibitem{Bretin2}
V. Bretin, S. Stock, Y. Seurin, and J. Dalibard, {\it Phys. Rev. Lett.} {\bf 92}, 040404 (2004). 
\bibitem{Baym}
G. Baym, {\it Phys. Rev. Lett.} {\bf 91}, 110402 (2003); {\it Phys. Rev.} {\bf A69}, 043618 (2004).  
\bibitem{Mizushima2}
T. Mizushima, Y. Kawaguchi, K. Machida, T. Ohmi, T. Isoshima, and M.M. Salomaa, {\it Phys. Rev. Lett.} {\bf 92}, 060407 (2004). 
\bibitem{Baksmaty}
L.O. Baksmaty, S.J. Woo, S. Choi, and N.P. Bigelow, {\it Phys. Rev. Lett.} {\bf 92}, 160405 (2004); S.J. Woo, L.O. Baksmaty, S. Choi, and N.P. Bigelow, {\it ibid} {\bf 92}, 170402 (2004). 
\bibitem{Cozzini}
M. Cozzini, L.P. Pitaevskii, and S. Stringari, {\it Phys. Rev. Lett.} {\bf 92}, 220401 (2004). 
\bibitem{Gifford}
S.A. Gifford and G. Baym, {\it Phys. Rev.} {\bf A70}, 033602 (2004). 
\bibitem{TPSimula}
A.L. Fetter, {\it Phys. Rev.} {\bf A68}, 063617 (2003); T.P. Simula, A.A. Penckwitt, and R.J. Ballagh, {\it Phys. Rev. Lett.} {\bf 92}, 060401 (2004). 
\bibitem{Fetteranh}
A.L. Fetter, {\it Phys. Rev.} {\bf A64}, 063608 (2001). 
\bibitem{Lundh2}
E. Lundh, {\it Phys. Rev.} {\bf A65}, 043604 (2002). 
\bibitem{Kasamatsu2}
K. Kasamatsu, M. Tsubota, and M. Ueda, {\it Phys. Rev.} {\bf A66}, 053606 (2002). 
\bibitem{Fischer}
U.W. Fischer and G. Baym, {\it Phys. Rev. Lett.} {\bf 90}, 140402 (2003). 
\bibitem{Kavoulakis}
G.M. Kavoulakis and G. Baym, {\it New Jour. Phys.} {\bf 5}, 51.1 (2003).
\bibitem{Aftalion2}
A. Aftalion and I. Danaila, {\it Phys. Rev.} {\bf A69}, 033608 (2004). 
\bibitem{Lunchatt}
E. Lundh, A. Collin, and K-A. Suominen, {\it Phys. Rev. Lett.} {\bf 92}, 070401 (2004); G.M. Kavoulakis, A.D. Jackson, and G. Baym {\it Phys. Rev.} {\bf A70}, 043603 (2004). 
\bibitem{adJack}
A.D. Jackson, G.M. Kavoulakis, and E. Lundh, {\it Phys. Rev.} {\bf A69}, 053619 (2004); A.D. Jackson, G.M. Kavoulakis, {\it Phys. Rev.} {\bf A70}, 023601 (2004). 
\bibitem{Saito}
H. Saito and M. Ueda, {\it Phys. Rev. Lett.} {\bf 93}, 220402 (2004).
\bibitem{Cooper}
N.R. Cooper, N.K. Wilkin, and J.M.F. Gunn, {\it Phys. Rev. Lett.} {\bf 87}, 120405 (2001). 
\bibitem{Paredes}
B. Paredes, P. Fedichev, J.I. Cirac, and P.Zollar, {\it Phys. Rev. Lett.} {\bf 87}, 010402 (2001). 
\bibitem{Sinova}
J. Sinova, C.B. Hanna, and A.H. MacDonald, {\it Phys. Rev. Lett.} {\bf 89}, 030403 (2002), {\bf 90}, 120401 (2003). 
\bibitem{Regnault}
N. Regnault and Th. Jolicoeur, {\it Phys. Rev. Lett.} {\bf 91}, 030402 (2003); {\it Phys. Rev.} {\bf B69}, 235309 (2004). 
\bibitem{Cazalilla}
M.A. Cazalilla, {\it Phys. Rev.} {\bf A67}, 063613 (2003). 
\bibitem{Nakajima}
T. Nakajima and M. Ueda, {\it Phys. Rev. Lett.} {\bf 91}, 140401 (2003). 
\bibitem{Ghosh}
T.K. Ghosh and G. Baskaran, {\it Phys. Rev.} {\bf A69}, 023603 (2004). 
\bibitem{UWEFI}
U.R. Fischer, {\it Phys. Rev. Lett.} {\bf 93}, 160403 (2004). 
\bibitem{Pismen}
L.M. Pismen, {\it Vortices in nonlinear fields} (Oxford University Press, Oxford, 1999).
\bibitem{Myatt} 
C.J. Myatt, E.A. Burt, R.W. Ghrist, E.A. Cornell, and C.E. Wieman, {\it Phys. Rev. Lett}. {\bf 78}, 586 (1997).
\bibitem{Hall1}
D.S. Hall, M.R. Matthews, J.R. Ensher, C.E. Wieman, and E.A. Cornell, {\it Phys. Rev. Lett.} {\bf 81}, 1539 (1998). 
\bibitem{Stenger}
J. Stenger, S. Inouye, D.M. Stamper-Kurn, H.-J Miesner, A.P. Chikkatur, and W. Ketterle, {\it Nature (London)} {\bf 396}, 345 (1998). 
\bibitem{Miesner}
H.-J. Miesner, D.M. Stamper-Kurn, J. Stenger, S Inouye, A.P. Chikkatur, and W. Ketterle, {\it Phys. Rev. Lett.} {\bf 82}, 2228 (1999). 
\bibitem{Barrett}
M.D. Barrett, J.A. Sauer, and M.S. Chapman, {\it Phys. Rev. Lett.} {\bf 87}, 010404 (2001). 
\bibitem{Schmaljohann}
H. Schmaljohann, M. Erhard, J. Kronj\"{a}ger, M. Kottke, S. van Staa, L. Cacciapuoti, J.J. Arlt, K. Bongs, and K. Sengstock, {\it Phys. Rev. Lett.} {\bf 92}, 040402 (2004).
\bibitem{Chang}
M.-S. Chang, C.D. Hamley, M.D. Barrett, J.A. Sauer, K.M. Fortier, W. Zhang, L. You, and M.S. Chapman, {\it Phys. Rev. Lett.} {\bf 92}, 140403 (2004).
\bibitem{Kuwamoto}
T. Kuwamoto, K. Araki, T. Eno, and T. Hirano, {\it Phys. Rev.} {\bf A69}, 063604 (2004).
\bibitem{Modugno2}
G. Modugno, M. Modugno, F. Riboli, G. Roati, and M. Inguscio, {\it Phys. Rev. Lett.} {\bf 89}, 190404 (2002). 
\bibitem{Ho2}
T.-L. Ho and V.B. Shenoy, {\it Phys. Rev. Lett.} {\bf 77}, 3276 (1996).
\bibitem{VLeonhardt}
U. Leonhardt and G.E. Volovik, {\it JETP Lett.} {\bf 72}, 46 (2000). 
\bibitem{Chui}
S. T. Chui, V.N. Ryzhov, and E.E. Tareyeva, {\it Phys. Rev.} {\bf A63}, 023605 (2001).
\bibitem{Jezek}
D.M. Jezek, P. Capuzzi, and H.M. Cataldo, {\it Phys. Rev.} {\bf A64}, 023605 (2001). 
\bibitem{Ripoll3}
J.J. Garc\'{i}a-Ripoll and V.M. P\'{e}rez-Garc\'{i}a, {\it Phys. Rev. Lett.} {\bf 84}, 4264 (2000).
\bibitem{Ripoll31}
V.M. P\'{e}rez-Garc\'{i}a and J.J. Garc\'{i}a-Ripoll, {\it Phys. Rev.} {\bf A62}, 033601 (2000).
\bibitem{Skrybin}
D.V. Skryabin, {\it Phys. Rev.} {\bf A63}, 013602 (2000). 
\bibitem{Ohberg}
P. \"{O}hberg and L. Santos, {\it Phys. Rev.} {\bf A66}, 013616 (2002).
\bibitem{Ripoll4}
J.J. Garc\'{i}a-Ripoll, V.M. P\'{e}rez-Garc\'{i}a, and F. Sols, {\it Phys. Rev.} {\bf A66}, 021602(R) (2002). 
\bibitem{Mueller2}
E.J. Mueller and T.-L. Ho, {\it Phys. Rev. Lett.} {\bf 88}, 180403 (2002). 
\bibitem{Kasamatsu3} 
K. Kasamatsu, M. Tsubota, and M. Ueda, {\it Phys. Rev. Lett.} {\bf 91}, 150406 (2003).  
\bibitem{Park2}
Q.H. Park and J.H. Eberly, {\it Phys. Rev.} {\bf A70}, 021602(R) (2004). 
\bibitem{Dotten}
Z. Dutton and J. Ruostekoski, {\it Phys. Rev. Lett.} {\bf 93}, 193602 (2004).  
\bibitem{Kasamatsupre}
K. Kasamatsu, M. Tsubota and M. Ueda, {\it Phys. Rev. Lett.} {\bf 93}, 250406 (2004).  
\bibitem{Kasamatsuskyrm}
K. Kasamatsu, M. Tsubota and M. Ueda, cond-mat/0411544.
\bibitem{Kawaja2}
U. Al Khawaja and H.T.C. Stoof, {\it Nature (London)} {\bf 411}, 918 (2001); {\it Phys. Rev.} {\bf A64}, 043612 (2001). 
\bibitem{Ruostekoski}
J. Ruostekoski and J.R. Anglin, {\it Phys. Rev. Lett.} {\bf 86}, 3934 (2001).
\bibitem{Battye} 
R.A. Battye, N.R. Cooper, and P.M. Sutcliffe, {\it Phys. Rev. Lett.} {\bf 88}, 080401 (2002).
\bibitem{Savage}
C.M. Savage and J. Ruostekoski, {\it Phys. Rev. Lett.} {\bf 91}, 010403 (2003); J. Ruostekoski, {\it Phys. Rev.} {\bf A70}, 041601(R) (2004). 
\bibitem{Ohmi}
T. Ohmi and J. Machida, {\it J. Phys. Soc. Jpn.} {\bf 67}, 1822 (1998)
\bibitem{Ho3} 
T.L. Ho, {\it Phys. Rev. Lett.} {\bf 81}, 742 (1998). 
\bibitem{Yip}
S.-K. Yip, {\it Phys. Rev. Lett.} {\bf 83}, 4677 (1999).
\bibitem{Marzlin}
K.P. Marzlin, W. Zhang, and B.C. Sanders, {\it Phys. Rev.} {\bf A62}, 013602 (2000). 
\bibitem{HPu}
H. Pu, S. Raghavan, and N.P. Bigelow, {\it Phys. Rev.} {\bf A63}, 063603 (2001).
\bibitem{Tuchiya}
S. Tuchiya and S. Kurihara, {\it J. Phys. Soc. Jpn.} {\bf 70}, 1182 (2001).
\bibitem{Isoshima}
T. Isoshima and K. Machida, {\it Phys. Rev.} {\bf A66}, 023602 (2002); T. Isoshima and K. Machida, and T. Ohmi {\it J. Phys. Soc. Jpn.} {\bf 70}, 1604 (2001). 
\bibitem{Mizushima3} 
T. Mizushima, K. Machida, and T. Kita, {\it Phys. Rev. Lett.} {\bf 89}, 030401 (2002); {\it Phys. Rev.} {\bf A66}, 053610 (2002)
\bibitem{Martikainen}
J.-P. Martikainen, A. Collin, and K.-A. Suominen, {\it Phys. Rev.} {\bf A66}, 053604 (2002). 
\bibitem{Kita}
T. Kita, T. Mizushima, and K. Machida, {\it Phys. Rev.} {\bf A66}, 061601(R) (2002). 
\bibitem{Zhai}
H. Zhai, W.Q. Chen, Z. Xu, and L. Chang, {\it Phys. Rev.} {\bf A68}, 043602 (2003). 
\bibitem{Bulgakov}
E.V. Bulgakov and A.F. Sadreev, {\it Phys. Rev. Lett.} {\bf 90}, 200401 (2003). 
\bibitem{Reijnders}
J.W. Reijnders, F.J.M. van Lankvelt, K. Schoutens, and N. Read, {\it Phys. Rev.} {\bf A69}, 023612 (2004). 
\bibitem{Mueller}
E.J. Mueller, {\it Phys. Rev.} {\bf A69}, 033606 (2004). 
\bibitem{Mizushima4}
T. Mizushima, N. Kobayashi, and K. Machida, {\it Phys. Rev.} {\bf A70}, 043613 (2004). 
\bibitem{Leanhardt2}
A.E. Leanhardt, Y. Shin, D. Kielpinski, D.E. Pritchard, and W. Ketterle, {\it Phys. Rev. Lett.} {\bf 90}, 140403 (2003). 
\bibitem{Stoofspin}
H.T.C. Stoof, E. Vliegen, and U. Al Khawaja, {\it Phys. Rev. Lett.} {\bf 87}, 120407 (2001).
\bibitem{Martispini}
J.-P. Martikainen, A. Collin, and K.-A. Suominen, {\it Phys. Rev. Lett.} {\bf 88}, 090404 (2002).
\bibitem{Changspin}
D.E. Chang, {\it Phys. Rev. A} {\bf 66}, 025601 (2002).
\bibitem{Ruostekoskispin}
J. Ruostekoski and J.R. Anglin, {\it Phys. Rev. Lett.} {\bf 91}, 190402 (2003).
\bibitem{Savage2}
C.M. Savage and J. Ruostekoski, {\it Phys. Rev.} {\bf A68}, 043604 (2003).
\bibitem{Onsager}
O. Penrose and L. Onsager, {\it Phys. Rev.} {\bf 104}, 576 (1956).
\bibitem{Puspil}
H. Pu, C.K. Law, J.H. Eberly, and N.P. Bigelow, {\it Phys. Rev.} {\bf A59}, 1533 (1999).
\bibitem{Mottonenspil}
M. M\"{o}tt\"{o}nen, T. Mizushima, T. Isoshima, M.M. Salomaa, and K. Machida, {\it Phys. Rev.} {\bf A68}, 023611 (2003).
\bibitem{Saitospil}
H. Saito and M. Ueda, {\it Phys. Rev.} {\bf A69}, 013604 (2004).
\bibitem{Kawaspil}
Y. Kawaguchi and T. Ohmi, {\it Phys. Rev.} {\bf A70}, 043610 (2004).
\bibitem{Shinspil}
Y. Shin, M. Saba, M. Vengalattore, T.A. Pasquini, C. Sanner, A.E. Leanhardt, M. Prentiss, D.E. Pritchard, and W. Ketterle, {\it Phys. Rev. Lett.} {\bf 93}, 160406 (2004).
\bibitem{Feynman}
R.P. Feynman, {\it in Progress in Low Temperature Physics} edited by C.J. Gorter (North-Holland, Amsterdam, 1955) vol.1, Chap. 2.
\bibitem{Baym3}
G. Baym and C.J. Pethick, {\it Phys. Rev. Lett.} {\bf 76}, 6 (1996). 
\bibitem{JacksonAdam}
B. Jackson, J.F. McCann, and C.S. Adams, {\it Phys. Rev.} {\bf A61}, 013604 (1999). 
\bibitem{Riboli}
F. Riboli and M. Modugno, {\it Phys. Rev.} {\bf A65}, 063614 (2002).
\bibitem{Mudrich}
M. Mudrich, S. Kraft, K. Singer, R. Grimm, A. Mosk, and M. Weidemuller, {\it Phys. Rev. Lett.} {\bf 88}, 253001 (2002).
\bibitem{Burke}
J.P. Burke, Jr. J.L. Bohn, B.D. Ersy, and C.H. Greene, {\it Phys. Rev. Lett.} {\bf 80}, 2097 (1998). 
\bibitem{Shchesnovich}
V.S. Shchesnovich, A.M. Kamchatnov, and R.A. Kraenkel, {\it Phys. Rev.} {\bf A69}, 033601 (2004). 
\bibitem{IBloch}
I. Bloch, M. Greiner, O. Mandel, T.W. H\"{a}nsch, and T. Esslinger, {\it Phys. Rev.} {\bf A64}, 021402(R) (2001).
\bibitem{Esry2}
B.D. Esry, C.H. Greene, J.P. Burke, Jr. and J.L. Bohn, {\it Phys. Rev. Lett.} {\bf 78}, 3594 (1998). 
\bibitem{Bashkin}
E.P. Bashkin and A.V. Vagov, {\it Phys. Rev.} {\bf A56}, 6208 (1997). 
\bibitem{Ohberg1}
P. \"{O}hberg and S. Stenholm, {\it Phys. Rev.} {\bf A57}, 1272 (1998). 
\bibitem{Pu}
H. Pu and N.P. Bigelow, {\it Phys. Rev. Lett.} {\bf 80}, 1130 (1998). 
\bibitem{Tim}
E. Timmermans, {\it Phys. Rev. Lett.} {\bf 81}, 5718 (1998). 
\bibitem{Ao}
P. Ao and S.T. Chui, {\it Phys. Rev.} {\bf A58}, 4836 (1998); S.T. Chui and P. Ao, {\it Phys. Rev.} {\bf A59}, 1473 (1999). 
\bibitem{Esry}
B.D. Esry and C.H. Greene, {\it Phys. Rev.} {\bf A59}, 1457 (1999); {\it Nature (London)} {\bf 392}, 434 (1998). 
\bibitem{Ohberg3}
P. Ohberg, {\it Phys. Rev.} {\bf A59}, 634 (1999). 
\bibitem{Trippenbach}
M. Trippenbach, K. G\'{o}ral, K. Rza\'{z}ewski, B. Malomed, and Y.B. Band, {\it J. Phys.} {\bf B33}, 4017 (2000). 
\bibitem{KasamatsuMQT}
K. Kasamatsu, Y. Yasui, and M. Tsubota, {\it Phys. Rev.} {\bf A64}, 053605 (2001). 
\bibitem{Svidzinsky}
A.A. Svidzinsky and S.T. Chui, {\it Phys. Rev.} {\bf A68}, 013612 (2003). 
\bibitem{Modugnopre}
G. Modugno, G. Ferrari, G. Roati, R.J. Brecha, A. Simoni, and M. Inguscio, {\it Science} {\bf 294}, 1320 (2001).
\bibitem{Simoni}
A. Simoni, F. Ferlain, G. Roati, G. Modugno, and M. Inguscio, {\it Phys. Rev. Lett.} {\bf 90}, 163202 (2003).
\bibitem{Erhard}
M. Erhard, H. Schmaljohann, J. Kronj\"{a}ger, K. Bongs, and K. Sengstock, {\it Phys. Rev.} {\bf A69}, 032705 (2004).
\bibitem{Stan}
C. A. Stan, M. W. Zwierlein, C. H. Schunck, S. M. F. Raupach, and W. Ketterle, {\it Phys. Rev. Lett.} {\bf 93}, 143001 (2004). 
\bibitem{Inouye}
S. Inouye, J. Goldwin, M. L. Olsen, C. Ticknor, J. L. Bohn, and D. S. Jin, {\it Phys. Rev. Lett.}, {\bf 93}, 183201 (2004).
\bibitem{Julienne}
P.S. Julienne, F.H. Mies, E. Tiesinga, and C.J. Williams, {\it Phys. Rev. Lett.}, {\bf 78}, 1880 (1997).
\bibitem{Hall2}
D.S. Hall, M.R. Matthews, C.E. Wieman, and E.A. Cornell, {\it Phys. Rev. Lett.} {\bf 81}, 1543 (1998). 
\bibitem{Matthews3}
M.R. Matthews, B.P. Anderson, P.C. Haljan, D.S. Hall, M.J. Holland, J.E. Williams, C.E. Wieman, and E.A. Cornell, {\it Phys. Rev. Lett.} {\bf 83}, 3358 (1999). 
\bibitem{Williams2}
J. Williams, R. Walser, J. Cooper, E. Cornell, and M. Holland, {\it Phys. Rev.} {\bf A59}, R31 (1999), {\bf A61}, 033612 (2000). 
\bibitem{Ohbergjos}
P. \"{O}hberg and S. Stenholm, {\it Phys. Rev.} {\bf A59}, 3890 (1999).
\bibitem{Park}
Q.H. Park and J.H. Eberly, {\it Phys. Rev. Lett.} {\bf 85}, 4195 (2000), {\bf A61}. 
\bibitem{Leggett}
A.J. Leggett, {\it Rev. Mod. Phys.} {\bf 73}, 307 (2001). 
\bibitem{Rejan}
R. Rajaraman, {\it Soliton and Instantons} (North-Holland, Amsterdam, 1989).
\bibitem{Kuklov}
A.B. Kuklov and J.L. Birman, {\it Phys. Rev. Lett.} {\bf 85}, 5488 (2000). 
\bibitem{Sinova2}
J. Sinova, S.M. Girvin, T. Jungwirth, and K. Moon, {\it Phys. Rev.} {\bf B61}, 2749 (2000). 
\bibitem{Son}
D.T. Son, and M.A. Stephanov, {\it Phys. Rev.} {\bf A65}, 063621 (2002). 
\bibitem{Zhang}
Y. Zhang, W.D. Li, L. Li and H.J.W. M\"{u}ller-Kirsten, {\it Phys. Rev.} {\bf A66}, 043622 (2002). 
\bibitem{Davis}
R.L. Davis and E.P.S. Shellard, {\it Phys. Lett.} {\bf B207}, 404 (1988); {\it Nucl. Phys.} {\bf B323}, 209 (1989).
\bibitem{Anderson2}
B.P. Anderson, P.C. Haljan, C.E. Wieman, and E.A. Cornell, {\it Phys. Rev. Lett.} {\bf 85}, 2857 (2000). 
\bibitem{McGee}
S.A. McGee and M.J. Holland, {\it Phys. Rev.} {\bf A63}, 043608 (2001). 
\bibitem{Klausen}
N.N. Klausen, J.L. Bohn, and C.H. Greene, {\it Phys. Rev.} {\bf A64}, 053602 (2001). 
\bibitem{Hooft}
G. 't Hooft, {\it Nucl. Phys.} {\bf B79}, 276 (1974); A.M. Polyakov, {\it JETP Lett.} {\bf 20}, 194 (1974).
\bibitem{Volker}
V. Schweikhard, I. Coddington, P. Engels, S. Tung, and E.A. Cornell, {\it Phys. Rev. Lett.} {\bf 93}, 210403 (2004). 
\bibitem{Ho4}
T.L. Ho, {\it Phys. Rev. Lett.} {\bf 87}, 060403 (2001). 
\bibitem{Baym2}
G. Baym and C.J. Pethick, {\it Phys. Rev.} {\bf A69}, 043619 (2004). 
\bibitem{NRCooper}
N.R. Cooper, S. Komineas, and N. Read, {\it Phys. Rev.} {\bf A70}, 033604 (2004). 
\bibitem{Watanabe}
G. Watanabe, G. Baym and C.J. Pethick, {\it Phys. Rev. Lett.} {\bf 93}, 190401 (2004).
\bibitem{AftalionLLL}
A. Afalion, X. Blanc, and J. Dalibard, {\it Phys. Rev. A} {\bf 71}, 023611 (2005). 
\bibitem{elliptic}
K. Chandrasekharan, {\it Elliptic Fuctions} (Springer-Verlag, New York, 1985), Chap. IV and V.
\bibitem{Crecipe}
W.H. Press {\it et al., Numerical Recipes in C} (Cambridge University Press, Cambridge, 1988). 
\bibitem{sheethe}
\"{U}. Parts, E.V. Thuneberg, G.E. Volovik, J.H. Koivuniemi, V.M.H. Ruutu, M. Heinil\"{a}, J.M. Karim\"{a}ki, and M. Krusius, {\it Phys. Rev. Lett.} {\bf 72}, 3839 (1994). 
\bibitem{Zhai2}
H. Zhai, Q. Zhou, R. L\"{u} and L. Chang, {\it Phys. Rev.} {\bf A69}, 063609 (2004). 
\bibitem{Truscott}
A.G. Truscott, K.E. Strecker, W.I. McAlexander, G.B. Partridge, and R.G. Hulet, {\it Nature(London)} {\bf 291}, 2570 (2001). 
\bibitem{Schreck}
F. Schreck, L. Khaykovich, K.L. Corwin, G. Ferrari, T. Bourdel, J. Cubizolles, and C. Salomon, {\it Phys. Rev. Lett.} {\bf 87}, 080403 (2001).
\bibitem{Hadzibabic}
Z. Hadzibabic, C.A. Stan, K. Dieckmann, S. Gupta, M.W. Zwierlein, A. G\"{o}rlitz, and W. Ketterle, {\it Phys. Rev. Lett.} {\bf 88}, 160401 (2002).
\bibitem{Roatimix}
G. Roati, F. Riboli, G. Modugno, M. Ingusio, {\it Phys. Rev. Lett.} {\bf 89}, 150403 (2002).
\bibitem{Modugnomix}
G. Modugno, G. Roati, F. Riboli, F. Ferlaino, R.J. Brecha, M. Ingusio, {\it Science} {\bf 297}, 2240 (2002).
\bibitem{Goldwin}
J. Goldwin, S. Inouye, M.L. Olsen, B. Newman, B.D. DePaola, and D.S. Jin, {\it Phys. Rev.} {\bf A70}, 021601 (2004).
\bibitem{Edward}
D.O. Edward and M. Pettersen, J. Low Temp. Phys. {\bf 87}, 473 (1992). 
\bibitem{StanFesh}
C.A. Stan, M.W. Zwierlein, C.H. Schunck, S.M.F. Raupach, and W. Ketterle, {\it Phys. Rev. Lett.} {\bf 93}, 143001 (2004).
\bibitem{InouyeFesh}
S. Inouye, J. Goldwin, M.L. Olsen, C. Ticknor, J.L. Bohn, and D.S. Jin, {\it Phys. Rev. Lett.} {\bf 93}, 183201 (2004).
\bibitem{Jezekmix}
D.M. Jezek, M. Barranco, M. Guilleumas, R. Mayol, and M. Pi, {\it Phys. Rev.} {\bf A70}, 043630 (2004). 
\bibitem{Koashi}
M. Koashi and M. Ueda, {\it Phys. Rev. Lett.} {\bf 84}, 1066 (2000). 
\bibitem{Ciobanu}
C.V. Ciobanu, S.K. Yip, amd T.L. Ho, {\it Phys. Rev.} {\bf A61}, 033607 (2000). 
\bibitem{Ueda}
M. Ueda and M. Koashi, {\it Phys. Rev.} {\bf A65}, 063602 (2002). 
\bibitem{Mottonen}
M. M\"{o}tt\"{o}nen, N. Matsumoto, M. Nakahara, and T. Ohmi, {\it J. Phys. Condens. Matter} {\bf 14}, 13481 (2002).
\bibitem{Kawaguchi}
Y. Kawaguchi, M. Nakahara, and T. Ohmi, {\it Phys. Rev.} {\bf A70}, 043605 (2004). 
\end{thebibliography}
\end{document}